%% file: main.tex
\documentclass[journal ]{new-aiaa}
\usepackage[utf8]{inputenc}
\usepackage{textcomp}

\usepackage{pgfplots}
\pgfplotsset{width=10cm,compat=1.9}
\usepgfplotslibrary{external}
\usetikzlibrary{external}
\usepackage{graphicx}
\usepackage{caption}
\usepackage{subcaption}
\usepackage{amsmath}
\usepackage[version=4]{mhchem}
\usepackage{siunitx}
\usepackage{longtable,tabularx}
\usepackage{pgffor}        % \foreach loops (from TikZ/PGF)
\usepackage{titlesec}
\renewcommand{\theparagraph}{\roman{paragraph}.)}

\titleformat{\paragraph}[hang]
  {\normalfont\normalsize\itshape}
  {\theparagraph~#1.} % Note the space after numbering
  {0pt} % No additional space
  {} % No additional formatting needed (title is included automatically here)

\titlespacing*{\paragraph}{0pt}{1ex plus 0.5ex minus 0.2ex}{1em}

\title{Computational and Experimental Comparison of CLF5605 and roamx-0201 Martian Helicopter Rotor Airfoils}

\author{Lidia Caros\footnote{Research Associate, Department of Aeronautics, Imperial College London, lidia.caros-roca19@alumni.imperial.ac.uk.}, 
Witold J. F. Koning\footnote{ROAMX Computational Lead, Analytical Mechanics Associates, witold.koning@nasa.gov.}, 
Takayuki Nagata\footnote{Assistant Professor, Department of Aerospace Engineering, Nagoya University, takayuki.nagata@mae.nagoya-u.ac.jp.}, 
Keisuke Asai\footnote{Professor Emeritus, Department of Aerospace Engineering, Tohoku University, asai@aero.mech.tohoku.ac.jp.},
Oliver Buxton\footnote{Professor, Department of Aeronautics, Imperial College London, o.buxton@imperial.ac.uk.},
Natalia Perez Perez\footnote{ROAMX Experimental Lead, Analytical Mechanics Associates,  nataliaperez@alumni.stanford.edu.},  \\
Ethan A. Romander\footnote{Aerospace Engineer, Aeromechanics Branch, NASA Ames Research Center, ethan.romander@nasa.gov.},
Taku Nonomura\footnote{Professor, Department of Aerospace Engineering, Nagoya University, nonomura@nagoya-u.jp.},
Haley V. Cummings\footnote{ROAMX Principal Investigator, Aeromechanics Branch, NASA Ames Research Center, haley.cummings@nasa.gov.}, 
Peter Vincent\footnote{Professor, Department of Aeronautics, Imperial College London, p.vincent@imperial.ac.uk.}
}
\affil{Imperial College London, London, England, UK}
\affil{NASA Ames Research Center, Moffett Field, California, USA}
\affil{Analytical Mechanics Associates, Hampton, Virginia, USA}
\affil{Tohoku University, Sendai, Miyagi, Japan}

\begin{document}

\maketitle

\begin{abstract}
This study compares aerodynamic performance of the CLF5605 rotor airfoil --- which flew on Ingenuity from 2021 to 2024 --- with that of a new optimized roamx-0201 airfoil designed for Martian conditions at NASA Ames. Specifically, performance is studied at a Reynolds number of $20,000$ and a Mach number of $0.60$, across a range of angles of attack, using three independent state-of-the-art methodologies: implicit large eddy simulations (ILES) using NASA's OVERFLOW solver, direct numerical simulations (DNS) using the high-order GPU-accelerated PyFR solver, and experimental testing in the Mars Wind Tunnel at Tohoku University. Discrepancies between results obtain using the various methodologies are analyzed and explained. Across all methodologies it can be seen that the roamx-0201 airfoil is able to achieve a given lift with less drag compared to the CLF5605 airfoil. Moreover, OVERFLOW and PyFR results show that the roamx-0201 airfoil has superior stall characteristics, and can achieve a maximum lift $\approx 20\%$ higher than that achieved by the CLF5605 airfoil. The work provides a strong body of evidence to support further studies into use of rotors based on the optimized roamx-0201 airfoil for future Mars helicopter missions.
\end{abstract}

\section{Introduction}

In April 2021 NASA successfully flew the Ingenuity helicopter on Mars \cite{Marshelicoptertechnologydemonstrator}, marking a significant milestone in planetary exploration. Ingenuity went on to complete a further 71 flights, unequivocally demonstrating that powered flight on the red planet was possible, and paving the way for future helicopter-based exploration of Mars. Consequently, NASA has now begun scoping a next-generation of Martian helicopters --- such as the Mars Science Helicopter \cite{johnson2020mars}, and the Mars Chopper concept \cite{withrow2025larger} --- with increased range, flight duration, and payload capacity compared to Ingenuity, such that they could explore further and faster, and carry scientific instruments and/or surface samples, enabling a wide range of currently intractable science and discovery missions.

A key determinant of the overall performance of a Martian helicopter is the aerodynamic efficiency of the rotor blades, which in turn depends significantly on the rotor blade airfoil profiles that are employed. The atmospheric conditions on Mars --- and in particular the low atmospheric density --- mean rotors must operate in a low-Reynolds number compressible regime for which terrestrial rotor blades have not been designed \cite{koning2019improved}. This has led to interest in a range of novel airfoil concepts \cite{koning2018generation,koning2019lowRe}, including corrugated airfoils \cite{koning2024elisa, herrero2021effects} inspired by dragonfly wing profiles \cite{okamoto1996aerodynamic}, polygonal airfoils including triangular airfoils \cite{Suwa2012,munday2015nonlinear,caros2021,caros2023,caros2024}, and thin cambered plates \cite{koning2020optimization,koning2024elisa}. One such design is the optimized roamx-0201 airfoil \cite{koning2025roamxgeom,koning2024elisa}, developed for Martian conditions of $Re=20,000$ and $Ma=0.60$ using the Evolutionary aLgorithm for Iterative Studies of Aeromechanics (ELISA) \cite{koning2024elisa}, as part of the Rotor Advancement for Mars eXploration (ROAMX) project \cite{cummings2022roamxintro}.

This study will compare aerodynamic performance of the CLF5605 rotor airfoil shown in Fig. \ref{fig:geometries}(a) \cite{doi:10.2514/6.2019-0620,koning2024ingenuitygeom} --- which flew on Ingenuity from 2021 to 2024 --- with that of the optimized roamx-0201 airfoil shown in Fig. \ref{fig:geometries}(b), at a chord-based $Re=20,000$ and $Ma=0.60$ across a range of angles of attack $\alpha$. Specifically, three independent state-of-the-art methodologies will be employed: Implicit Large Eddy Simulations (ILES) using NASA's OVERFLOW solver \cite{nichols,doi:10.2514/6.2011-3851}, Direct Numerical Simulations (DNS) using the high-order GPU-accelerated PyFR solver \cite{pyfr,WITHERDEN2025109567}, and experimental testing in the Mars Wind Tunnel (MWT) at Tohoku University \cite{anyoji2011developmentA}. Section \ref{compmeth} details the computational methodologies, section \ref{expmeth} details the experimental methodologies, section \ref{results} presents the results and discussion, and finally section \ref{conc} draws conclusions.

\section{Computational Methodologies}
\label{compmeth}

\input{Figures-tex/geometries}

\subsection{Solvers and Numerical Methods}

OVERFLOW \cite{nichols,doi:10.2514/6.2011-3851} is a compressible Navier--Stokes based on the finite difference method. For the present study OVERFLOW 2.3d was used. Inviscid fluxes were computed using an improved Harten, Lax, and van Leer flux scheme (HLLE++) with a fifth-order WENOM upwind reconstruction \cite{doi:10.2514/1.36849} and viscous fluxes were computed using second-order central differencing. Time was advanced using a second-order backward differencing scheme (BDF2), with the dual time-stepping approach as described in \cite{doi:10.2514/6.1993-3360,doi:10.2514/6.1995-78}. ILES with OVERFLOW were undertaken using CPU nodes on the Pleiades supercomputer provided by the NASA High-End Computing Program through the NASA Advanced Supercomputing Division at Ames Research Center.

PyFR \cite{pyfr,WITHERDEN2025109567} is an open-source computational fluid dynamics solver based on the high-order Flux Reconstruction approach of Huynh \cite{huynh_reconstruction_2009}. For the present study PyFR v1.15.0 and v2.0.3 were used. Fourth-order polynomials were used to represent the solution within each element of the mesh, thus nominally achieving fifth-order accuracy in space. A Rusanov Riemann solver was used to calculate the inter-element inviscid fluxes, the Local Discontinuous Galerkin approach was used to calculate the inter-element viscous fluxes, an explicit RK45 scheme \cite{KENNEDY2000177} was employed to advance the solution in time, anti-aliasing was employed to stabilize the simulations, and all runs were performed using double precision arithmetic. DNS with PyFR were undertaken using Nvidia P100 GPUs on Piz Daint and Nvidia GH200 GPUs on Alps at the Swiss National Supercomputing Centre (CSCS) under project s1271.

Both OVERFLOW and PyFR have previously been validated in a low-Reynolds number compressible regime, and used to simulate flow over Martian rotor blade airfoil concepts \cite{caros2021,caros2023,caros2024,koning2019lowRe,koning2020optimization,doi:10.2514/1.C037023}.

\subsection{Governing Equations}

OVERFLOW and PyFR were used to solve the compressible Navier--Stokes equations for a calorically perfect gas with constant viscosity, which can be written in 3D as
    \begin{equation}
        \frac{\partial \mathbf{u}}{\partial t} + \boldsymbol{\nabla} \cdot \mathbf{f} = 0,
    \end{equation}
where $\mathbf{u} = \mathbf{u}(\mathbf{x},t) = (\rho,\rho v_x, \rho v_y, \rho v_z, E)$ is the solution with $\rho$ the fluid density, $\mathbf{v} = (v_x,v_y,v_z)$ the fluid velocity and $E$ the total energy per unit volume, and where $\mathbf{f} = \mathbf{f(u, \boldsymbol{\nabla}u)} = \mathbf{f^i-f^v}$ is the flux, with $\mathbf{f^i}$ the inviscid flux given by
\begin{equation}
    \mathbf{f^i} = 
    \begin{Bmatrix}
        \rho v_x    &    \rho v_y    &   \rho v_z \\
        \rho v_x^2 + p   &    \rho v_y v_x   &   \rho v_z v_x \\
        \rho v_x v_y     &    \rho v_y^2 + p
        &   \rho v_z v_y \\
        \rho v_x v_z      &    \rho v_y v_z   & \rho v_z^2 + p \\
        v_x (E+p)        &    v_y (E+p)   &   v_z (E+p)
   \end{Bmatrix},
\end{equation}
where $p$ is the pressure, which for a calorically perfect gas is given by
\begin{equation} \label{eq:idealgas}
    p = (\gamma-1)\left(E - \frac{1}{2}\rho\|\mathbf{v}\|^2 \right),
\end{equation}
with $\gamma = c_p/c_v$, where $c_p$ and $c_v$ are specific heat capacities at constant pressure and volume, respectively, and $\mathbf{f^v}$ is the viscous flux given by
\begin{equation}
    \mathbf{f^v} = 
     \begin{Bmatrix}
    0   &   0   &  0 \\
    \mathscr{T}_{xx} & \mathscr{T}_{yx} & \mathscr{T}_{zx} \\
    \mathscr{T}_{xy} & \mathscr{T}_{yy} & \mathscr{T}_{zy} \\
    \mathscr{T}_{xz} & \mathscr{T}_{yz} & \mathscr{T}_{zz} \\
    v_i \mathscr{T}_{ix} + \kappa \partial_x T
    &
    v_i \mathscr{T}_{iy} + \kappa \partial_y T
    &
    v_i \mathscr{T}_{iz} + \kappa \partial_z T
   \end{Bmatrix},
\end{equation}
where the stress-energy tensor is given by
\begin{equation}
    \mathscr{T}_{ij} = \mu (\partial_i v_j + \partial_j v_i) - \frac{2}{3}\mu \delta_{ij} \boldsymbol{\nabla}	\cdot \mathbf{v},
\end{equation}
with $\mu$ the dynamic viscosity, $\kappa = \mu c_p/Pr$, with $Pr$ the Prandtl number, and $T$ is the temperature, which for a calorically perfect gas is given by
\begin{equation}
T = \frac{1}{c_v}\frac{1}{\gamma -1}\frac{p}{\rho}.
\end{equation}
On Mars $\gamma=1.29$ and $Pr=1.18$. However, in this study values of $\gamma=1.4$ and $Pr=0.72$ were employed in order to match the experimental conditions in the Tohoku University MWT.

\subsection{Two-dimensional (2D) Simulations}

\subsubsection{Domain}

Both OVERFLOW and PyFR were used to undertake 2D simulations of flow over the CLF5605 and roamx-0201 airfoils. Figs.~\ref{fig:OVERFLOW-2D-domain} and \ref{fig:PyFR-2D-domain} show 2D domains for OVERFLOW and PyFR simulations, respectively. In all cases, each airfoil is defined to have a chord of $1.0$. 

For OVERFLOW simulations $\alpha$ was changed by varying the freestream inflow angle. The origin in the streamwise-vertical $xy$-plane was located at the airfoil quarter chord position. The dimensions of the computational domain for OVERFLOW simulations were $x \in [-205,205]$ in the streamwise direction and $y \in [-205,205]$ in the vertical direction. For PyFR simulations $\alpha$ was changed by rotating each airfoil about its trailing edge. The origin in the streamwise-vertical $xy$-plane was located at the leading edge of each airfoil when $\alpha = 0^{\circ}$. The dimensions of the computational domain were $x \in [-20,46]$ in the streamwise direction and $y \in [-20,20]$ in the vertical direction.

\subsubsection{Meshes}

For OVERFLOW simulations, overset structured meshes were used, consisting of near-body and off-body components. Specifically, a near body O-mesh was overset with cartesian blocks in the wake, extending a distance of 5 downstream, and a coarse cartesian background mesh. Fig.~\ref{fig:OVERFLOW-2D-domain} shows meshes used for the CLF5605 and roamx-0201 airfoils, with each having $\sim 3\times 10^5$ mesh points.

For PyFR simulations, meshes were comprised of structured and unstructured regions. Specifically, structured quadrilateral boundary layer meshes were located adjacent to the airfoil upper and lower surfaces, extending a distance of $0.1$ normally from the surface into the domain. Around the leading and trailing edges, a refined unstructured mesh was located adjacent to the surface, extending a distance of $0.1$ normally from the surface into the domain. Refined regions were extended a distance of $0.5$ normally into the domain on the upper surface to provide enough resolution for the separated flow at the highest angles of attack. An unstructured refined region was also located in the wake region, extending a distance of $6$ downstream of the airfoil. The remainder of the domain was then tessellated with an unstructured mix of quadrilaterals and triangles. Fig.~\ref{fig:PyFR-2D-domain} shows meshes used for the CLF5605 and roamx-0201 airfoils at $\alpha = 6^{\circ}$, which each had $\sim 1$--$2\times 10^4$ elements, and $\sim$ 4--$5\times 10^5$ degrees of freedom per equation in total when fourth-order solution polynomials were used to represent the solution with each element. 

Mesh independence was demonstrated for both OVERFLOW and PyFR (see \hyperref[Appendix]{Appendix A}).

\input{Figures-tex/meshes}

\subsubsection{Boundary Conditions}

For both OVERFLOW and PyFR, a characteristic boundary condition was applied at the inflow, outflow, bottom, and top planes in order to achieve freestream flow in the $x$-direction at the inflow plane with a chord-based $Re=20,000$ and $Ma=0.60$ for all simulations. Also, an impermeable no-slip adiabatic boundary condition was prescribed at the airfoil surface. Finally, for PyFR simulations a sponge region was applied adjacent to the outlet of the domain to prevent spurious reflections traveling upstream.

\subsubsection{Startup Process and Data Extraction}

Both OVERFLOW and PyFR simulations were initiated at $t=0$ with spatially constant flow in the $x$-direction. Simulations were then advanced a period $t_t$ to remove initial transients and then for a further $t_e$, henceforth referred to as the Data Extraction Period during which data was extracted for analysis. For OVERFLOW $t_t=620t_c$, $t_e=60t_c$ and for PyFR $t_t=100t_c$, $t_e=100t_c$ with in all cases $t_c= 1/v_\infty$ where $v_\infty$ is the free-stream velocity magnitude at the inflow plane.

\subsection{Three-dimensional Spanwise Periodic (3D-SP) Simulations}

\subsubsection{Domain}

Both OVERFLOW and PyFR were used to undertake 3D-SP simulations of flow around each airfoil profile. Domains for each simulation were obtained by extruding the respective 2D domain a distance $0.5$ in the spanwise direction. Spanwise domain size independence was demonstrated for both OVERFLOW and PyFR (see \hyperref[span-indep]{Appendix B}). 

\subsubsection{Meshes}

For OVERFLOW, 3D-SP meshes were obtained by extruding 2D meshes, similar to those used for each respective 2D simulation, in the spanwise direction. These 3D-SP meshes had $\ 62.8\times 10^6$ mesh points for the CLF5605 and $\ 62.4\times 10^6$ mesh points for the roamx-0201 airfoil.

For PyFR, 3D-SP meshes were obtained by extruding the respective 2D meshes in the spanwise direction, generating hexahedra and triangular prisms throughout the domain. These 3D-SP meshes had $\sim$ 0.8--$1 \times 10^6$ elements, and $\sim 0.1\times 10^9$ degrees of freedom per equation in total when fourth-order solution polynomials were used to represent the solution with each element. For the $\alpha = 6^{\circ}$ case, it was verified a posteriori that $3.5 \eta > \Delta$ in the wake for 6 chords downstream of the airfoil (see \hyperref[DNS-resolution]{Appendix C}). Where,
\begin{equation}\label{eq:pointspace}
\Delta = \frac{\sqrt[3]{V}}{P+1}
\end{equation}
is an estimate of the local solution point spacing, with $V$ the local element volume and $P$ the polynomial order used to represent the solution within each element of the mesh, and
\begin{equation}\label{eq:kolmogorov}
   \eta = \bigg(\frac{\nu^3}{\varepsilon} \bigg)^{1/4}
\end{equation}
is the Kolmogorov length scale, with $\nu$ the kinematic viscosity and
\begin{equation}
    \varepsilon = 2\nu S_{ij} S_{ij}
\end{equation} is the dissipation rate, where $S_{ij}$ is the fluctuating rate-of-strain tensor. This is within the threshold required to achieve DNS resolution, given the simulations are nominally fifth-order accurate in space \cite{soton354999}.

\subsubsection{Boundary Conditions}

For both OVERFLOW and PyFR, a characteristic boundary condition was applied at the inflow, outflow, bottom, and top planes in order to achieve freestream flow in the $x$-direction at the inflow plane with a chord-based $Re=20,000$ and $Ma=0.60$ for all simulations. Also, an impermeable no-slip adiabatic boundary condition was prescribed at the airfoil surface, and periodic boundary conditions were applied in the span-wise direction.

\subsubsection{Startup Process and Data Extraction}

Both OVERFLOW and PyFR simulations were initiated and advanced in analogous way to the 2D simulations, with for OVERFLOW $t_t=532.5t_c$, $t_e=60t_c$ and for PyFR $t_t=90t_c$, $t_e=60t_c$.

\subsection{Virtual Wind Tunnel (VWT) Simulations}

\subsubsection{Domain}

PyFR was used to undertake VWT simulations of flow around roamx-0201 at two angles of attack, $\alpha = 4.5^{\circ}$ and $\alpha = 6^{\circ}$, in the entire working section of the MWT, thus accounting for blockage effects, and the interaction of the side-wall boundary layer with the extruded airfoil. As per the 2D and 3D-SP simulations, the airfoil was defined to have a chord of $1.0$. The origin in the streamwise-vertical $xy$-plane was located at the leading edge of each airfoil when $\alpha = 0^{\circ}$. When changing the angle of attack, the airfoil was rotated about a point at 25$\%$ chord to agree with the experimental setup. The dimensions of the computational domain were $x \in [-3.75,4.25]$ in the streamwise direction, $y \in [-1.496,1.504]$ in the vertical direction at the inlet, and $z \in [0,2]$ in the span-wise direction, with the top and bottom walls inclined at an angle of $1.3^{\circ}$ to the streamwise direction. This configuration matches the working section of the Tohoku University MWT, in which the experiments were undertaken.

\subsubsection{Meshes}

\begin{figure}[h!]
  \centering
        \begin{subfigure}[t]{0.98\linewidth}
        \includegraphics[width=\linewidth]{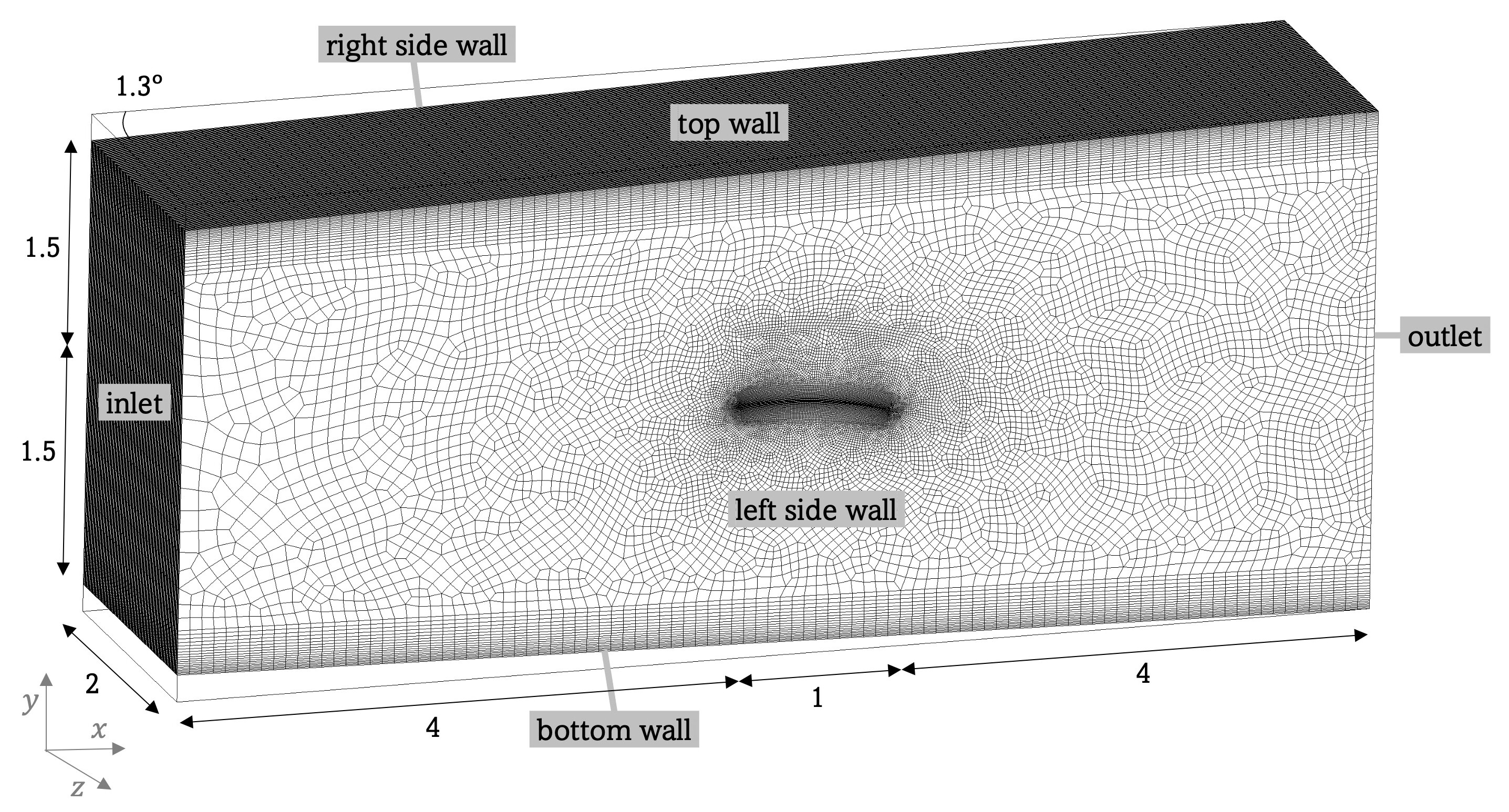}
        \end{subfigure}
        \caption{View of the computational mesh used for the virtual wind tunnel (VWT) with PyFR for roamx-0201 at $\alpha = 6^{\circ}$. The origin in the streamwise-vertical $xy$-plane is located at the leading edge of the airfoil when $\alpha = 0^{\circ}$.}
        \label{fig:PyFR-VWT-domain}
\end{figure}

The virtual wind tunnel meshes were obtained by extruding meshes analogous to the 2D streamwise-vertical $xy$-plane meshes in the span-wise $z$ direction. The streamwise-vertical $xy$-plane meshes were comprised of structured and unstructured regions. Specifically, structured quadrilateral boundary layer meshes were located adjacent to the airfoil upper and lower surfaces, extending a distance of $0.1$ normally from the surface into the domain. Around the leading and trailing edges, a refined unstructured mesh was located adjacent to the surface, extending a distance of $0.1$ normally from the surface into the domain. Refined regions were extended a distance of $0.5$ normally into the domain on the upper surface to provide enough resolution for the separated flow at the highest angles of attack. Structured quadrilateral meshes were also located adjacent to the top and bottom walls, extending a distance of $0.3$ normally into the domain. The remainder of the streamwise-vertical $xy$-plane meshes are tessellated with an unstructured mix of quadrilaterals and triangles, with a refinement in the wake of the airfoil. Extrusion of the 2D meshes in the span-wise $z$ direction generates hexahedra and triangular prisms throughout the domain. Fig.~\ref{fig:PyFR-VWT-domain} shows the external surface of the mesh for roamx-0201 at $\alpha = 6^{\circ}$. Meshes had $\sim 5 \times 10^6$ elements, and $\sim 625\times 10^6$ degrees of freedom per equation in total when fourth-order solution polynomials are used to represent the solution within each element.

Finally, we note that mesh cells adjacent to the wind tunnel walls are sized such that $\delta_\nu > \Delta $, where $\delta_\nu$ is the local diffusive length scale of a Blasius boundary layer solution. 

\subsubsection{Boundary Conditions}

A fixed uniform total pressure and total temperature was applied at the inflow plane, and a characteristic boundary condition was applied at the outflow plane, in order to achieve freestream flow in the $x$-direction at the inflow plane with a chord-based $Re=20,000$ and $Ma=0.60$. Also, an impermeable no-slip adiabatic boundary condition was prescribed at the airfoil surface and the wind tunnel walls. Note that no boundary layer and no turbulence was injected at the inflow. 

\subsubsection{Startup Process and Data Extraction}

Simulations were initiated and advanced in analogous way to the 2D and 3D-SP simulations, with $t_t=80t_c$ and $t_e=20t_c$. 

\subsection{Extracted Quantities}

Instantaneous lift and drag coefficients were extracted, defined as
 \begin{equation}\label{eq:coefficients}
       C_L = \frac{F_L}{q_\infty}\;, \hspace{1cm} C_D = \frac{F_D}{q_\infty}\;,
        \end{equation}
respectively, where $F_L$ and $F_D$ are the sum of the instantaneous pressure and viscous forces on the airfoil per unit area in the $y$ and $x$ directions, respectively, and $q_{\infty}$ was obtained via
\begin{equation}\label{eq:qinf}
        q_{\infty} = \frac{1}{2}\rho_\infty v_\infty^2
 \end{equation}
where $\rho_\infty$ is the free-stream density. From these, time-averaged lift and drag coefficients $\overline{C}_L$ and $\overline{C}_D$, could be obtained by time-averaging $C_L$ and $C_D$, respectively, over the Data Extraction Period. Additionally, time-averaged pressure coefficients were extracted, defined as \begin{equation}
    \overline{C}_p = \frac{\overline{p}-p_{\infty}}{q_{\infty}}
\end{equation}
where $\overline{p}$ is the obtained by time-averaging $p$ over the Data Extraction Period. Finally, instantaneous snapshots of density gradient magnitude $|\boldsymbol{\nabla}\rho|$ in the $xy$-plane, and Q-criterion iso-surfaces in the volume, were extracted.

\section{Experimental Methodology}
\label{expmeth}

\subsection{Mars Wind Tunnel}

\begin{figure}[hbt!]
\centering
\includegraphics[]{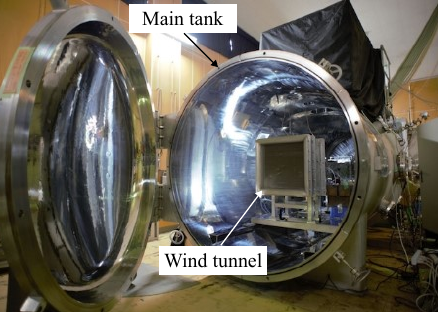}
\caption{The MWT at Tohoku University, Japan. The facility is now located at Nagoya University, Japan.}
\label{fig:mwt}
\end{figure}

The Tohoku University MWT \cite{anyoji2011development,anyoji2011developmentA} was used for the experimental measurements. This wind tunnel is housed in a vacuum chamber, as shown in Fig.~\ref{fig:mwt}, and hence experiments can be conducted under low-pressure conditions. The wind tunnel is driven by an ejector system \cite{anyoji2021supersonic}, due to the inability of a fan to generate high-subsonic flow under low-pressure conditions. The ejector system is located downstream of the test section and induces a flow in the wind tunnel via supersonic jets. The pressure in the chamber increases due to the operation of the ejector. Therefore, an appropriate amount of gas is exhausted into a separately installed buffer tank, and the total pressure inside the wind tunnel can be maintained for a finite period. The amount of exhaust gas is adjusted by a PID-controlled butterfly valve installed in the flow path connecting the main tank and buffer tank. The test section of the wind tunnel is 100~mm wide and 150~mm high.

Various fundamental studies of low-Reynolds-number compressible flow over an airfoil \cite{anyoji2015effects,munday2015nonlinear,nagata2025compressibility}, plate \cite{mangeol2017compressibility,kusama2021investigation}, and circular cylinder \cite{nagata2020experimental,nagata2025schlieren} have been conducted in the MWT. Surface flow visualization for a rotor blade, using the pressure-sensitive paint (PSP) and sublimation method, has also been conducted \cite{nagata2022visualization}. In addition, the experimental report and data for the CLF5605 and roamx-0201 airfoils is presented in \cite{koning2024experimental}.

\subsection{CLF5605 Models}

\begin{figure}[h!]
  \centering
        \begin{subfigure}[t]{0.33\linewidth}
        \includegraphics[width=\linewidth]{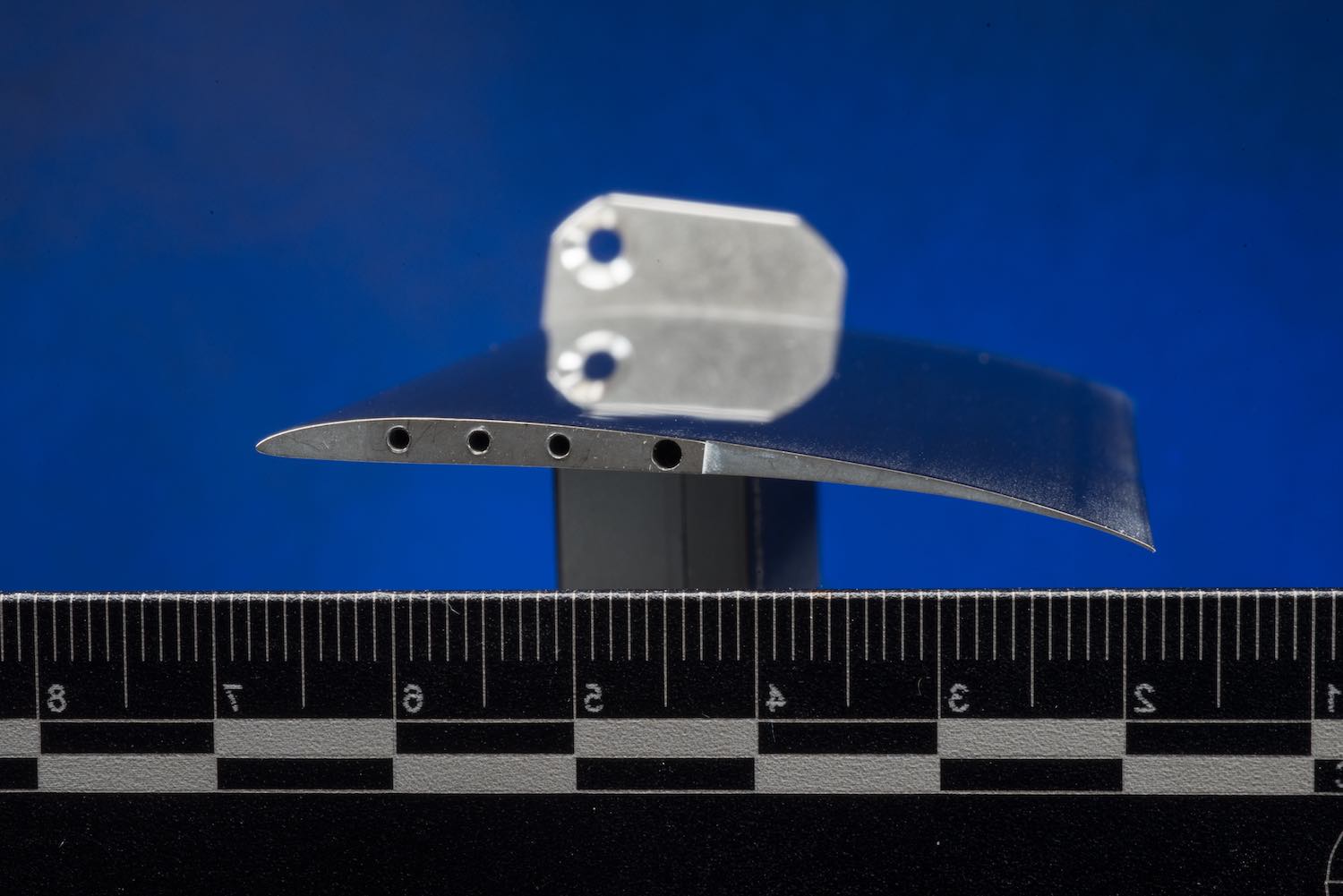}
        \caption{clf5605-us-fp}
        \end{subfigure}
        \begin{subfigure}[t]{0.33\linewidth}
        \includegraphics[width=\linewidth]{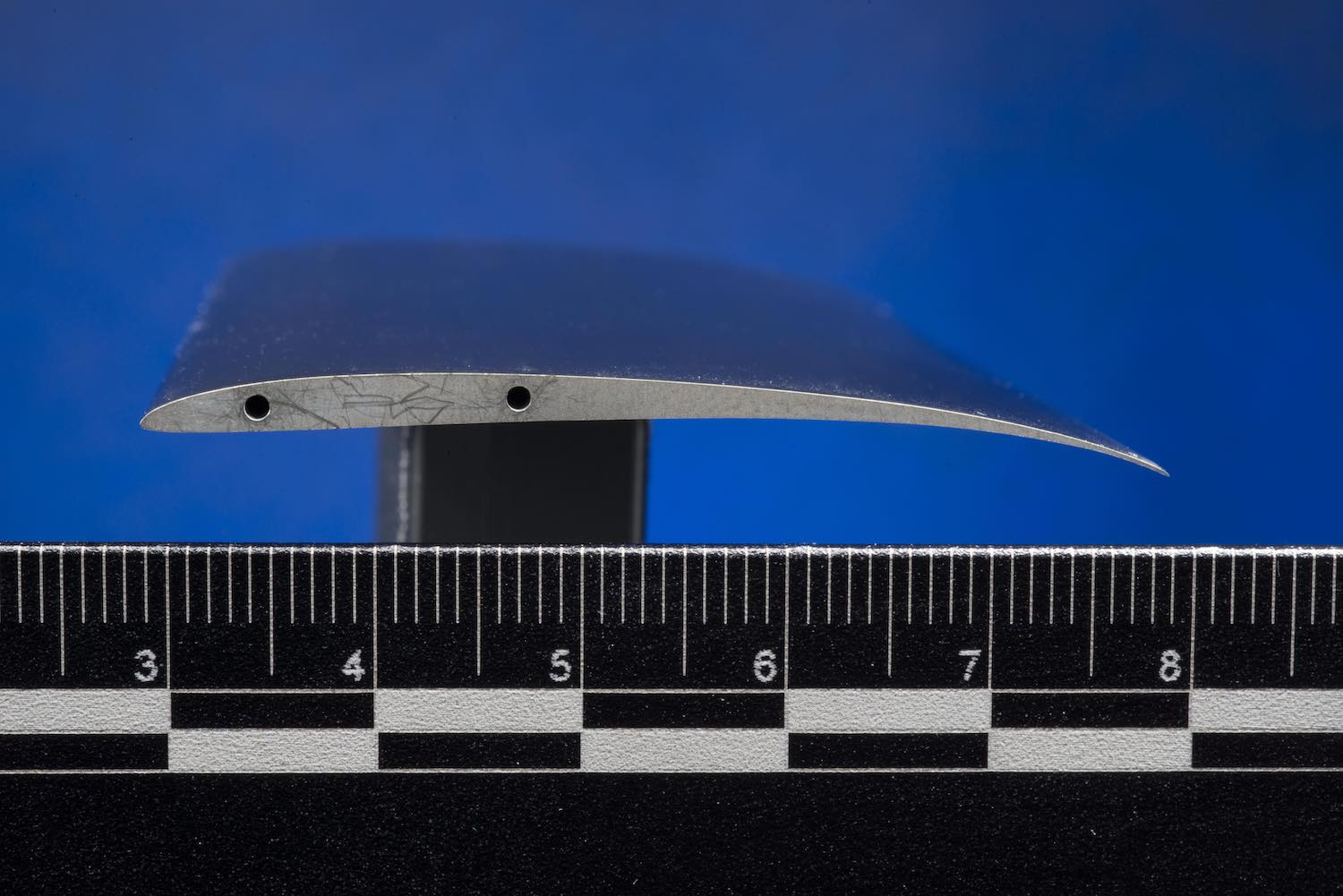}
        \caption{clf5605-us-s}
        \end{subfigure}
        \begin{subfigure}[t]{0.33\linewidth}
        \includegraphics[width=\linewidth]{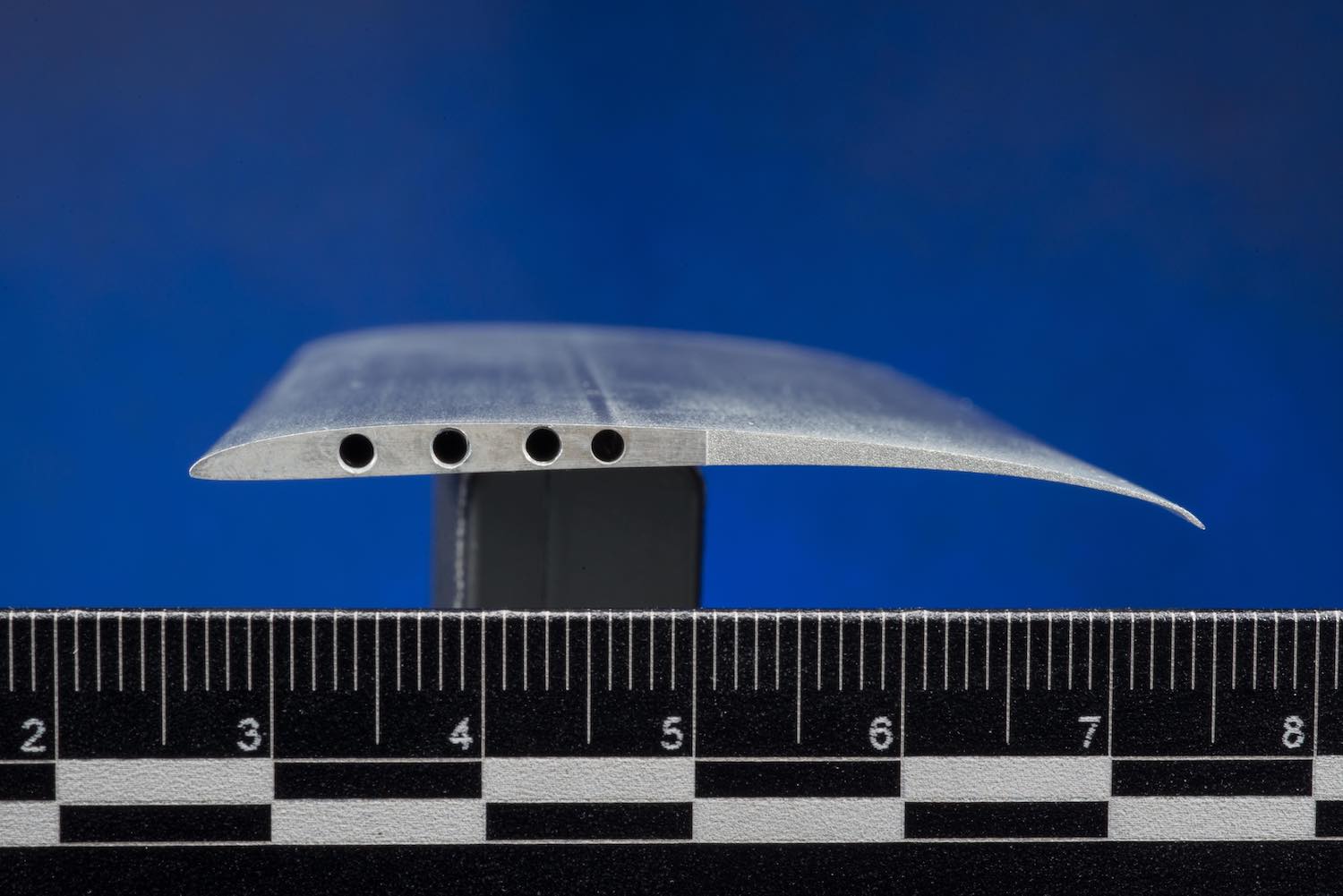}
        \caption{clf5605-jp-f}
        \end{subfigure}
        \caption{Images of CLF5605 models.}
        \label{fig:models-clf5605}
\end{figure}

Three models of the CLF5605 profile were produced, as shown in Fig.~\ref{fig:models-clf5605}. Models clf5605-us-s and clf5605-us-fp were fabricated in the United States from AISI 4130 using Electrical Discharge Machining (EDM) and used for schlieren measurements, force and PSP measurements, respectively, and model clf5605-jp-f was fabricated in Japan from A5052 using an EDM method and used for force measurements. We note that the trailing edge of each model was slightly thicker than that of the target profile due to manufacturing limitations, and there were also differences between the trailing edge profiles between models.

\subsection{roamx-0201 Models}

\begin{figure}[h!]
  \centering
        \begin{subfigure}[t]{0.33\linewidth}
        \includegraphics[width=\linewidth]{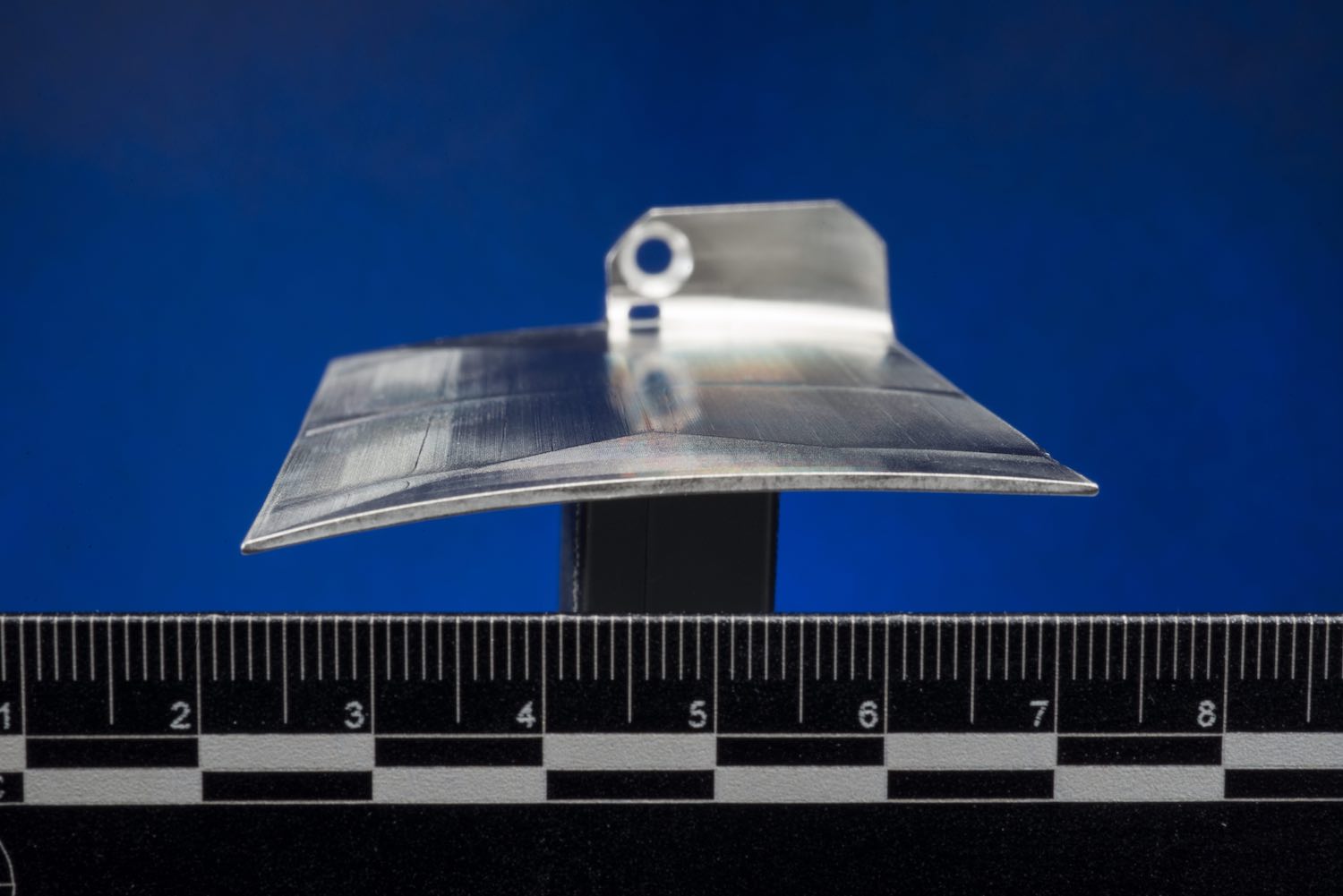}
        \caption{roamx-0201-us-f}
        \end{subfigure}
        \begin{subfigure}[t]{0.33\linewidth}
        \includegraphics[width=\linewidth]{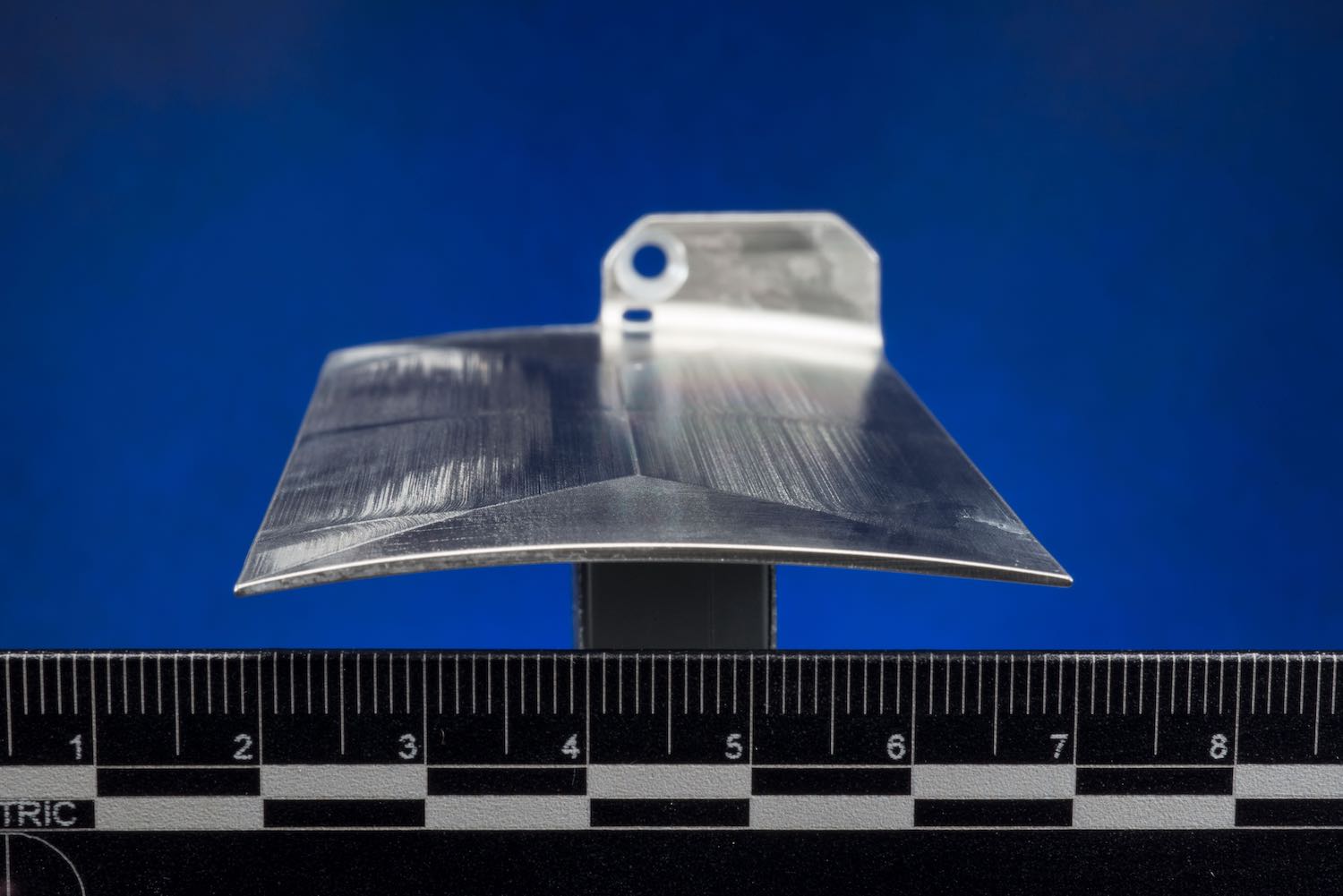}
        \caption{roamx-0201-us-pl}
        \end{subfigure}
        \begin{subfigure}[t]{0.33\linewidth}
        \includegraphics[width=\linewidth]{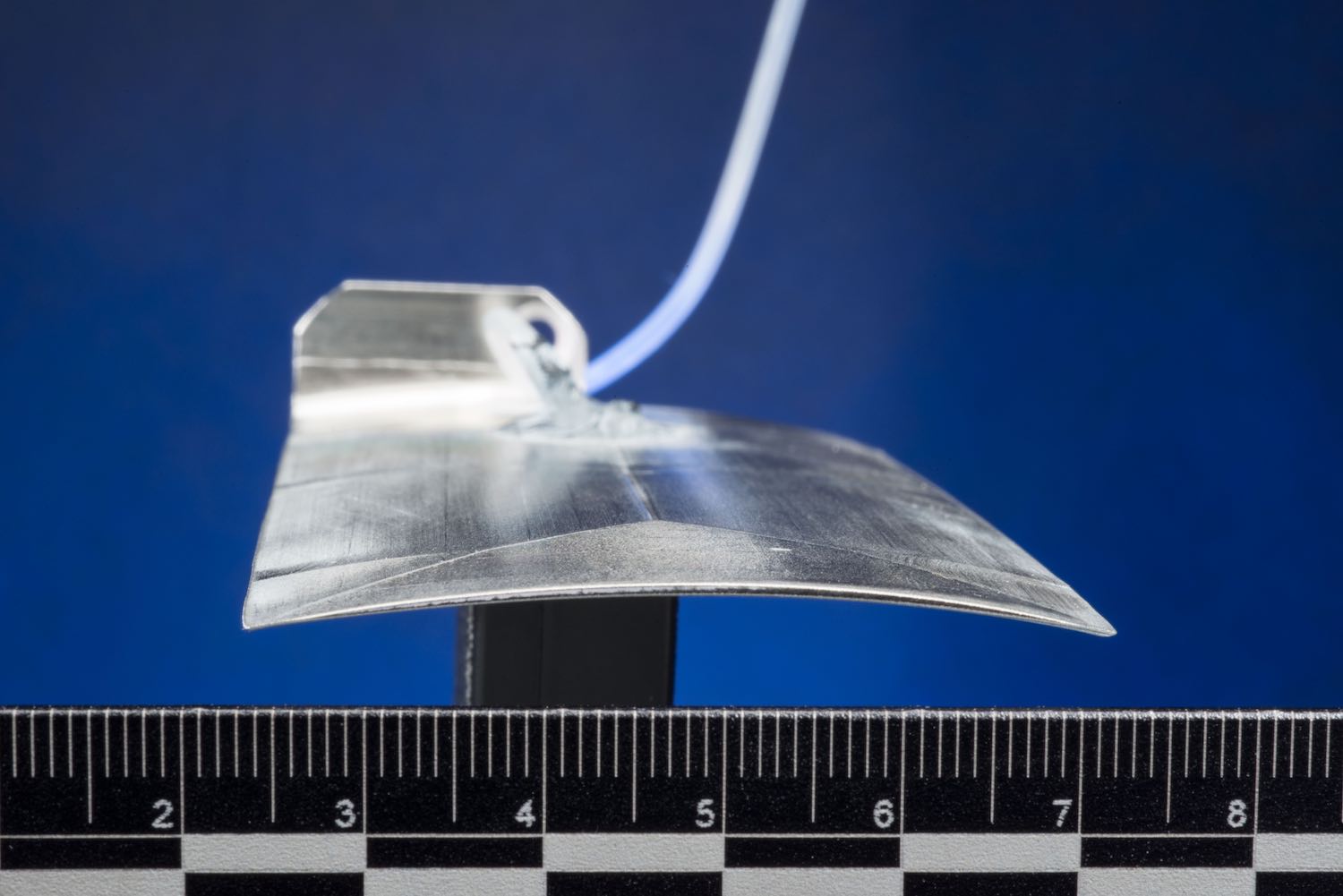}
        \caption{roamx-0201-us-pu}
        \end{subfigure}
        
        \begin{subfigure}[t]{0.33\linewidth}
        \includegraphics[width=\linewidth]{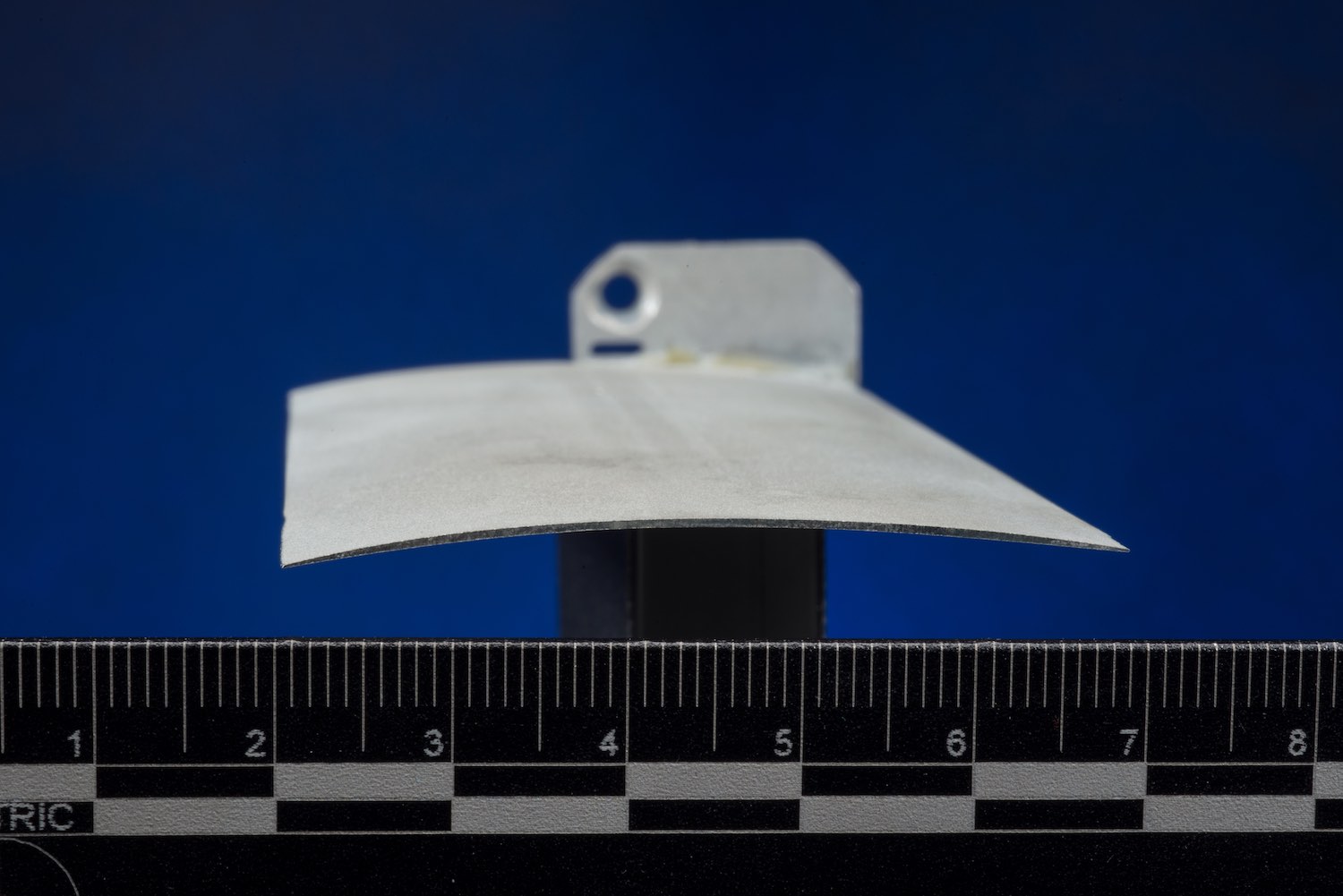}
        \caption{roamx-0201-jp-fp}
        \end{subfigure}
        \begin{subfigure}[t]{0.33\linewidth}
        \includegraphics[width=\linewidth]{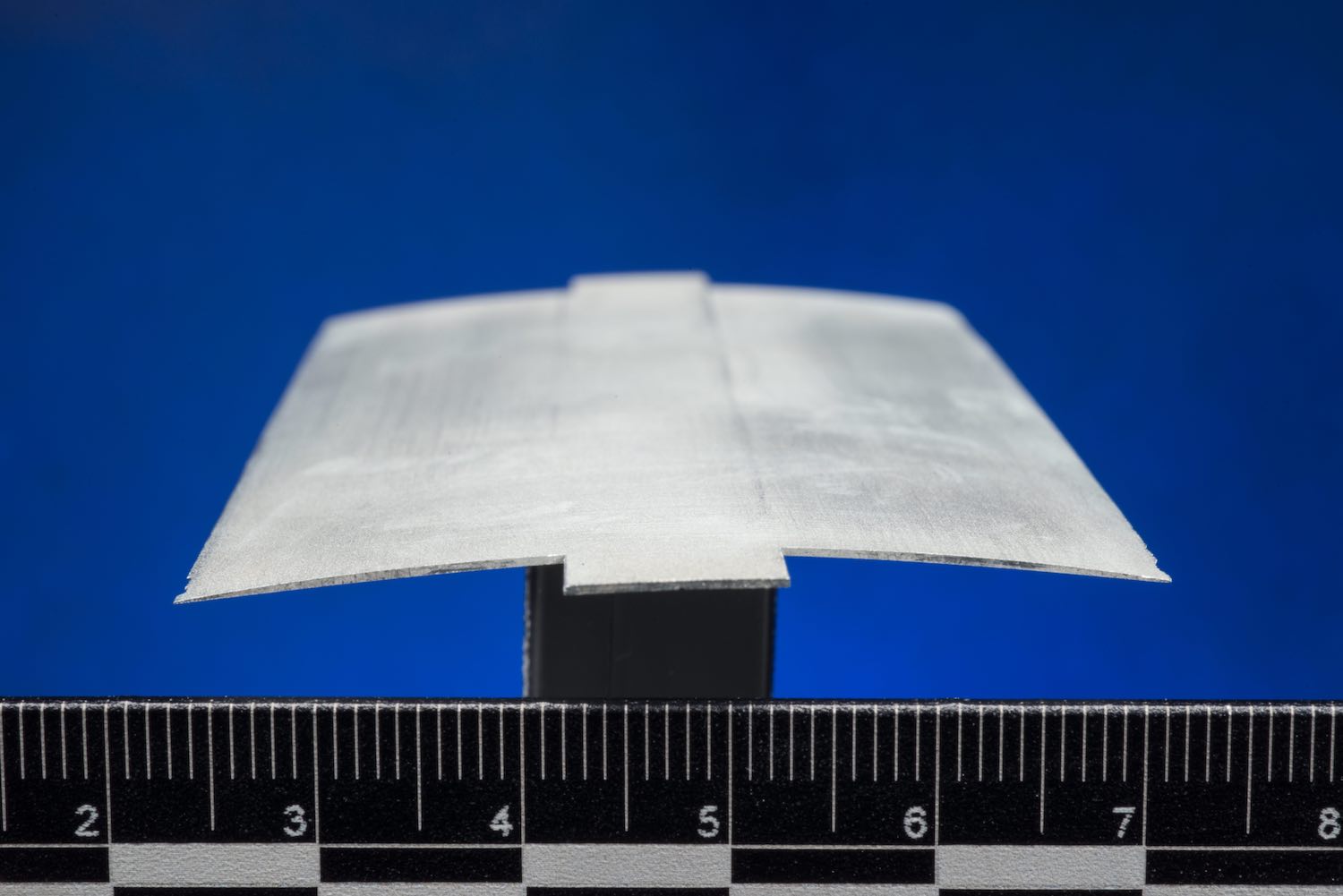}
        \caption{roamx-0201-jp-s}
        \end{subfigure}
        \caption{Images of roamx-0201 models.}
        \label{fig:models-roamx0201}
\end{figure}

Five models of the roamx-0201 profile were produced as shown in Fig.~\ref{fig:models-roamx0201}. Models roamx-0201-us-f, roamx-0201-us-pu, and roamx-0201-us-pl were fabricated in the United States from AISI 4130 using a computer numerical control (CNC) mill and were used for force measurements, upper surface PSP measurements, and lower surface PSP measurements, respectively, and models roamx-0201-jp-s, and roamx-0201-jp-fpu were fabricated in Japan from A5052 using EDM and used for schlieren measurements, and force and upper surface PSP measurements, respectively. We note that the roamx-0201-us-f, roamx-0201-us-pu and roamx-0201-us-pl models were slightly thicker than the target profile due to manufacturing limitations. In addition we note that the roamx-0201-us models have machining marks due to the CNC milling process, and these marks and hence surface texture vary between models. We alos note that the roamx-0201-jp-s and roamx-0201-jp-fpu models have a slightly rougher surface than those fabricated in the United States. This is because the models fabricated in Japan did not undergo a polishing process after EDM processing. Omitting this polishing step allows for a very thin and sharp profile. Finialy, we note that the roamx-0201-jp-s and roamx-0201-jp-fpu models are very thin and made of A5052, so it is possible that some deformation occurred during the experimental campaign.

The roamx-0201-based PSP assemblies require separate models for the upper and lower sides since the mounting bracket is part of the assembly, and hence the model cannot be reversed to illuminate the lower surface. This is in contrast to CLF5605-based PSP models which have sufficient thickness to fix the mounting bracket with screws, so it can be detached. The roamx-0201-jp-fpu model was initially only intended for force measurements, but was later also used for upper surface PSP measurements.

\subsection{Operating Conditions}

The target flow condition was $Ma=0.60$ at $Re=20,000$, and the angle of attack was varied in the range of $-2.0^\circ \leq \alpha \leq 6.0^\circ$. Due to time constraints, schlieren visualizations and PSP measurements were conducted at fewer freestream flow conditions than the force measurements, as shown in Table~\ref{tab:exp_conditions}.

\begin{table}[hbt!]
\caption{\label{tab:exp_conditions} List of experimental conditions.}
\centering
\begin{tabular}{cccc}
\hline
$\alpha$ [$^\circ$] & Schlieren visualization & Force measurement & PSP measurement \\ \hline
$-2.0$ & $\checkmark$ & $\checkmark$ & $\checkmark$ \\
0.0    & $\checkmark$ & $\checkmark$ & $\checkmark$ \\
2.0    & $\checkmark$ & $\checkmark$ & $\checkmark$ \\
2.5    &              & $\checkmark$ &         \\
3.0    & $\checkmark$ & $\checkmark$ & $\checkmark$ \\
3.5    &              & $\checkmark$ &         \\
4.0    & $\checkmark$ & $\checkmark$ & $\checkmark$ \\
4.5    &              & $\checkmark$ &         \\
5.0    & $\checkmark$ & $\checkmark$ & $\checkmark$ \\
5.5    &              & $\checkmark$ &         \\
6.0    & $\checkmark$ & $\checkmark$ & $\checkmark$ \\
\hline
\end{tabular}
\end{table}

\subsection{Instrumentation}

\subsubsection{Schlieren visualization}

\begin{figure}[hbt!]
\centering
\includegraphics[]{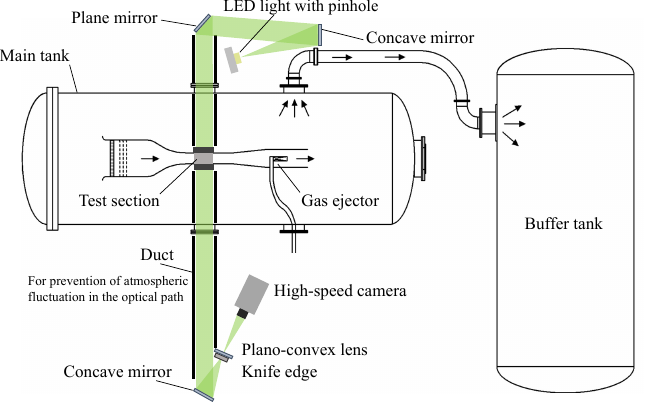}
\caption{Schematic diagram of optical system used for Schlieren visualization.}
\label{fig:schlieren_setup}
\end{figure}

The setup for schlieren visualization is shown in Fig.~\ref{fig:schlieren_setup}. The Schlieren optical system uses a white LED (PFBR-600SW-LL, CCS) as a light source. A pinhole (F70, Surugaseiki) with a diameter of 2.0~mm was used, and a point light source was created. Concave mirrors with a focal length of 1500~mm were used as the first and second collimators, and a plano-convex lens with a focal length of 1000~mm was used as the imaging lens. The knife edge was installed horizontally, and the lower half of the light flux was cutoff at the focal point of the second collimator, which visualized the density gradient field perpendicular to the freestream. The time series of schlieren images was captured by a high-speed camera (SA-X2, Photoron). The exposure time and frame rate were 2.5~\textmu s and 72,000~fps, respectively. In this experiment, parallel light was reflected by a plane mirror and passed through the test section due to space constraints in the laboratory. The airfoil model was mounted directly between the optical glasses of the wind tunnel side wall. The influence of noise caused by atmospheric fluctuations was reduced physically by covering the optical path with duct hoses both outside and inside the main tank. 

\subsubsection{Force measurement}

The lift and drag forces were measured using a three-component force balance system. This balance system consists of load cells (LC-4100-G-600 and LC-4101-K-003, A\&D) for drag and lift directions and a stepping motor (PK513PA-H50S, Oriental Motor LTD) for adjusting the angle of attack. The accuracy of the stepping motor is 0.0144$^\circ$. The zero angle of attack was determined by the inclinometer, and an airfoil mold was used for placing the inclinometer on the airfoil to ensure a reliable reference surface. The accuracy of the mold placement onto the airfoil was not quantified. However, we note that the roamx-0201-us models were slightly modified by the machining process, and thus, the mold and wing model did not fit perfectly. The accuracy of the inclinometer used to determine a $0^\circ$ pitch angle is 0.1$^\circ$.
The rated capacities of the load cells for the drag and lift directions are 6~N and 30~N, respectively, and the accuracies are $4.0\times 10^{-4}$~N and $1.4\times10^{-3}$, respectively. The measurement system samples signals from force sensors at 100~Hz and records the average value of 10 samples at 10~Hz.

\subsubsection{PSP measurement}

The time-averaged measurements of the surface pressure were performed with polymer-based PSP \cite{liu2021pressure}. Under low-pressure conditions, the normalized sensitivity (normalized by the ambient pressure) and time response of PSP is degraded \cite{nagata2020optimum,kasai2021frequency,kasai2024indexes}. Therefore, PSP specialized for low-pressure conditions must be used. The PSP used in the present study was composed of Pd(II) meso-tetra(pentafluorophenyl) porphyrin (PdTFPP) luminophore and (poly)trimethylsilylpropanoic [(poly)TMSP] polymer \cite{nagai2001poly}, which has high-pressure sensitivity under low-pressure conditions \cite{mori2006pressure,anyoji2014pressure}. The PC-PSP binder of \cite{sugioka2018polymer} with a reduced particle mass content of 85 wt\% without the luminophore was used as the white base coat. Although polymer-based PSPs for unsteady measurement under low-pressure conditions \cite{kasai2023evaluation} have been developed recently, the present study focused on obtaining time-averaged measurements.

\begin{figure}[hbt!]
\centering
\includegraphics[]{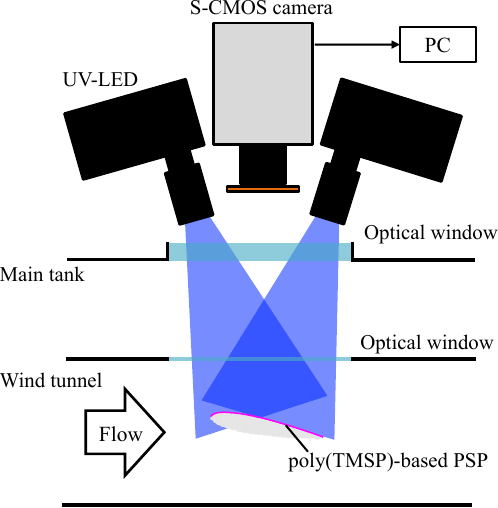}
\caption{Schematic diagram of an optical system used for PSP measurement.}
\label{fig:psp_setup}
\end{figure}

Fig.~\ref{fig:psp_setup} shows a schematic diagram of the experimental PSP measurement setup. Two UV-LEDs (IL-106, Hardsoft) were used as an excitation light, and a scientific CMOS camera (C13440-20CU, Hamamatsu Photonics) was used as a photo detector. A camera lens (Nikkor 105 mm f/2.0, Nikon) was used as an imaging lens with a $640\pm 50$~nm band-pass filter (PB640-100, Asahi Spectra) to cutoff the light other than PSP emission. Excitation of PSP and imaging of PSP emission were carried out through the optical windows of the main tank and the upper wall of the wind tunnel. The optical system was covered with a blackout curtain to prevent stray light. Exposure time for the imaging was set between 20-200~ms so that the image intensity becomes approximately 80\% of the full well-capacity of the image sensor. Wind-on, reference, and dark images were acquired. 100 images were then obtained for each experiment, and random noise was reduced by taking the average of each image set.

A single-point pressure measurement was simultaneously conducted by the pressure transducer (CCQ-093-5A, Kulite Semiconductor Products) and was used as a reference pressure for the PSP measurement. The signal from the pressure transducer was amplified by an amplifier (SA-570ST, TEAC) and was recorded by a data acquisition system (PCI6251, National Instruments) with a sampling frequency of 10~Hz. Although the pressure sensor is of a fast-responding type, it is not embedded in the model and is connected through a tube. The length of the tubing from the pressure tap to the sensor is around 15~cm. The measurement system samples the pressure signal at 100~Hz and records the average value of 10 samples at 10~Hz. Measurements on the upper and lower surfaces were carried out as individual experiments.

The clf5605-us-fp model uses pressure taps on the upper and lower surfaces of the airfoil for pressure calibration during PSP measurements. The pressure taps are routed out to the side of the airfoil profile to facilitate connection to the pressure gauges. The roamx-0201-us-pu and roamx-0201-us-pl models are too thin to route the pressure tap, and therefore use a through-hole, and a thin tube on the opposite side of the airfoil, which is used to obtain the reference pressure data. This naturally requires separate measurements for the upper and lower sides.

The Stern-Volmer coefficient and the temperature sensitivity of the PSP used in the present experiment were approximately 0.78 and $-1.41$\%/K, respectively, at the ambient pressure of 10~kPa and temperature of 283~K. A Stern-Volmer coefficient of 0.78 means a pressure sensitivity of 7.8\%/kPa. Note that the PSPs using PTMSP polymers have a contamination issue due to oil mist from the vacuum pump, and the pressure sensitivity of the PSP was gradually degraded during the experimental campaign due to a decrease in the oxygen permeability caused by oil mist contamination \cite{nagai2001poly,anyoji2014pressure}. The oil mist contamination gradually progressed with repeated experiments, and the lowest pressure sensitivity in the present experiment was approximately 6\%/kPa. In addition, the degree of contamination caused by oil mist adhesion depends on the local flow structure, and the pressure sensitivity and its distribution slightly change run by run. Hence, calibration tests were conducted in the wind tunnel before every experiment. The pressure in the main tank was changed through seven conditions, and the sensitivity distribution was obtained by correlating the pressure and intensity of the PSP emission at each pressure.

\section{Results and Discussion}
\label{results}

\subsection{Results}

\subsubsection{Lift and Drag}

Fig. \ref{fig:Cl-Cd-clf-5605} shows plots of $\overline{C}_L$ (a), $\overline{C}_D$ (b) and $\overline{C}_L/\overline{C}_D$ (c) as a function of $\alpha$ and $\overline{C}_L$ as a function of $\overline{C}_D$ (d) for the CLF5605 airfoil. Results are shown for 2D OVERFLOW, 2D PyFR, 3D-SP OVERFLOW, 3D-SP PyFR simulations, and MWT force balance experiments on models clf5605-us-fp and clf5605-jp-f.

Fig. \ref{fig:Cl-Cd-roamx-0201} shows plots of $\overline{C}_L$ (a), $\overline{C}_D$ (b) and $\overline{C}_L/\overline{C}_D$ (c) as a function of $\alpha$ and $\overline{C}_L$ as a function of $\overline{C}_D$ (d) for the roamx-0201 airfoil. Results are shown for 2D OVERFLOW, 2D PyFR, 3D-SP OVERFLOW, 3D-SP PyFR, VWT PyFR simulations, and MWT force balance experiments on models roamx-0201-us-f and roamx-0201-jp-fpu.

\input{Figures-tex/aero-forces}

\subsubsection{Surface Pressure}

Fig. \ref{fig:Cp-clf-5605} shows plots of $\overline{C}_P$ on both upper and lower surfaces of the CLF5605 airfoil as a function of $\tilde{x}=x/a$, where $a$ is the $x$-wise distance between the leading and trailing edge of the airfoil, for various $\alpha$. Results are shown for 2D OVERFLOW, 2D PyFR, 3D-SP OVERFLOW, 3D-SP PyFR simulations, and MWT PSP experiments on model clf5605-us-fp. 

Fig. \ref{fig:Cp-roamx-0201} shows plots of $\overline{C}_P$ on both upper and lower surfaces of the roamx-0201 airfoil as a function of $\tilde{x}=x/a$, where $a$ is the $x$-wise distance between the leading and trailing edge of the airfoil, for various $\alpha$. Results are shown for 2D OVERFLOW, 2D PyFR, 3D-SP OVERFLOW, 3D-SP PyFR, VWT PyFR simulations, and MWT PSP experiments on models roamx-0201-us-pu, roamx-0201-us-pl and roamx-0201-jp-fpu.

Note that for 3D-SP PyFR, 3D-SP OVERFLOW and VWT PyFR results, $\overline{C}_P$ is measured at the midspan, while for all MWT PSP experiments, $\overline{C}_P$ was obtained by taking a spanwise average with a 5\% span width (approximately imaging 80 pixels) around the model centerline.

Plots of $\overline{C}_P$ at all $\alpha$ are included in \hyperref[appendix-cp]{Appendix D}.

\input{Figures-tex/pressure-clf5605}

\input{Figures-tex/pressure-roamx0201}

\subsubsection{Schlieren, Instantaneous Density Gradient Magnitude, and Time-Averaged Line Integral Convolution}

Fig. \ref{fig:schlieren-clf} shows Schlieren images for MWT experiments using model clf5605-us-s (top) and images of instantaneous density gradient magnitude $|\boldsymbol{\nabla}\rho|$ (middle) and of time-averaged normalised velocity magnitude $|\mathbf{v}|/v_\infty$ with superimposed Line Integral Convolutions (LICs) (bottom) for 3D-SP OVERFLOW (blue) and 3D-SP PyFR (red) on CLF5605 at $\alpha = -2^\circ, 0^\circ, 5^\circ$ and $6^\circ$.

Fig. \ref{fig:schlieren-roamx} shows Schlieren images for MWT experiments using model roamx-0201-jp-s (top) and images of instantaneous density gradient magnitude $|\boldsymbol{\nabla}\rho|$ (middle) and of time-averaged normalised velocity magnitude $|\mathbf{v}|/v_\infty$ with superimposed LICs (bottom) for 3D-SP OVERFLOW (blue) and 3D-SP PyFR (red) on roamx-0201 at $\alpha = -2^\circ, 0^\circ, 5^\circ$ and $6^\circ$.

Schlieren images and images of instantaneous density gradient magnitude at all $\alpha$ are included in \hyperref[appendix-schlieren]{Appendix E}.

\input{Figures-tex/schlieren}

\subsubsection{Instantaneous Q-criterion Isosurfaces}

Fig. \ref{fig:qcrit-OF-PyFR-q3D-clf} shows instantaneous Q-criterion iso-surfaces coloured by normalised velocity magnitude $|\mathbf{v}|/|\mathbf{v}_\infty|$ for 3D-SP OVERFLOW (left) and 3D-SP PyFR (right) simulations of CLF5605 at various $\alpha$.

Fig. \ref{fig:qcrit-OF-PyFR-q3D-0201} shows instantaneous Q-criterion iso-surfaces coloured by normalised velocity magnitude $|\mathbf{v}|/|\mathbf{v}_\infty|$ for 3D-SP OVERFLOW (left) and 3D-SP PyFR (right) simulations of roamx-0201 at various $\alpha$.

Fig. \ref{fig:qcrit-VWT-roamx} shows instantaneous Q-criterion iso-surfaces coloured by normalised velocity magnitude $|\mathbf{v}|/|\mathbf{v}_\infty|$ for a VWT PyFR simulation of roamx-0201 at $\alpha=6^\circ$.

\input{Figures-tex/q-criterion}

\subsubsection{Flow Physics}

Across all methodologies, when $\alpha=-2^\circ$, both CLF5605 and roamx-0201 exhibit flow separation on the lower surface of the airfoil. For 3D-SP OVERFLOW and PyFR simulations, and MWT experiments, the separated flow breaks down in the spanwise direction, resulting in an almost zero lift force for the CLF5605 airfoil, and a near-zero or negative lift for the roamx-0201 airfoil. The flow separation and spanwise breakdown can be observed in the Schlieren/density gradient images in Figs. \ref{fig:schlieren-clf} and \ref{fig:schlieren-roamx} and in the Q-criterion isosurfaces in Figs. \ref{fig:qcrit-OF-PyFR-q3D-clf} and \ref{fig:qcrit-OF-PyFR-q3D-0201}.

Additionally, all methodologies show linear behaviour of the lift slope for a range of $\alpha$, for both CLF5605 and roamx-0201 airfoils. However, the range of $\alpha$ varies across methodologies and airfoils. For the CLF5605 airfoil, 2D and 3D-SP OVERFLOW and PyFR simulations show this behaviour for $0^\circ \leq \alpha \leq 4^\circ$ while MWT experiments on both models show it for all $\alpha$. For the roamx-0201 airfoil, 2D and 3D-SP OVERFLOW and PyFR simulations show this behaviour for $0^\circ \leq \alpha \leq 4^\circ$, MWT roamx-0201-us-f experiments for $-2^\circ \leq \alpha \leq 4^\circ$ and MWT roamx-0201-jp-fpu experiments for all $\alpha$. By looking at Schlieren/density gradient images in Figs. \ref{fig:schlieren-clf} and \ref{fig:schlieren-roamx} and Q-criterion isosurfaces in Figs. \ref{fig:qcrit-OF-PyFR-q3D-clf} and \ref{fig:qcrit-OF-PyFR-q3D-0201}, we can correlate the linearity of the lift slope with spanwise coherence of the flow. The linearity of the lift slope, for both airfoils, terminates as soon as the flow breaks down substantially in the spanwise direction.

For the CLF5605 airfoil, when $\alpha > 5^\circ$, 2D OVERFLOW and PyFR simulations show an increment in lift while 3D-SP OVERFLOW and PyFR simulations show stall behaviour. This stall behaviour for 3D-SP OVERFLOW and PyFR simulations can be attributed to the laminar boundary completely separating from the suction surface, with subsequent span-wise breakdown and a re-circulation region in a time-averaged sense above the trailing edge of the suction surface, as can be observed in the density gradient and LIC images in Fig. \ref{fig:schlieren-clf} and the Q-criterion isosurfaces in Fig. \ref{fig:qcrit-OF-PyFR-q3D-clf}.

For the roamx-0201 airfoil, when $\alpha > 4^\circ$, 2D and 3D-SP OVERFLOW and PyFR simulations, VWT PyFR simulations, and MWT roamx-0201-us-f experiments, show an increment in lift. This behavior can be attributed to separation from the sharp leading edge followed by a Kelvin--Helmholtz instability and subsequent roll-up of vortices close to the upper surface of the airfoil, resulting in a re-circulation region in a time-averaged sense adjacent to the front half of the upper surface, as can be observed in the density gradient and LIC images in Fig. \ref{fig:schlieren-roamx} and the Q-criterion isosurfaces in Fig. \ref{fig:qcrit-OF-PyFR-q3D-0201}. The resulting increase in suction on the upper surface of the airfoil can be observed in the $\overline{C}_P$ distributions in Fig. \ref{fig:Cp-roamx-0201}.

\subsection{Comparison between Methodologies}

\subsubsection{Comparison between OVERFLOW and PyFR}

Results obtained from OVERFLOW and PyFR are similar across both 2D and 3D-SP simulations, for both CLF5605 and roamx-0201 airfoils. This serves as a good cross-validation of the independent meshing and solver pipelines. Moreover, it suggest that the ILES resolution afforded by the OVERFLOW simulations is sufficient to capture the relevant flow physics.

\subsubsection{Comparison between 2D and 3D-SP}

For the CLF5605 airfoil, Fig. \ref{fig:Cl-Cd-clf-5605} shows discrepancies between 2D and 3D-SP results in $\overline{C}_L$ for all $\alpha$, with lower $\overline{C}_L$ values for 3D-SP, as well as discrepancies in $\overline{C}_D$ for $\alpha < 0^\circ$, with higher $\overline{C}_D$ values for 3D-SP, and $\alpha > 3^\circ$, with lower $\overline{C}_L$ values for 3D-SP. The $\overline{C}_L$ discrepancies correlate with discrepancies in $\overline{C}_P$ distributions in Fig. \ref{fig:Cp-clf-5605}, where 3D-SP show lower suction on the upper surface of the airfoil. These discrepancies, at the lowest and highest values of $\alpha$, are caused by flow separation and subsequent spanwise breakdown of the vortices for the 3D-SP simulations, which is not possible for the 2D simulations. In particular, and notably, the 2D simulations fail to capture the stall behavior of the CLF5605 airfoil for $\alpha > 5^\circ$ and show, instead, an increment in lift.

For the roamx-0201 airfoil, Fig. \ref{fig:Cl-Cd-roamx-0201} only shows discrepancies between 2D and 3D-SP results in $\overline{C}_L$ and $\overline{C}_D$ for $\alpha = -2^\circ$, which correlates with discrepancies in $\overline{C}_P$ distributions for this $\alpha$. $\overline{C}_L$ and $\overline{C}_D$ for all other $\alpha$ are in good agreement. However, for $\alpha=5^\circ$ and $\alpha=6^\circ$, Fig. \ref{fig:Cp-roamx-0201} shows that $\overline{C}_P$ distributions on the suction side of the airfoil differ between 2D and 3D-SP simulations, due to flow separation and subsequent spanwise breakdown of the vortices for the 3D-SP simulations, which is not possible for the 2D simulations. This apparent discrepancy is explained by the fact that at $\alpha=5^\circ$ and $\alpha=6^\circ$ the 2D simulations produce higher and lower suction regions on the airfoil that coincidentally add to the same force obtained by the 3D-SP simulations.

\subsubsection{Comparison of 3D-SP with MWT Experiments}

For the CLF5605 airfoil, Fig. \ref{fig:Cl-Cd-clf-5605} (a) shows that 3D-SP simulations and experimental data follow a similar trend for $\overline{C}_L$ when $0^\circ \leq  \alpha \leq 5^\circ$. However, $\overline{C}_L$ values for the MWT experiments are, in general, between 20\% and 40\% lower. For $\alpha = -2^\circ$, $\overline{C}_L$ values are closer. For $\alpha > 5^\circ$, the experiments do not seem to show stall behaviour of the CLF5605 airfoil, which is seen in 3D-SP simulations. Fig. \ref{fig:Cl-Cd-clf-5605} (b) shows that 3D-SP simulations and experimental data follow a similar trend for $\overline{C}_D$ for all $\alpha$, but with higher experimental values up to $\alpha = 3^\circ$.

For the roamx-0201 airfoil, Fig. \ref{fig:Cl-Cd-roamx-0201} shows that 3D-SP simulations and experimental data follow a similar trend for $\overline{C}_L$ and $\overline{C}_D$ for all angles of attack except $\alpha = -2^\circ$. In particular, MWT roamx-0201-us-fp experiments capture the non-linear increment in $\overline{C}_L$ and $\overline{C}_D$ from $\alpha=4.5^\circ$ caused by the separated shear layer and roll-up of vortices. Still, experimental values for $\overline{C}_L$ are in general between 30\% and 55\% lower, and $\overline{C}_D$ values are in general between 20\% and 30\% higher. Although for $\alpha \ge 5^\circ$, $\overline{C}_D$ values obtained from 3D-SP simulations and MWT roamx-0201-us-fp experiments overlap.

To aid understanding of discrepancies between 3D-SP simulations and MWT experiments, VWT PyFR simulations were undertaken for the roamx-0201 airfoil at $\alpha = 4.5^\circ$ and $\alpha = 6^\circ$. These simulations model flow in the entire working section of the MWT, thus accounting for blockage effects, and the interaction of the side-wall boundary layer with the extruded airfoil.

Fig. \ref{fig:qcrit-VWT-roamx}, shows a shapshot of Q-criterion isosurfaces coloured by normalised velocity magnitude $|\mathbf{v}|/|\mathbf{v}_\infty|$ for the $\alpha = 6^\circ$ case, looking along the length of the tunnel from the entrance of the test section. Vortical structures resulting from the interaction of the side-wall boundary layer with the extruded airfoil are clearly visible. $\overline{C}_L$ values obtained from VWT PyFR shown in Fig. \ref{fig:Cl-Cd-roamx-0201} (a) are closer to experimental values, reducing the discrepancy with MWT roamx-0201-us-f experiments to 15--20\% and the discrepancy with MWT roamx-0201-jp-fpu experiments to 40\%. However, $\overline{C}_D$ values from VWT PyFR shown in Fig. \ref{fig:Cl-Cd-roamx-0201} (b) remain similar to results from 3D-SP. The reduction in lift for the VWT PyFR can be attributed to the interaction of the side-wall boundary layer with the extruded airfoil, which results in further spanwise breakdown and less coherent roll-up of vortices. Fig. \ref{fig:time-series-avg-PyFR} shows time series of $C_L$ for the Data Extraction Period alongside snapshots of instantaneous density gradient magnitude $|\boldsymbol{\nabla}\rho|$ for 2D OVERFLOW and PyFR (a), mid-span 3D-SP OVERFLOW and PyFR (b) and mid-span VWT PyFR (c). These images illustrate how the coherence of the vortices is reduced as one moves from 2D to 3D-SP and then to VWT PyFR simulations. The coherence of the vortices correlates both with higher amplitude fluctuations of $C_L$ and with a higher $\overline{C}_L$. Additionally, 2D and 3D-SP simulations exhibit larger low-frequency components compared to VWT PyFR simulations, indicative of larger coherent structures with longer timescales. When comparing $C_L$ time-series, it is important to note that for 3D-SP, $C_L$ values are obtained over a span of 0.5 chords and for VWT, over a span of 2 chords.

\input{Figures-tex/time-series}

Possible reasons for the remaining discrepancies between VWT PyFR simulations and MWT experiments include manufacturing tolerances for the MWT airfoil models, possible inlet turbulence in the MWT that is not accounted for in the VWT PyFR simulations, and uncertainty around $\alpha$ calibration for the experiments.

\subsubsection{Comparison between MWT Experiments}

Differences in aerodynamic forces are observed between different models of the same geometry. Specifically, for the CLF5605 airfoil, clf5605-us-fp and clf5605-jp-f show differences of 10--30\% in $\overline{C}_L$ and 10-20\% in $\overline{C}_D$, and for the roamx-0201 airfoil, roamx-0201-us-f and roamx-0201-jp-fpu also show differences of 10-30\% in $\overline{C}_L$ and 10-20\% in $\overline{C}_D$.

Additionally, $\overline{C}_P$ distributions for the roamx-0201 airfoil also differ when comparing roamx-0201-us-pu/pl and roamx-0201-jp-fpu, where roamx-0201-jp-fpu does not show the separation bubble at $\alpha=6^\circ$ which was identified in the 3D-SP simulations.

Possible reasons for these discrepancies include manufacturing tolerances for the MWT airfoil models, spurious spanwise deformation of models when mounted in the wind tunnel, and uncertainty around $\alpha$ calibration for the experiments. Future work should look to make detailed measurements of the models in order to compare them quantitatively against the underlying target airfoil profiles, and each other.

\subsection{Comparison between roamx-0201 and CLF5605}

Fig. \ref{fig:Cl-Cd-compare} plots $\overline{C}_L$-$\overline{C}_D$ polars comparing the performance of the CLF5605 and the roamx-2021 airfoils. Across all methodologies, it can be seen visually that the roamx-0201 airfoil is able to achieve a given lift with less drag compared to the CLF5605 airfoil. Fig. \ref{fig:E-compare} plots $\overline{C}_L/\overline{C}_D$ as a function of angle of attack $\alpha$ comparing performance of the CLF5605 and the roamx-2021 airfoils. Across all methodologies, maximum $\overline{C}_L/\overline{C}_D$ achieved by the roamx-0201 airfoil is larger than that achieved by the CLF5605 airfoil, with an increase of $30.0\%$ for 2D OVERFLOW simulations, $28.8\%$ for 2D PyFR simulations, $41.5\%$ for 3D-SP OVERFLOW simulations, $42.9\%$ for 3D-SP PyFR simulations, $32.3\%$ for MWT experiments comparing clf5605-us-fp with roamx-0201-us-f, and $18.9\%$ for MWT experiments comparing clf5605-jp-f with roamx-0201-jp-fpu. Finally, we observe that the 3D-SP OVERFLOW and 3D-SP PyFR simulations predict that the CLF5605 airfoil will lose lift and stall when the angle of attack $\alpha>5^\circ$, but the roamx-0201 airfoil will achieve increased lift when $\alpha \ge 4.5^\circ$. This increase in lift is due to separation from the sharp leading edge of the roamx-0201, followed by a Kelvin--Helmholtz instability and subsequent roll-up of vortices close to the upper surface of the airfoil, resulting in a re-circulation region in a time-averaged sense adjacent to the front half of the upper surface. Consequently, the 3D-SP OVERFLOW and 3D-SP PyFR simulations predict that the roamx-0201 airfoil can achieve a maximum lift $\approx 20\%$ higher than that achieved by the CLF5605 airfoil.

\input{Figures-tex/polars}
\input{Figures-tex/cl_over_cd_plots}

\section{Conclusions}
\label{conc}

This study compares aerodynamic performance of the CLF5605 rotor airfoil --- which flew on Ingenuity in 2021 --- with that of a new optimized roamx-0201 airfoil designed for Martian conditions at NASA Ames. Specifically, performance was studied at a chord-based Reynolds number of 20,000 and a Mach number of 0.60, across a range of angles of attack, using three independent state-of-the-art methodologies: ILES using NASA's OVERFLOW solver, DNS using the high-order GPU-accelerated PyFR solver, and experimental testing in the MWT at Tohoku University. Discrepancies between results obtained using the various methodologies were analyzed and explained, including disagreement between the 2D and 3D-SP simulation results, between the 3D-SP simulation results and the MWT experimental results, and between the various MWT experimental results. However, across all methodologies, it can be seen that the roamx-0201 airfoil is able to achieve a given lift with less drag compared to the CLF5605 airfoil. Moreover, the 3D-SP OVERFLOW and PyFR simulations suggest that the roamx-0201 airfoil has superior stall characteristics, and can achieve a maximum lift $\approx 20\%$ higher than that achieved by the CLF5605 airfoil. The work provides a strong body of evidence to support further studies into the use of rotors based on the optimized roamx-0201 airfoil for future Mars helicopters, with the potential to deliver increased range, flight duration, and payload compared to Ingenuity. Such capabilities would allow next-generation vehicles to explore further and faster, and to carry scientific instruments and/or surface samples, enabling a wider set of Mars science and discovery missions.

\section*{Appendix} 
\label{Appendix}

\subsection*{A. Mesh Independence Study (OVERFLOW and PyFR)}
\label{2D-grid-indep}

Fig. \ref{fig:grid-independence} shows absolute percentage differences in time-averaged lift and drag coefficients with respect to the finest mesh, \( \lvert \Delta \overline{C}_L \rvert \% \) and \( \lvert \Delta \overline{C}_D \rvert \% \), plotted as a function of the number of elements \( N_e \) to assess mesh convergence for 2D OVERFLOW and 2D PyFR simulations of CLF5605 and roamx-0201 at \( \alpha = 6^\circ \). 

The absolute percentage differences in time-averaged lift and drag coefficients are defined as 
\begin{equation}
    \lvert \Delta \overline{C}_L \rvert \% = \left| \frac{\overline{C}_L^{(i)} - \overline{C}_L^{(\text{fine})}}{\overline{C}_L^{(\text{fine})}} \right| \times 100 \% \hspace{1cm} \text{and} \hspace{1cm} \lvert \Delta \overline{C}_D \rvert \% = \left| \frac{\overline{C}_D^{(i)} - \overline{C}_D^{(\text{fine})}}{\overline{C}_D^{(\text{fine})}} \right| \times 100 \%
\end{equation}
respectively, where \( \overline{C}_L^{(i)} \) and \( \overline{C}_D^{(i)} \) denote values obtained on mesh \( i \), and \( \overline{C}_L^{(\text{fine})} \) and \( \overline{C}_D^{(\text{fine})} \) refer to the values obtained from the finest mesh. For both OVERFLOW and PyFR, the 2D meshes for which both \( \lvert \Delta \overline{C}_L \rvert \% \) and \( \lvert \Delta \overline{C}_D \rvert \% \) were below 1\% was used for all simulations, corresponding to the second finest mesh tested.

\input{Figures-tex/Appendix/grid-independence}

\subsection*{B. Spanwise Domain Extent Independence Study (OVERFLOW and PyFR)}\label{span-indep}

Fig. \ref{fig:Cp-span-independence} shows plots of $\overline{C}_P$ on both upper and lower surfaces of the CLF5605 and roamx-0201 airfoils at $\alpha=6^\circ$ as a function of $\tilde{x}=x/a$ for 3D-SP OVERFLOW and 3D-SP PyFR with different spanwise extents of the domain, $d_z = 0.25$, $0.5$ and $0.75$. The plots demonstrate span independence from  $d_z = 0.5$, which is chosen as the spanwise extent of the domain for all 3D-SP simulations.
\input{Figures-tex/Appendix/span-independence}

\subsection*{C. DNS Resolution Study (PyFR)} \label{DNS-resolution}

Figs. \ref{fig:DNS-resolution-y} and \ref{fig:DNS-resolution-x} show plots of $\Delta/\eta$ along various mid-span lines for 3D-SP PyFR simulations at $\alpha=6^\circ$ for CLF5605 and roamx-0201, where $\Delta$ is an estimate of the solution point spacing given by Eq. (\ref{eq:pointspace}) and $\eta$ is an estimate of the Kolmogorov length scale given by Eq. (\ref{eq:kolmogorov}). It is shown that $3.5 \eta > \Delta$ in the wake for 6 chords downstream of the airfoil.
\input{Figures-tex/Appendix/DNS-resolution}

\subsection*{D. Surface Pressure for all $\alpha$} \label{appendix-cp}
Fig. \ref{fig:Cp-allAoA-clf-5605} shows plots of $\overline{C}_P$ on both upper and lower surfaces of the CLF5605 airfoil as a function of $\tilde{x}=x/a$, where $a$ is the $x$-wise distance between the leading and trailing edge of the airfoil, for all tested $\alpha$. Results are shown for 2D OVERFLOW, 2D PyFR, 3D-SP OVERFLOW, 3D-SP PyFR and MWT PSP experiments on model clf5605-us-fp. 

Fig. \ref{fig:Cp-allAoA-roamx-0201} shows plots of $\overline{C}_P$ on both upper and lower surfaces of the roamx-0201 airfoil as a function of $\tilde{x}=x/a$, where $a$ is the $x$-wise distance between the leading and trailing edge of the airfoil, for all tested $\alpha$. Results are shown for 2D OVERFLOW, 2D PyFR, 3D-SP OVERFLOW, 3D-SP PyFR, VWT PyFR and MWT PSP experiments on models roamx-0201-us-pu, roamx-0201-us-pl and roamx-0201-jp-fpu.

\input{Figures-tex/Appendix/all-cp}

\subsection*{E. Schlieren and Instantaneous Density Gradient for all $\alpha$} \label{appendix-schlieren}

Fig. \ref{fig:clf5605-schlieren-allAoA} shows Schlieren images for MWT experiments on model clf5605-us-s (left) and images of instantaneous density gradient magnitude $|\boldsymbol{\nabla}\rho|$ for 3D-SP OVERFLOW (middle) and PyFR (right) on CLF5605 at all tested $\alpha$.

Fig. \ref{fig:roamx0201-schlieren-allAoA} shows Schlieren images for MWT experiments on model roamx-0201-jp-s (left) and images of instantaneous density gradient magnitude $|\boldsymbol{\nabla}\rho|$ for 3D-SP OVERFLOW (middle) and PyFR (right) on roamx-0201 at all tested $\alpha$.
\input{Figures-tex/Appendix/all-schlieren}

\section*{Acknowledgments}

This paper is dedicated to the memory of Prof. Keisuke Asai, who sadly passed away on August 11th, 2025. The authors gratefully acknowledge the outstanding support of Miku Kasai, Yudai Kanzaki and Muku Miyagi in conducting the MWT experimental work. Resources supporting this work were provided by the NASA High-End Computing (HEC) Program through the NASA Advanced Supercomputing (NAS) Division at Ames Research Center. The authors are grateful for support from the Engineering and Physical Sciences Research Council via an EPSRC Fellowship (EP/R030340/1), and for compute allocation on Piz Daint and Alps at the Swiss National Supercomputing Centre (CSCS). The authors also thank Jon-Pierre Wiens and Charles Cornelison for the professional photography of the airfoil models.

\bibliography{sample}
\end{document}

%% file: Figures-tex/geometries.tex
\begin{figure}[htbp]\centering
\includegraphics[width=0.8\linewidth]{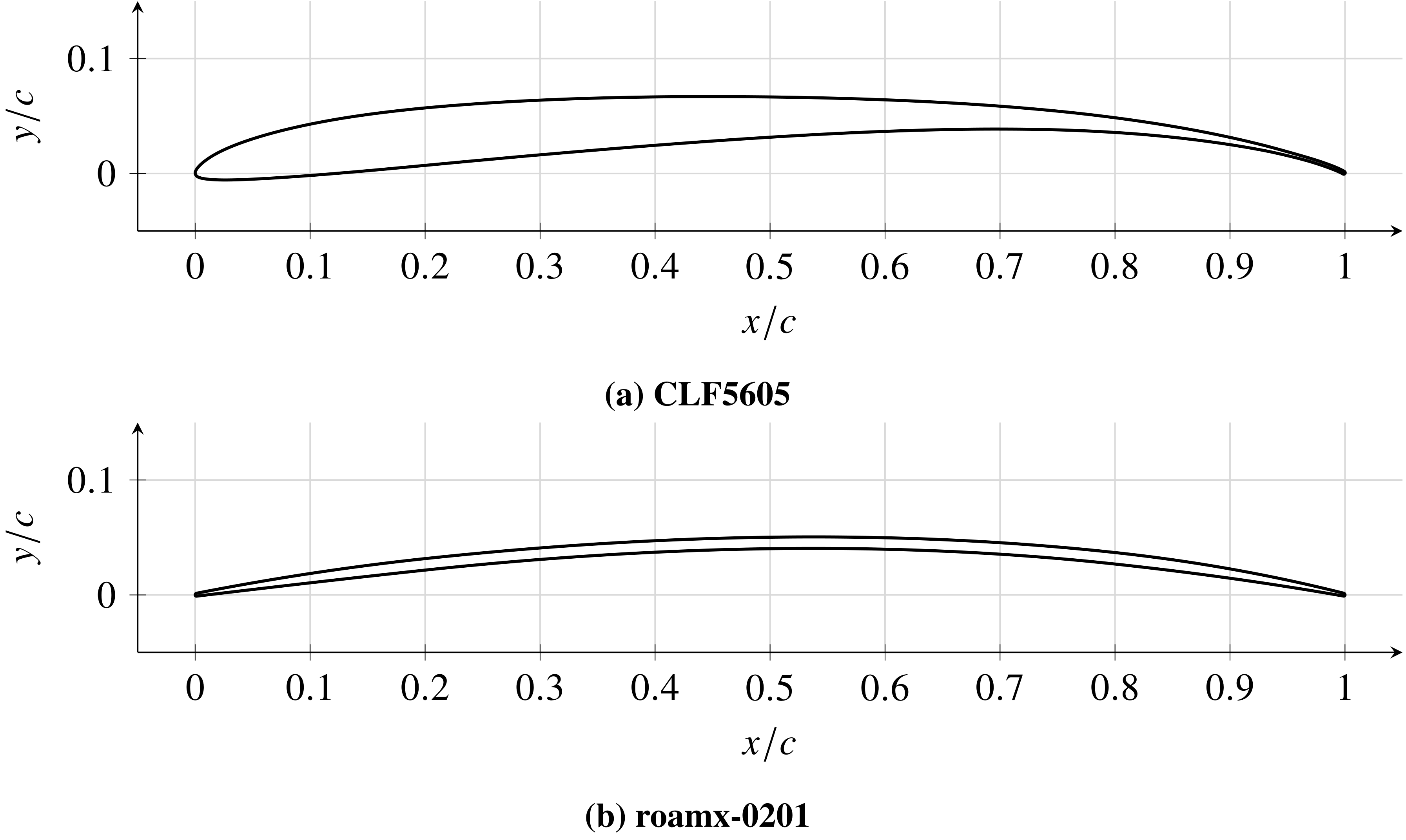}
\caption{CLF5605 (a) and roamx-0201 (b) airfoils.}
\label{fig:geometries}
\end{figure}

%% file: Figures-tex/meshes.tex
\begin{figure}[h!]
  \centering
        \begin{subfigure}[t]{0.98\linewidth}
        \includegraphics[width=\linewidth]{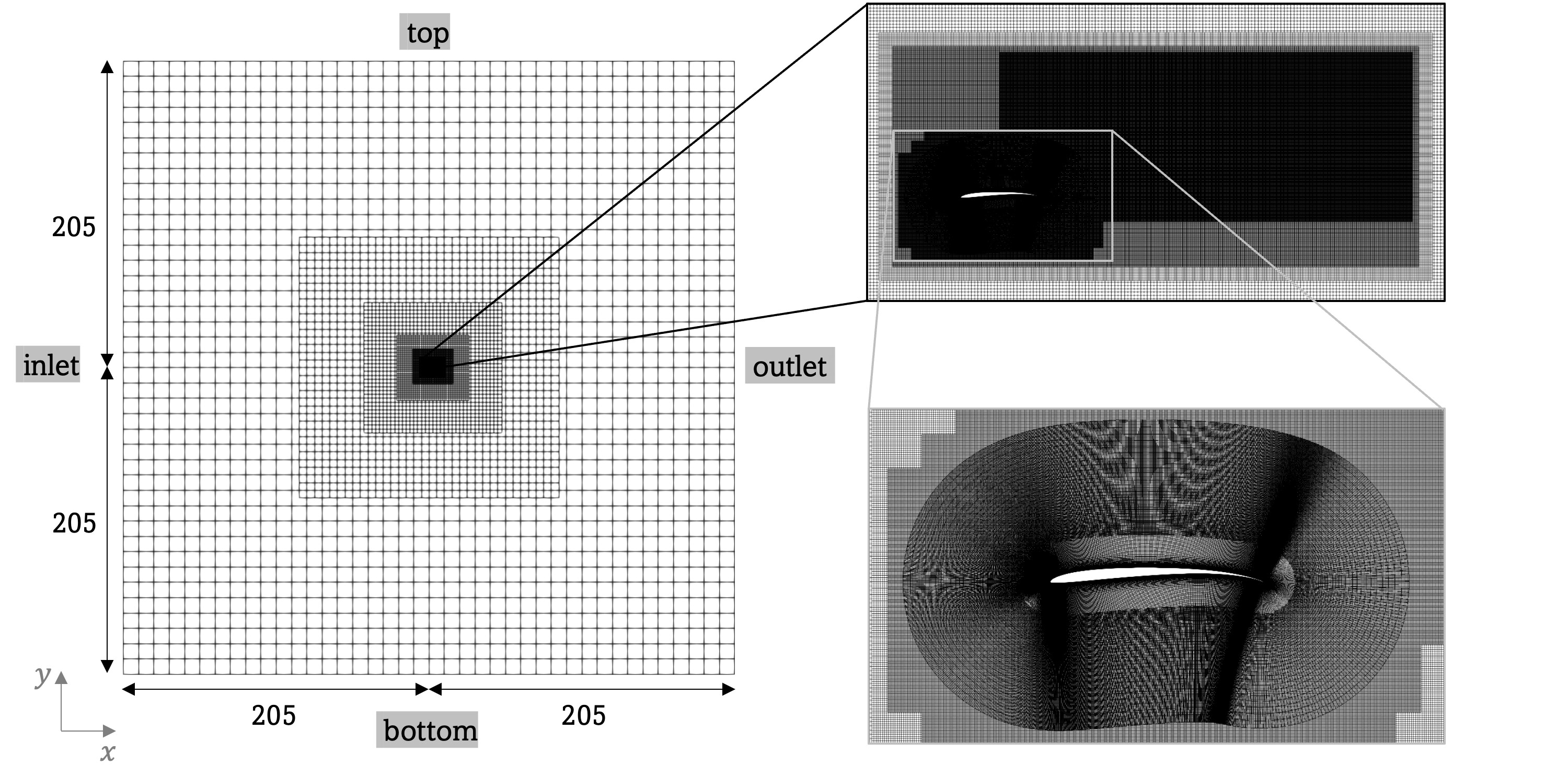}
        \caption{CLF5605}
        \end{subfigure}
        
        \vspace{0.6cm}
        \begin{subfigure}[t]{0.98\linewidth}
        \includegraphics[width=\linewidth]{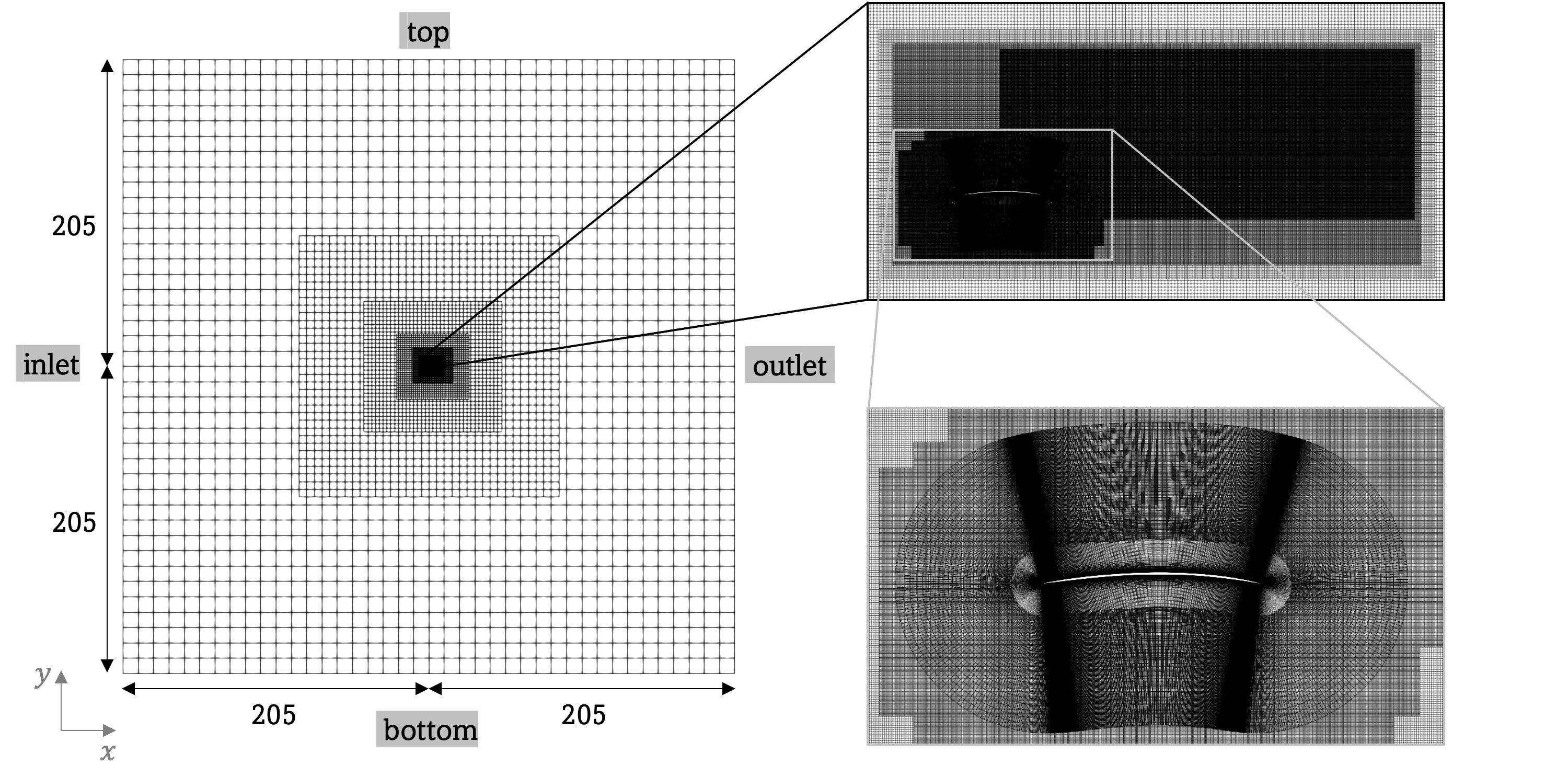}
        \caption{roamx-0201}
        \end{subfigure}
        \caption{Views of domain and mesh in the streamwise-vertical xy-plane used for 2D simulations with OVERFLOW for the CLF5605 (a) and roamx-0201 (b) airfoils.}
        \label{fig:OVERFLOW-2D-domain}
\end{figure}

\begin{figure}[h!]
  \centering
        \begin{subfigure}[t]{0.9\linewidth}
        \includegraphics[width=\linewidth]{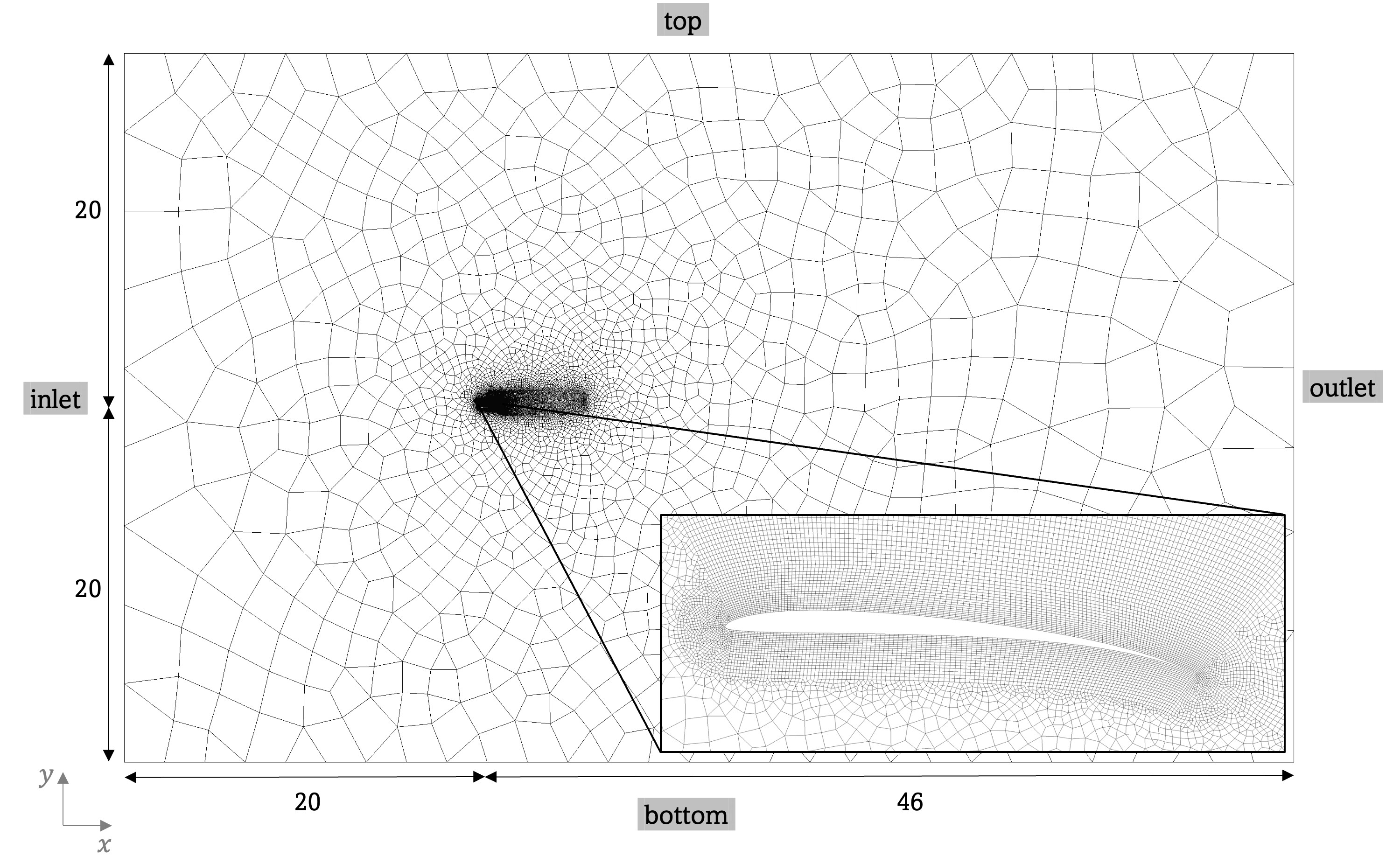}
        \caption{CLF5605}
        \end{subfigure}
        
        \vspace{0.6cm}
        \begin{subfigure}[t]{0.9\linewidth}
        \includegraphics[width=\linewidth]{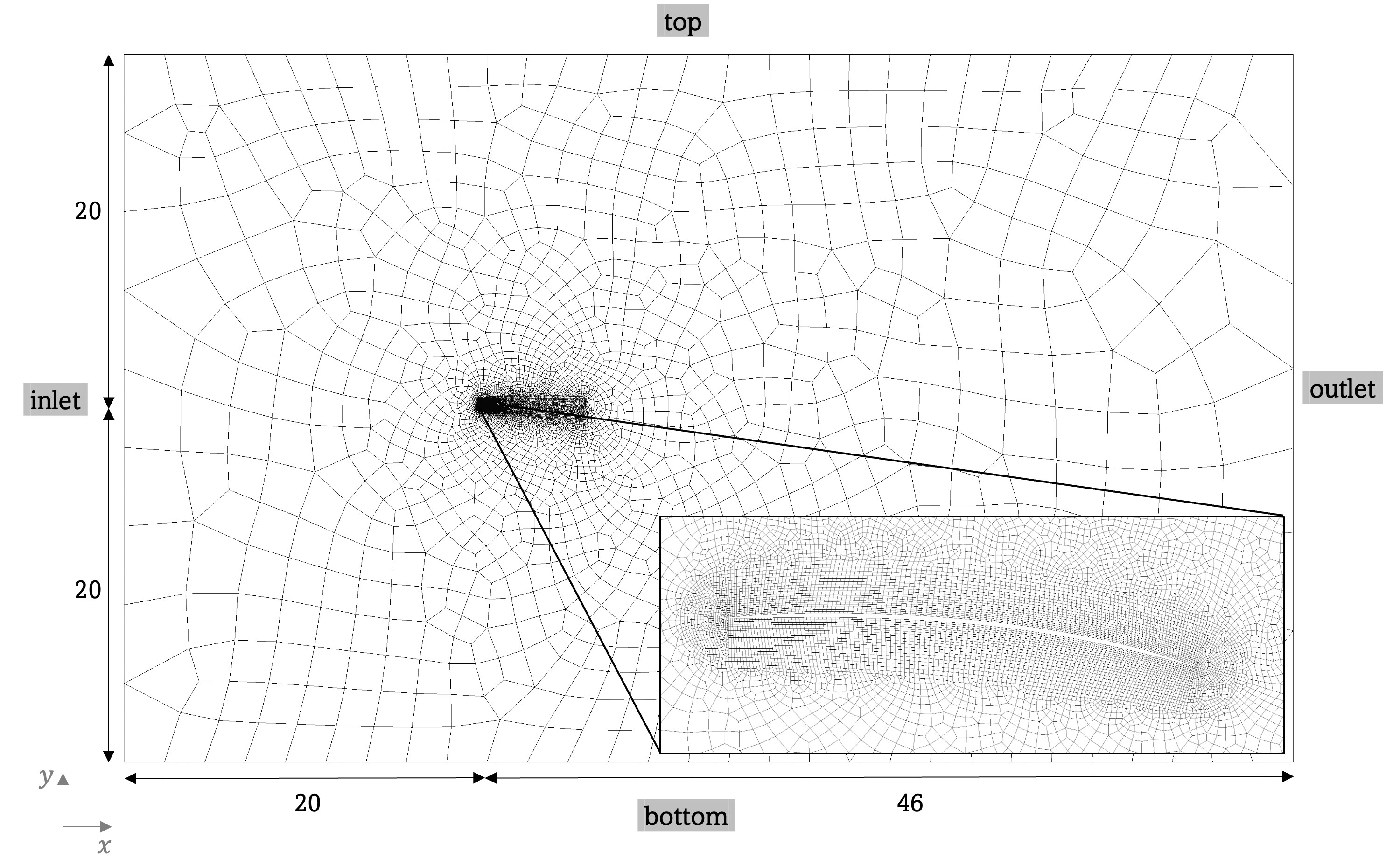}
        \caption{roamx-0201}
        \end{subfigure}
        \caption{Views of domain and mesh in the streamwise-vertical xy-plane used for 2D and 3D-SP simulations with PyFR for the CLF5605 (a) and roamx-0201 (b) airfoils when $\alpha = 6^{\circ}$.}
        \label{fig:PyFR-2D-domain}
\end{figure}

%% file: Figures-tex/aero-forces.tex
%\subsection{CLF5605}
% ---------------------------------------------------------------- clf5605 Cl - AoA, Cd - AoA

\begin{figure}[htbp]\centering
\includegraphics[width=\linewidth]{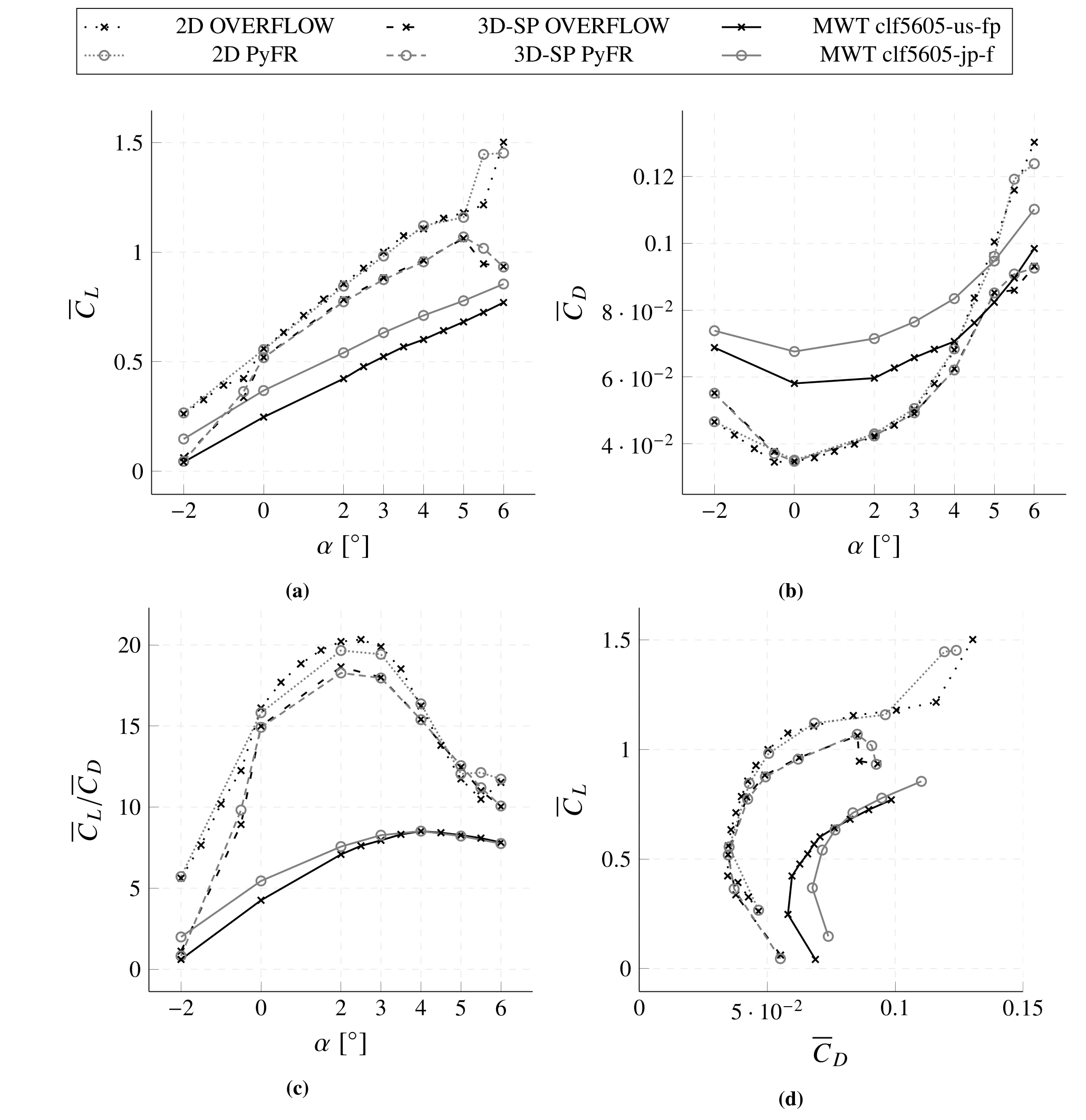}
\caption{Plots of time-averaged lift coefficient $\overline{C}_L$ (a), drag coefficient $\overline{C}_D$ (b), lift-to-drag-ratio $\overline{C}_L/\overline{C}_D$ (c) as a function of angle of attack $\alpha$ and lift coefficient $\overline{C}_L$ as a function of drag coefficient $\overline{C}_D$ (d) for the CLF5605 airfoil. }
\label{fig:Cl-Cd-clf-5605}
\end{figure}

% ---------------------------------------------------------------- roamx-0201 Cl - AoA, Cd - AoA

\begin{figure}[htbp]\centering
\includegraphics[width=\linewidth]{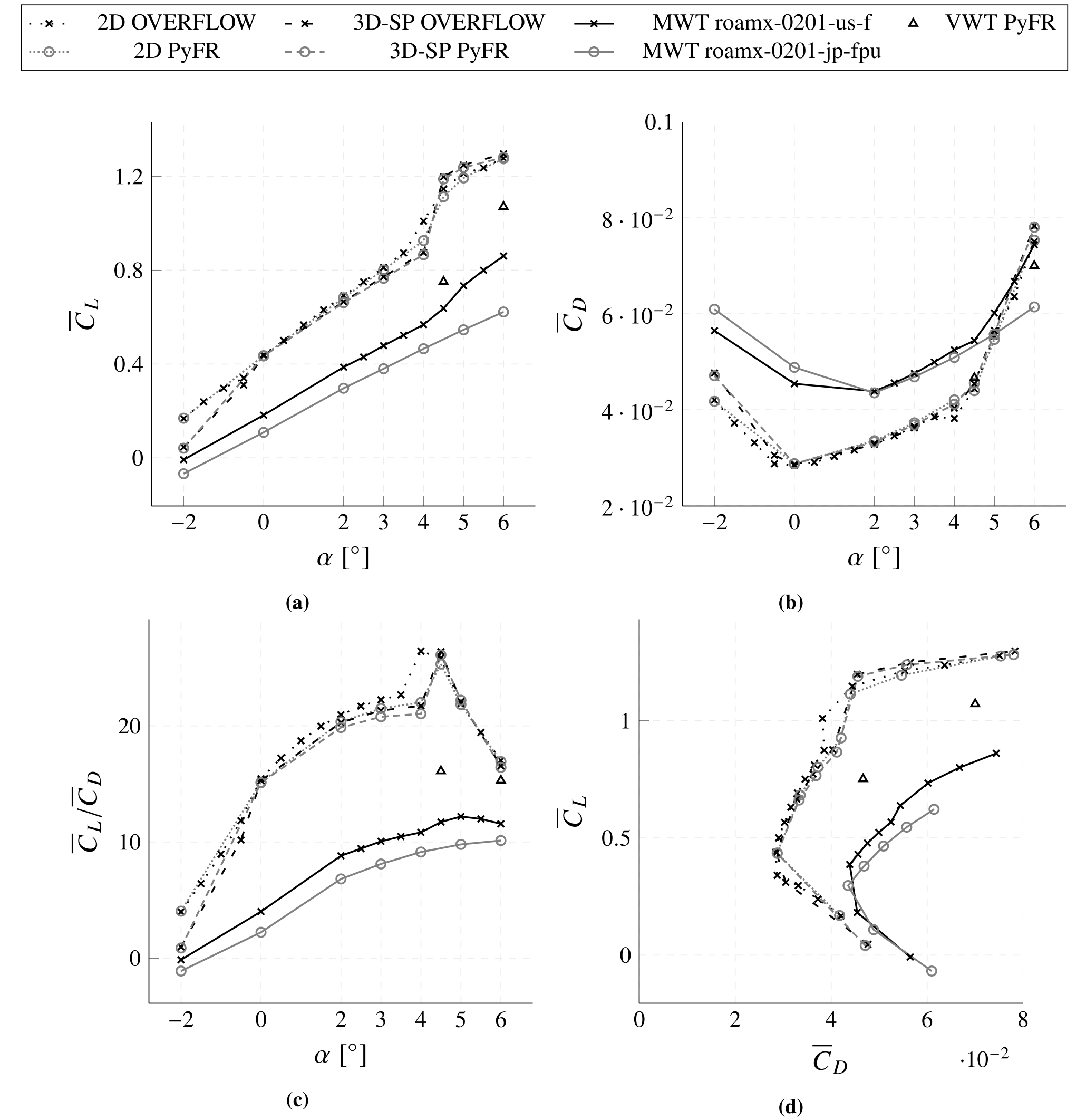}
\caption{Plots of time-averaged lift coefficient $\overline{C}_L$ (a), drag coefficient $\overline{C}_D$ (b), lift-to-drag-ratio $\overline{C}_L/\overline{C}_D$ (c) as a function of angle of attack $\alpha$ and lift coefficient $\overline{C}_L$ as a function of drag coefficient $\overline{C}_D$ (d) for the roamx-0201 airfoil.}
\label{fig:Cl-Cd-roamx-0201}
\end{figure}

%% file: Figures-tex/pressure-clf5605.tex
% ---------------------------------------------------------------- clf-5605 Cp 

\begin{figure}[!h]\centering
\includegraphics[width=\linewidth]{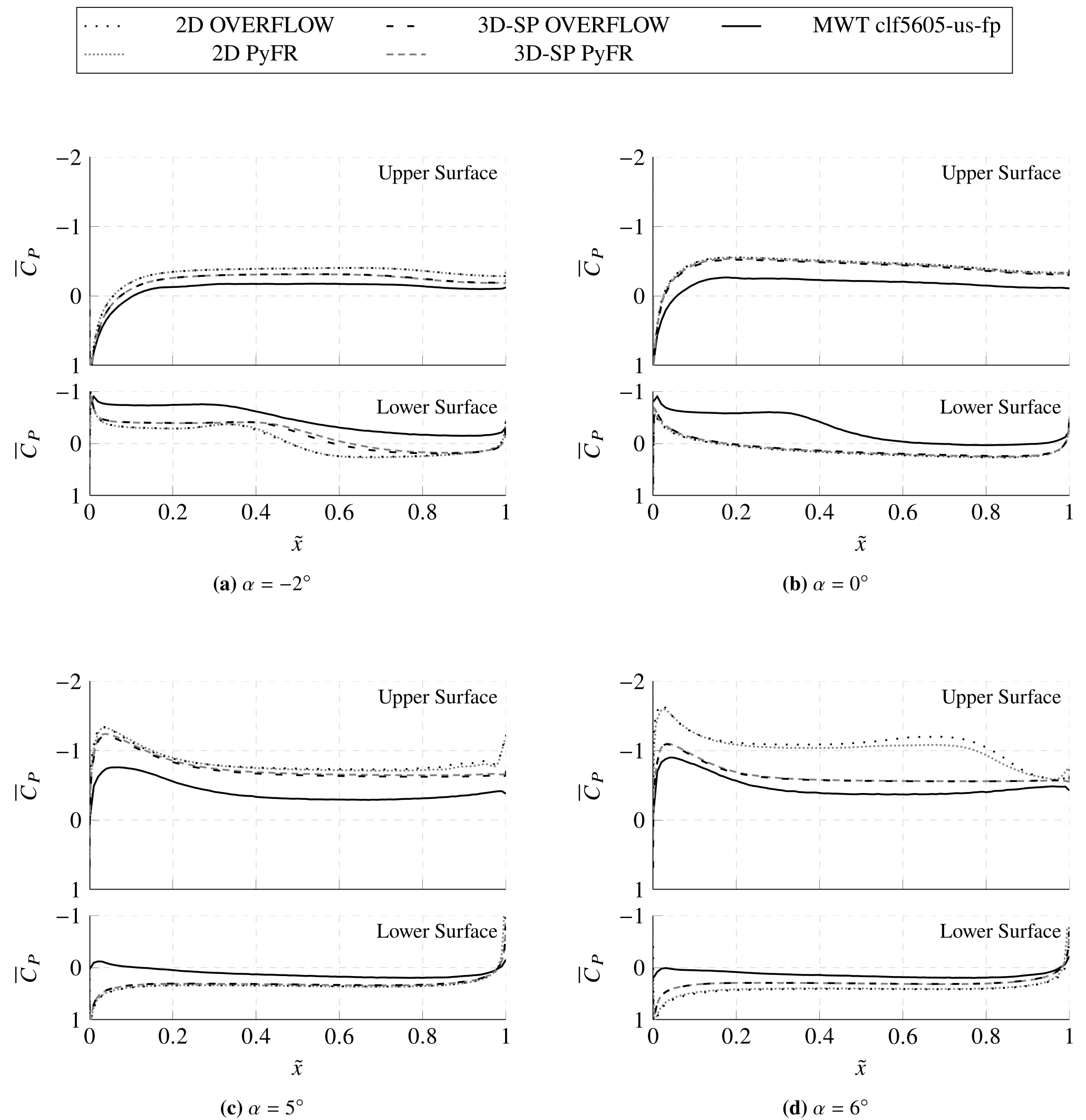}
\caption{Plots of time-averaged pressure coefficient $\overline{C}_P$ as a function of $\tilde{x}$ for different angles of attack for the CLF5605 airfoil. }
\label{fig:Cp-clf-5605}
\end{figure}

%% file: Figures-tex/pressure-roamx0201.tex
% ---------------------------------------------------------------- roamx-0201 Cp

\begin{figure}[!h]\centering
\includegraphics[width=\linewidth]{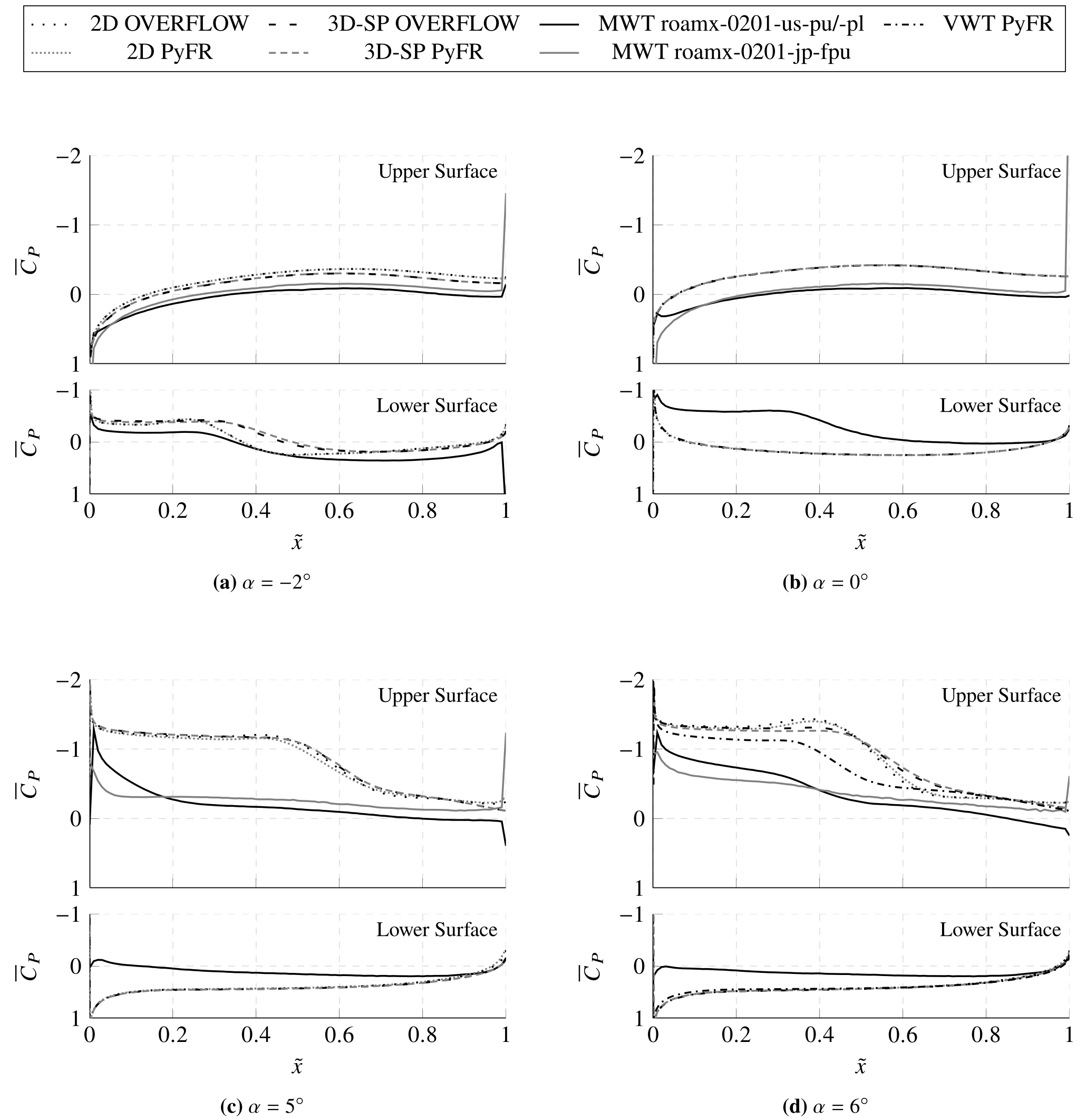}
\caption{Plots of time-averaged pressure coefficient $\overline{C}_P$ as a function of $\tilde{x}$ for different angles of attack for the roamx-0201 airfoil. }
\label{fig:Cp-roamx-0201}
\end{figure}

%% file: Figures-tex/schlieren.tex
\begin{figure}[h!]
  \centering
        \vspace{-1cm}
        \begin{subfigure}[t]{0.46\linewidth}
        \includegraphics[width=\linewidth]{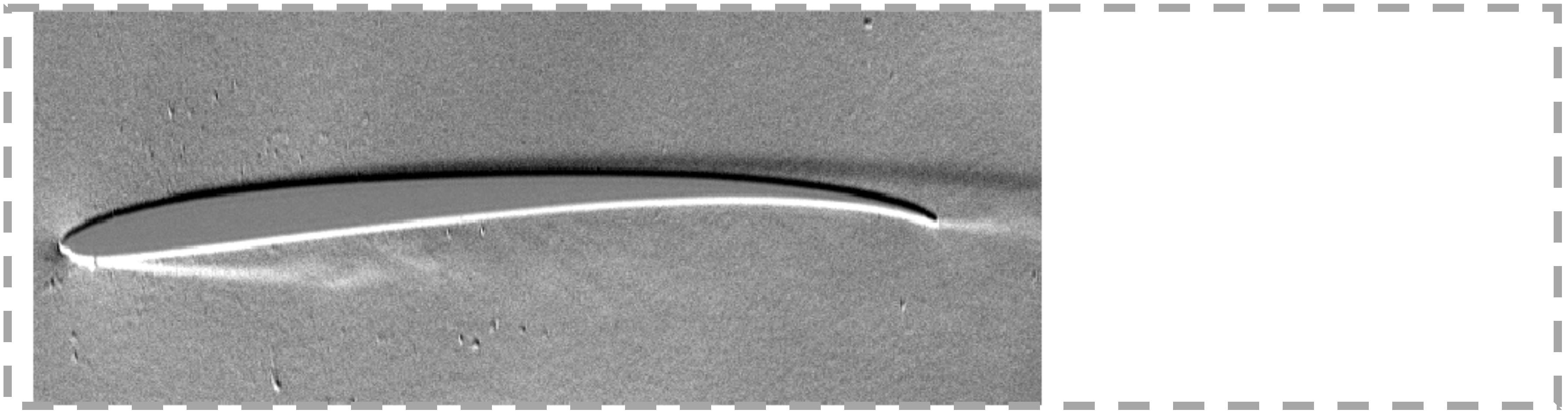}
        \includegraphics[width=\linewidth]{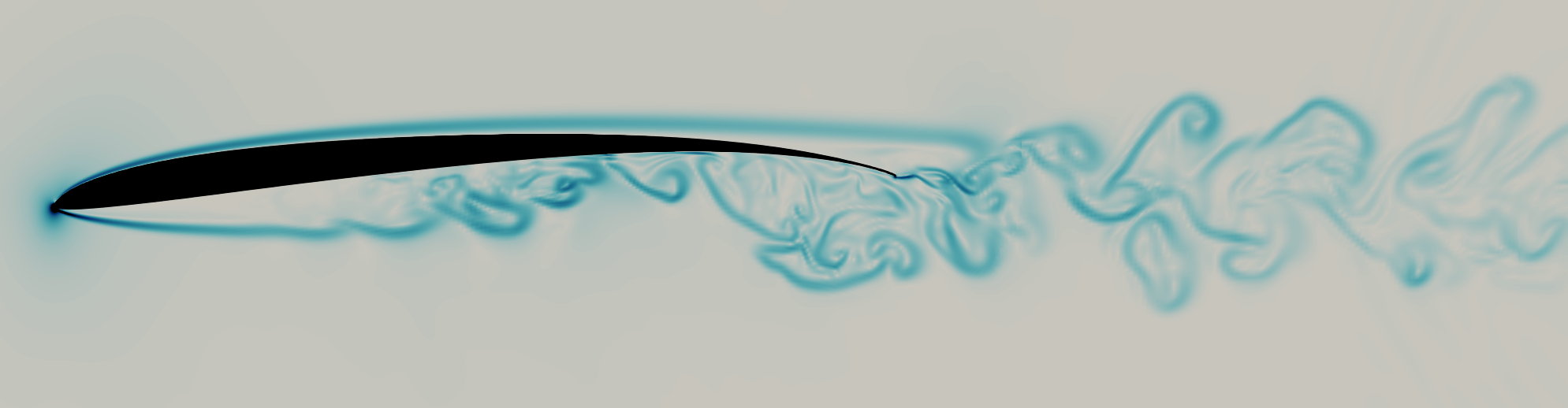}
        \includegraphics[width=\linewidth]{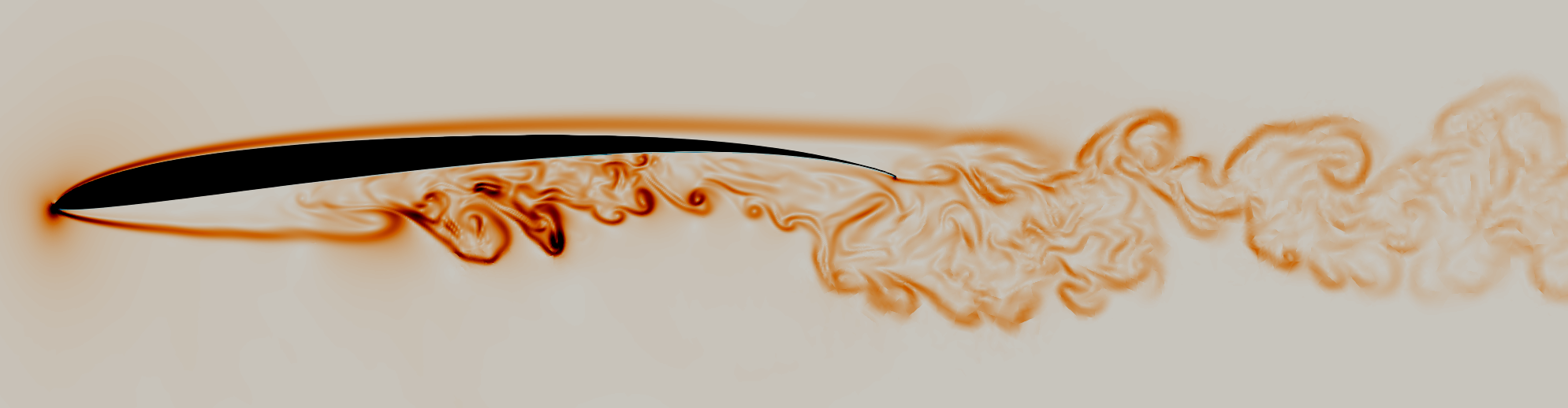}
        \includegraphics[width=\linewidth]{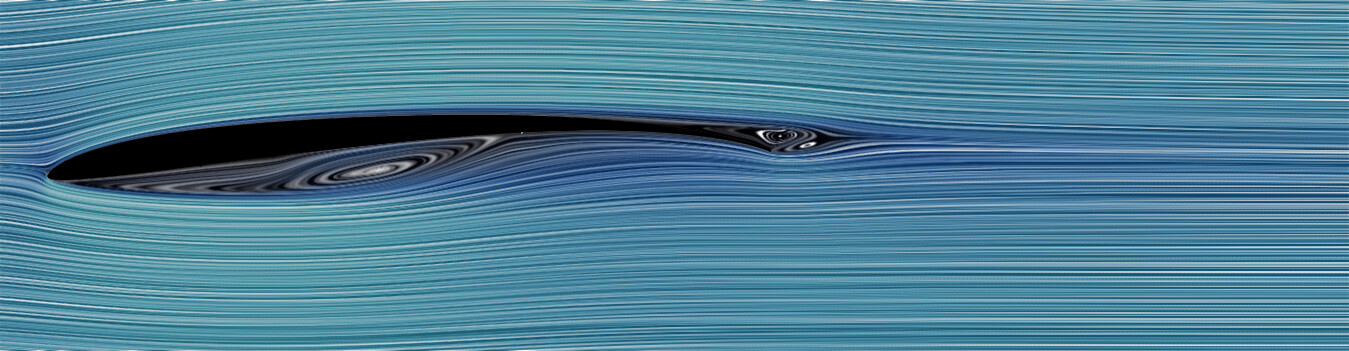}
        \includegraphics[width=\linewidth]{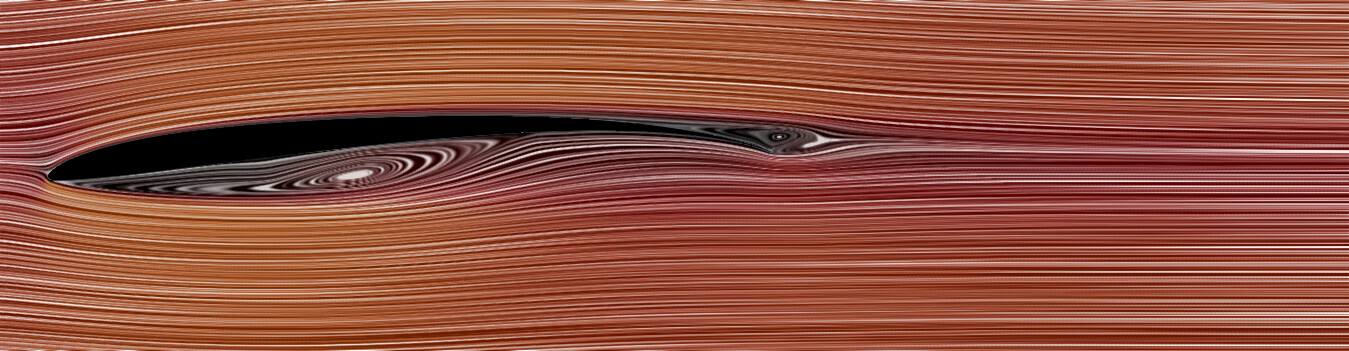}
        \caption{$\alpha = -2^\circ$}
        \end{subfigure}  
        \hspace{0.25cm}
        \begin{subfigure}[t]{0.46\linewidth}
        \includegraphics[width=\linewidth]{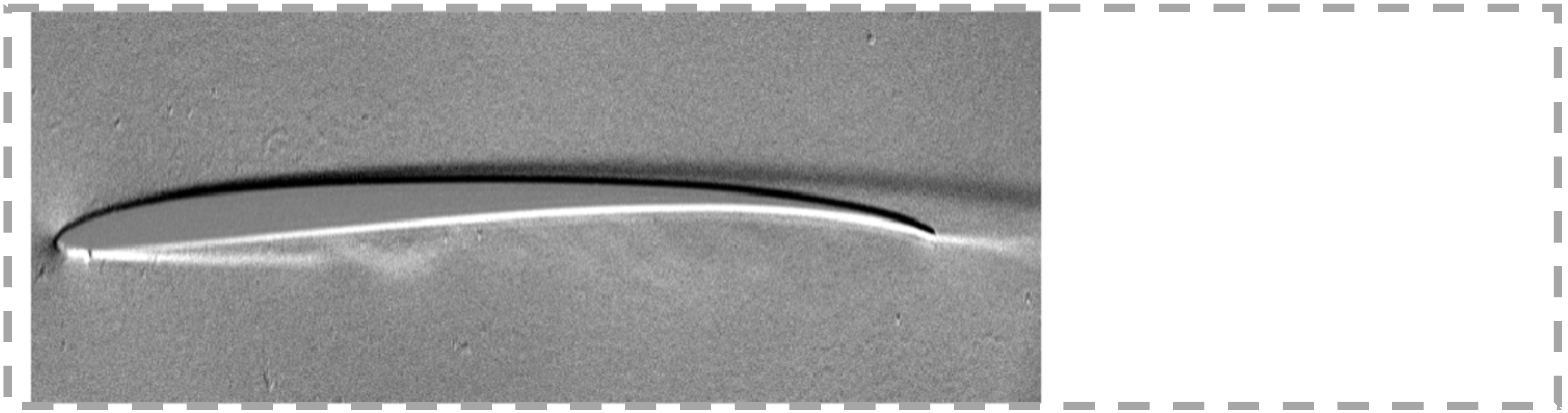}
        \includegraphics[width=\linewidth]{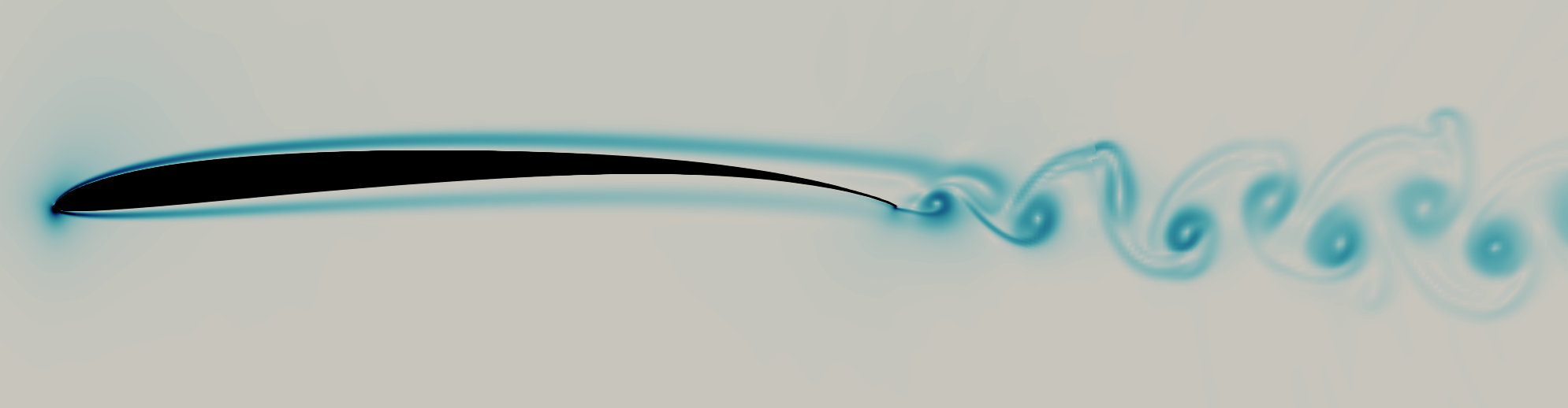}
        \includegraphics[width=\linewidth]{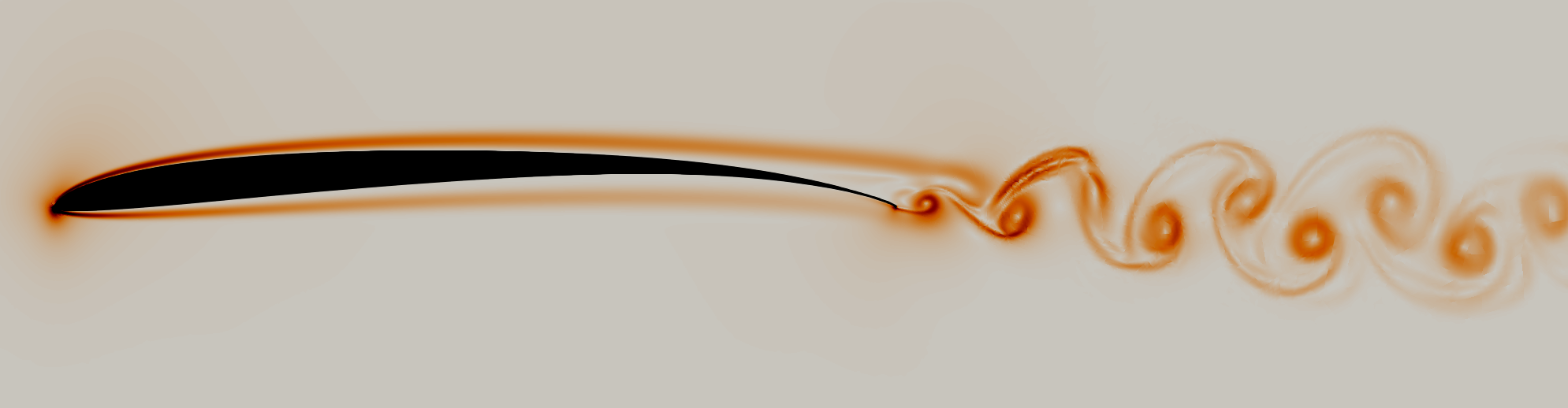}
        \includegraphics[width=\linewidth]{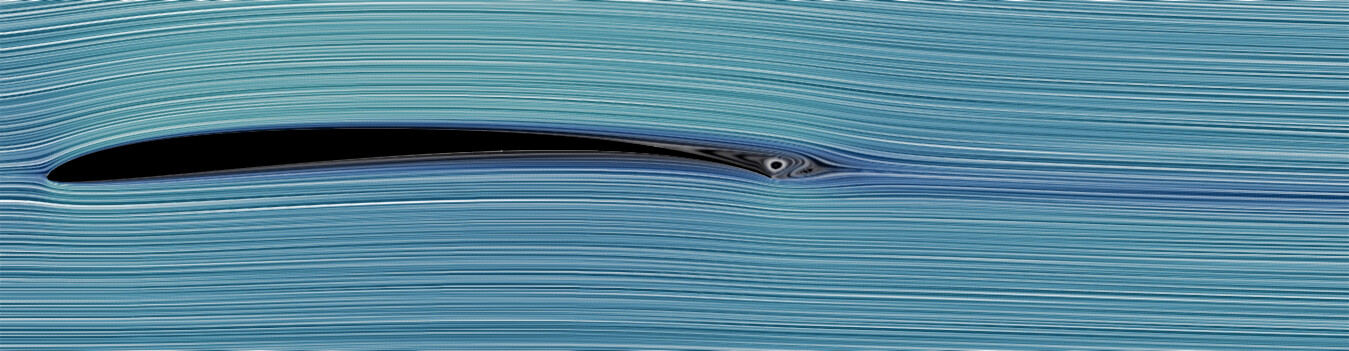}
        \includegraphics[width=\linewidth]{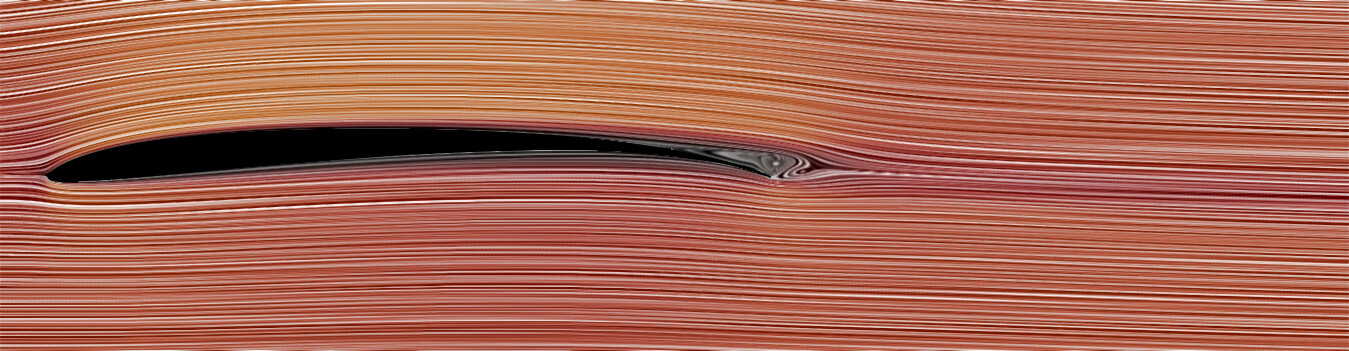}
        \caption{$\alpha = 0^\circ$}
        \end{subfigure}
        \begin{subfigure}[t]{0.46\linewidth}
        \includegraphics[width=\linewidth]{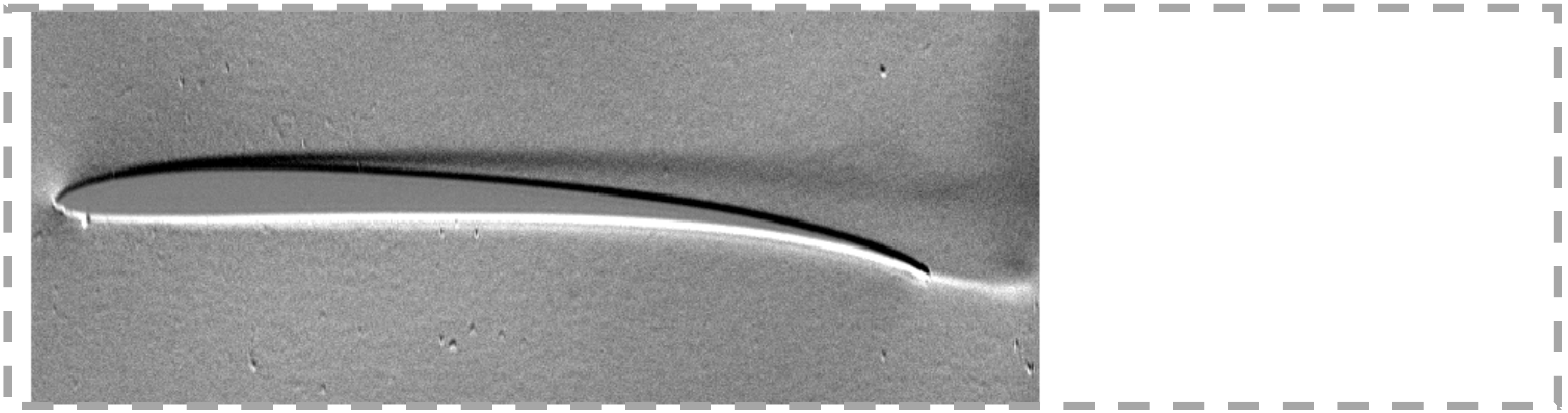}
        \includegraphics[width=\linewidth]{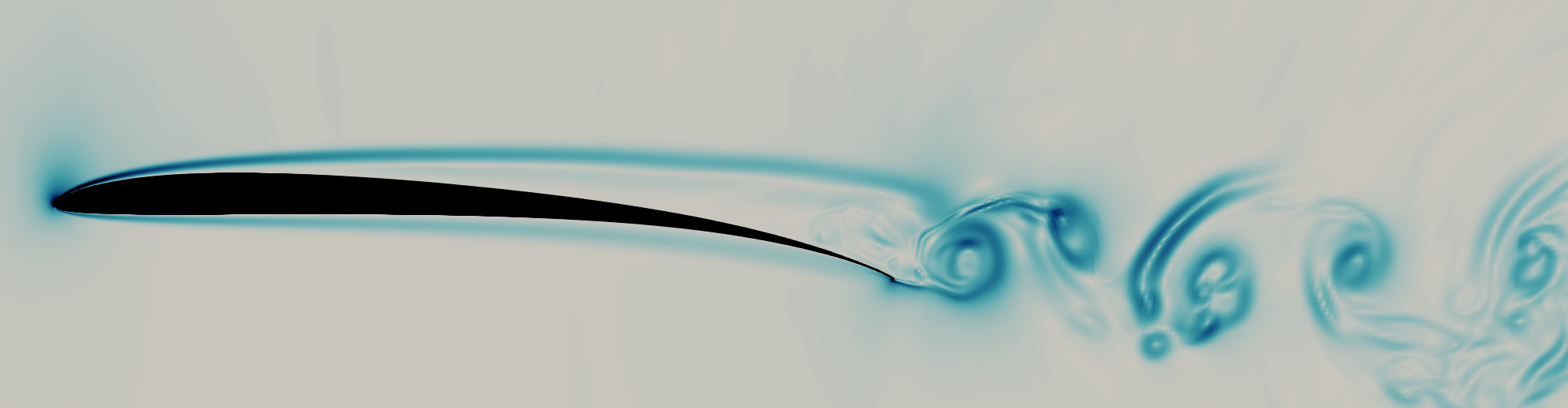}
        \includegraphics[width=\linewidth]{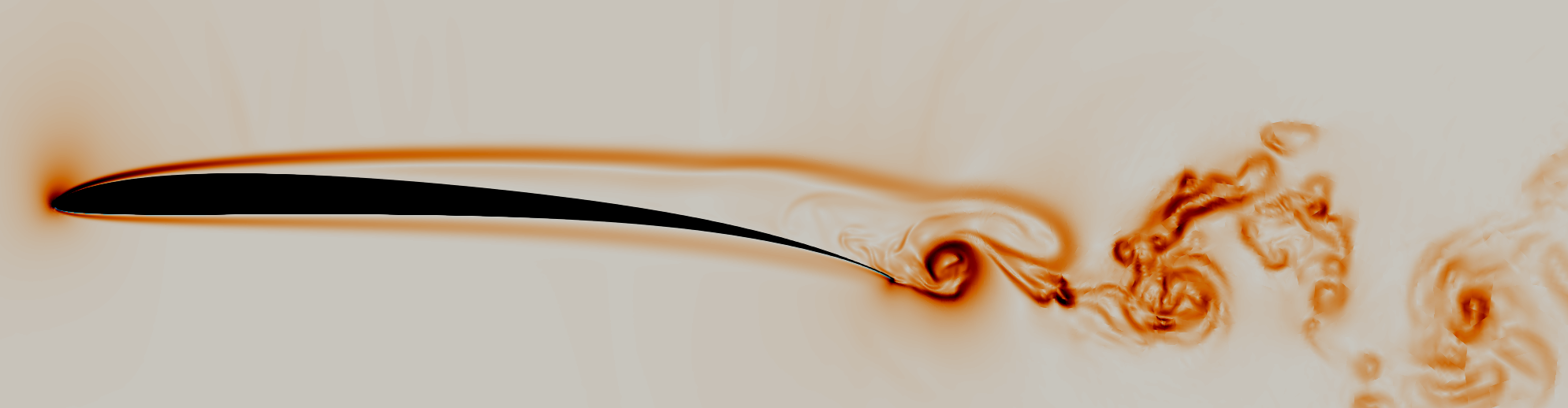}
        \includegraphics[width=\linewidth]{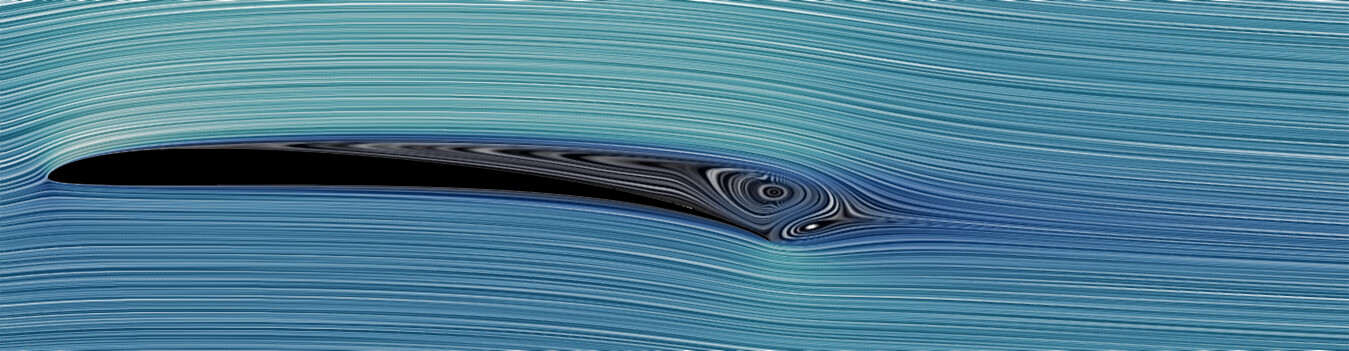}
        \includegraphics[width=\linewidth]{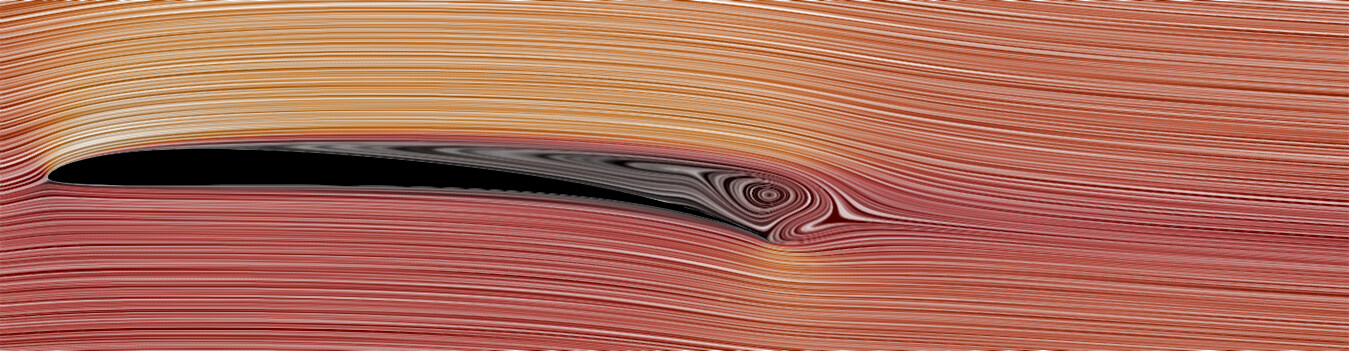}
        \caption{$\alpha = 5^\circ$}
        \end{subfigure} 
        \hspace{0.25cm}
        \begin{subfigure}[t]{0.46\linewidth}
        \includegraphics[width=\linewidth]{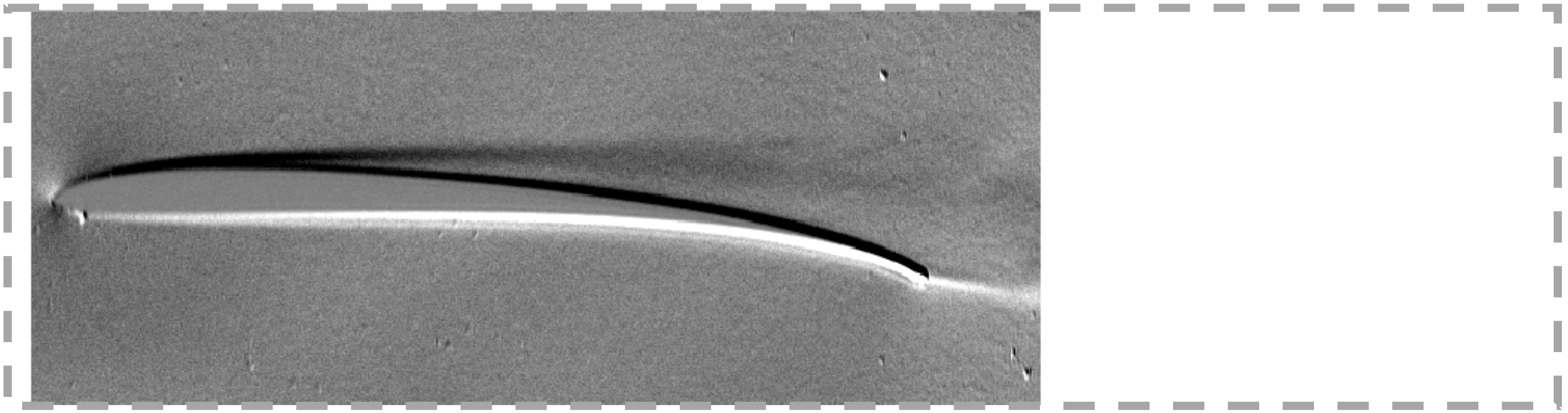}
        \includegraphics[width=\linewidth]{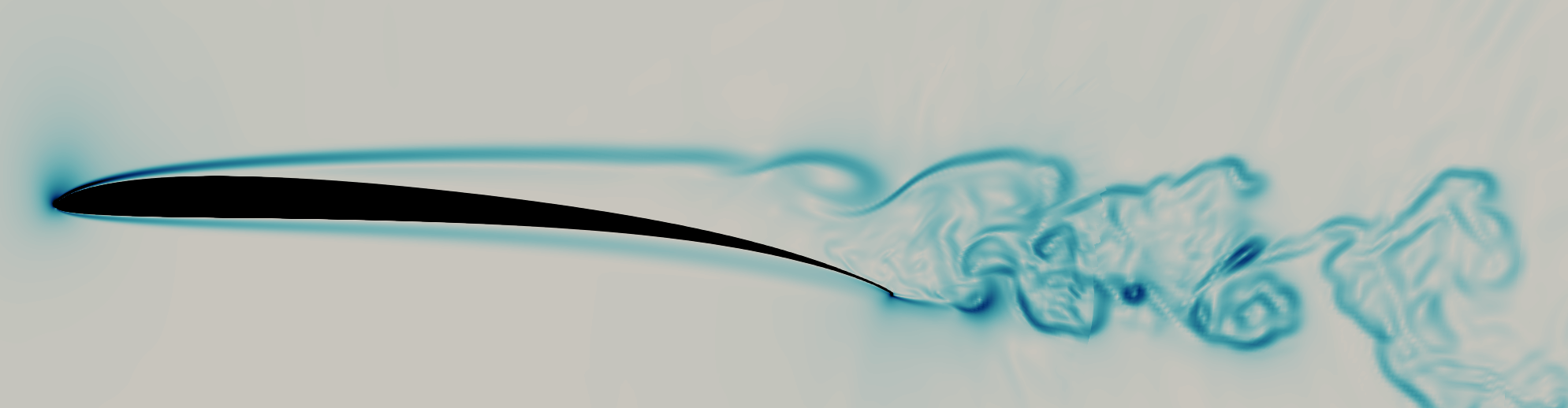}
        \includegraphics[width=\linewidth]{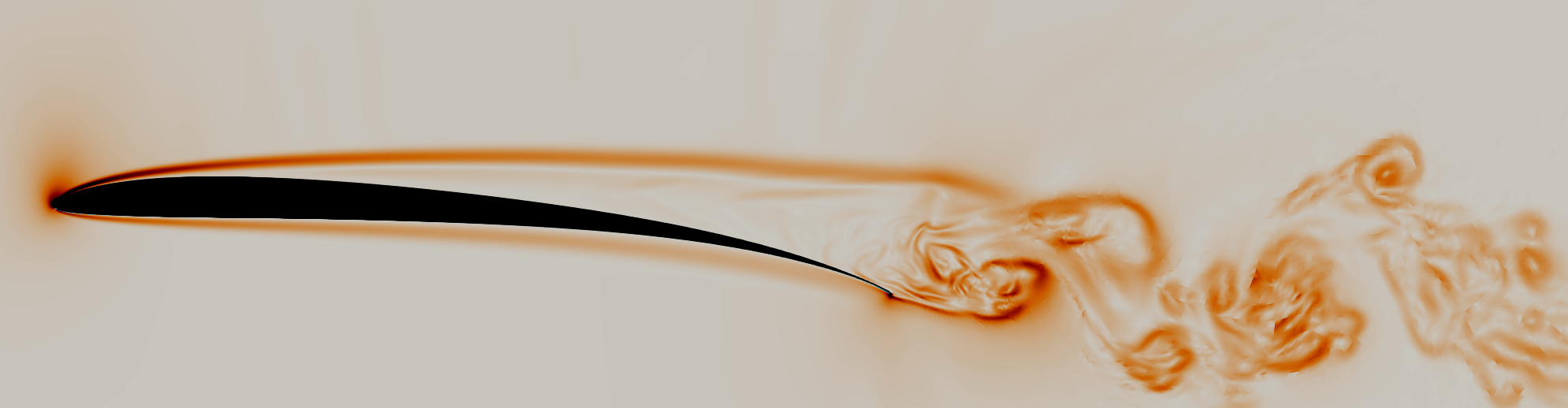}
        \includegraphics[width=\linewidth]{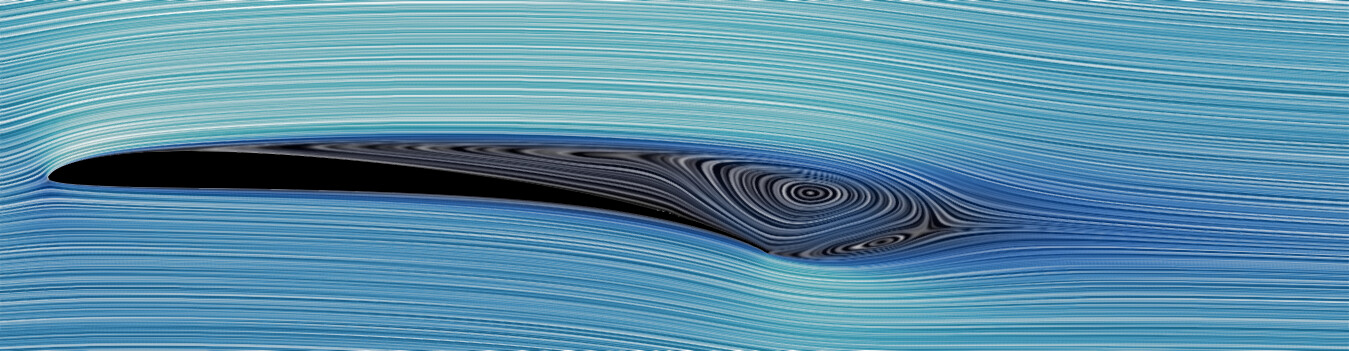}
        \includegraphics[width=\linewidth]{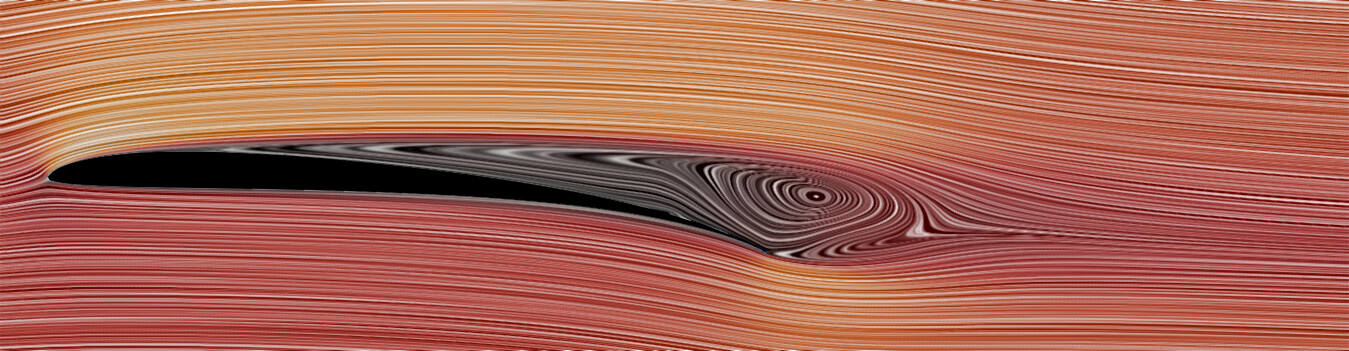}
        \caption{$\alpha = 6^\circ$}
        \end{subfigure}
        \caption{Schlieren images for MWT experiments on clf5605-us-s (top) and images of instantaneous density gradient magnitude $|\boldsymbol{\nabla}\rho|$ (middle) and of time-averaged normalised velocity magnitude $|\mathbf{v}|/v_\infty$ with superimposed LICs (bottom) for 3D-SP OVERFLOW (blue) and 3D-SP PyFR (red) on CLF5605 for different $\alpha$.}
        \label{fig:schlieren-clf}
\end{figure}

\begin{figure}[h!]
  \centering
        \vspace{-1cm}
        \begin{subfigure}[t]{0.46\linewidth}
        \includegraphics[width=\linewidth]{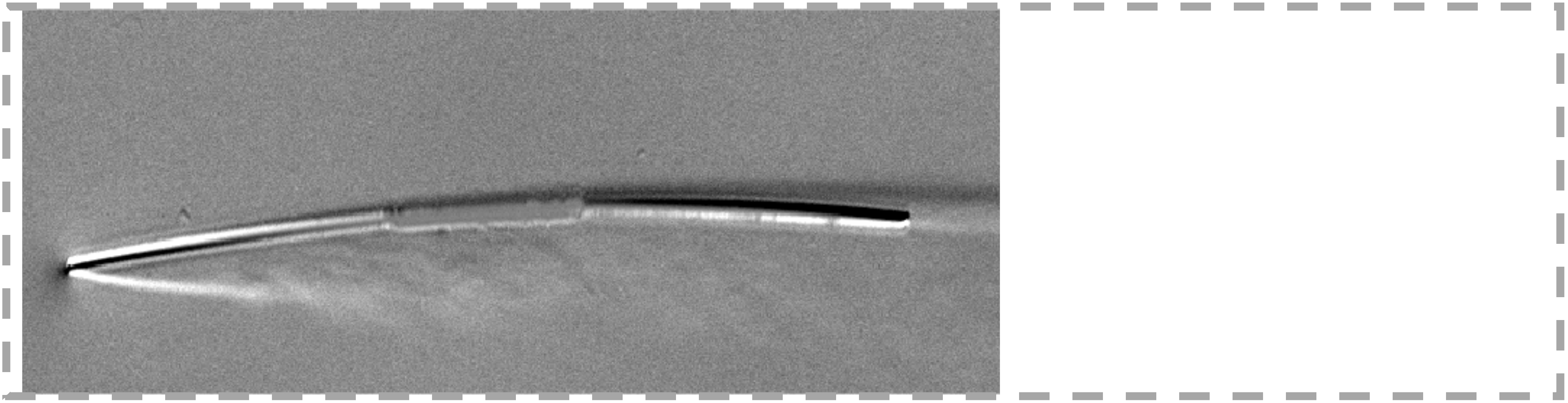}
        \includegraphics[width=\linewidth]{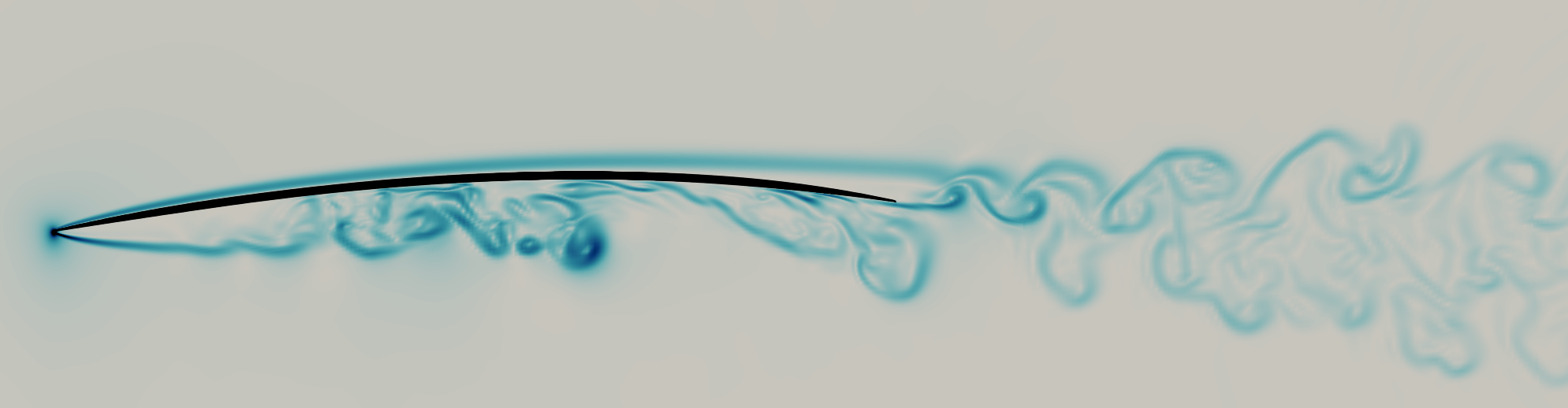}
        \includegraphics[width=\linewidth]{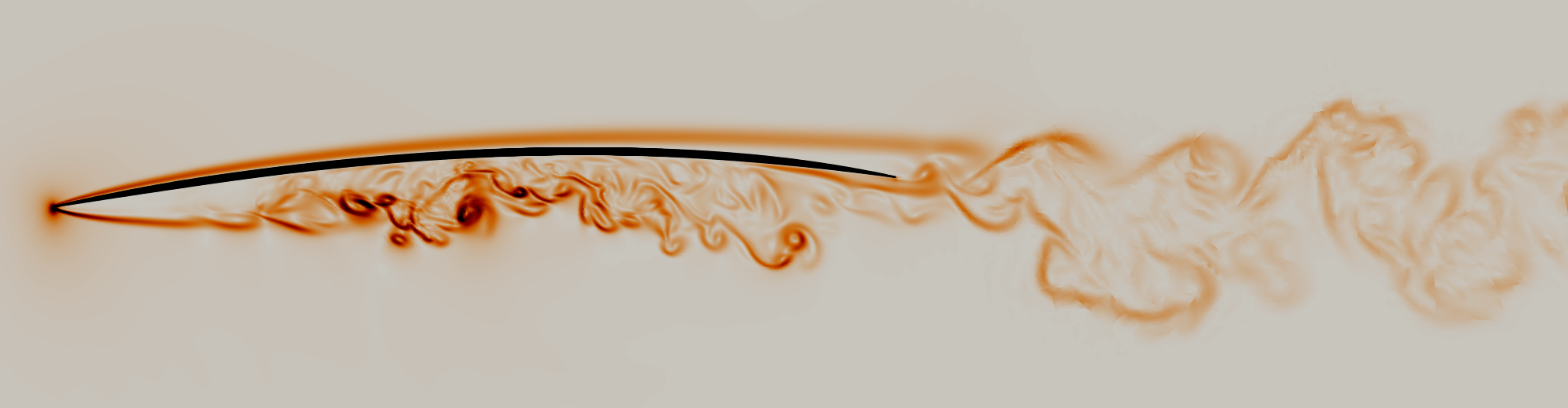}
        \includegraphics[width=\linewidth]{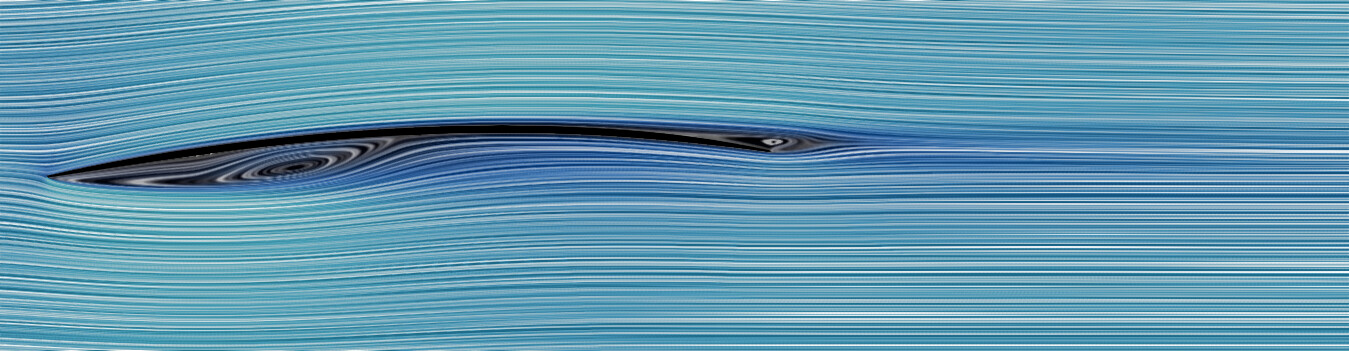}
        \includegraphics[width=\linewidth]{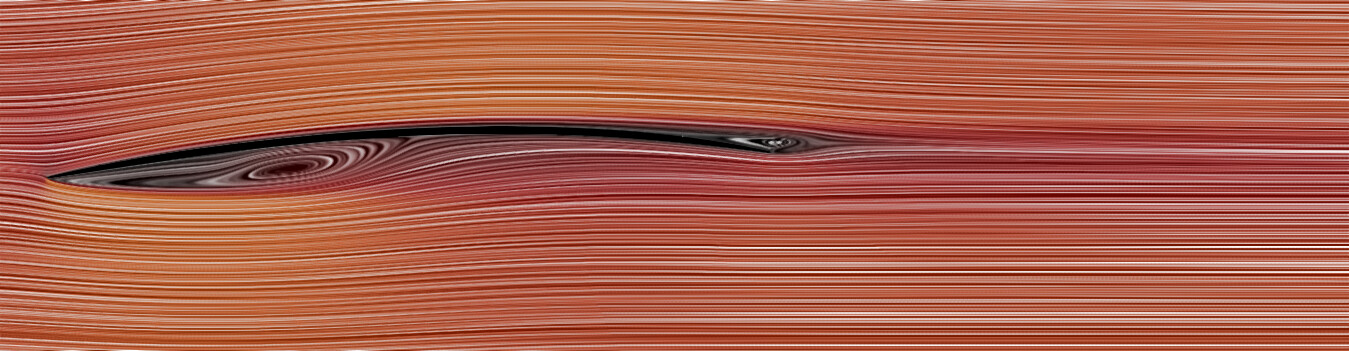}
        \caption{$\alpha = -2^\circ$}
        \end{subfigure} 
        \hspace{0.25cm}
        \begin{subfigure}[t]{0.46\linewidth}
        \includegraphics[width=\linewidth]{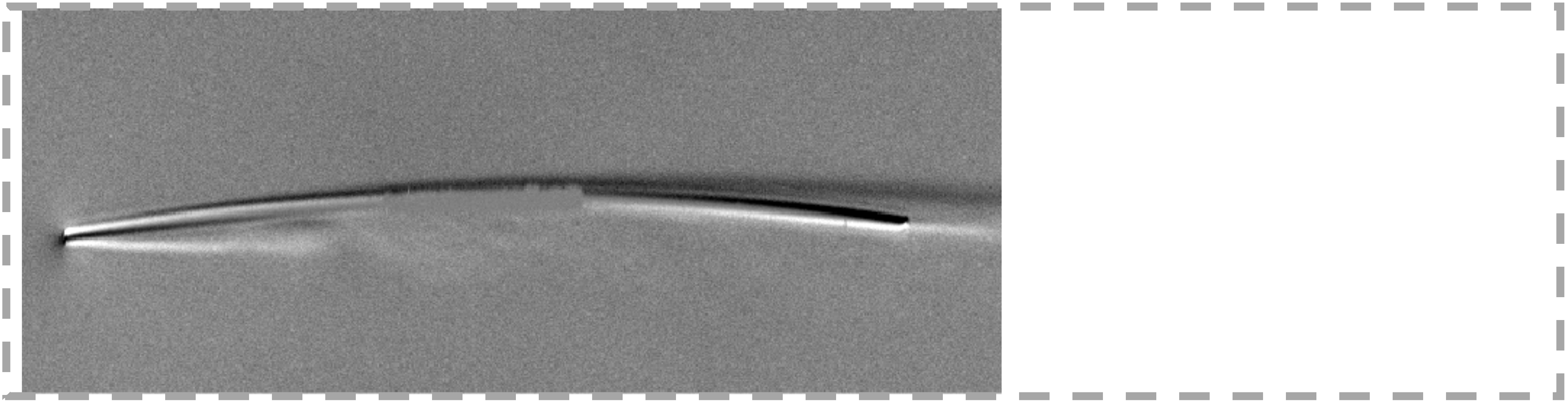}
        \includegraphics[width=\linewidth]{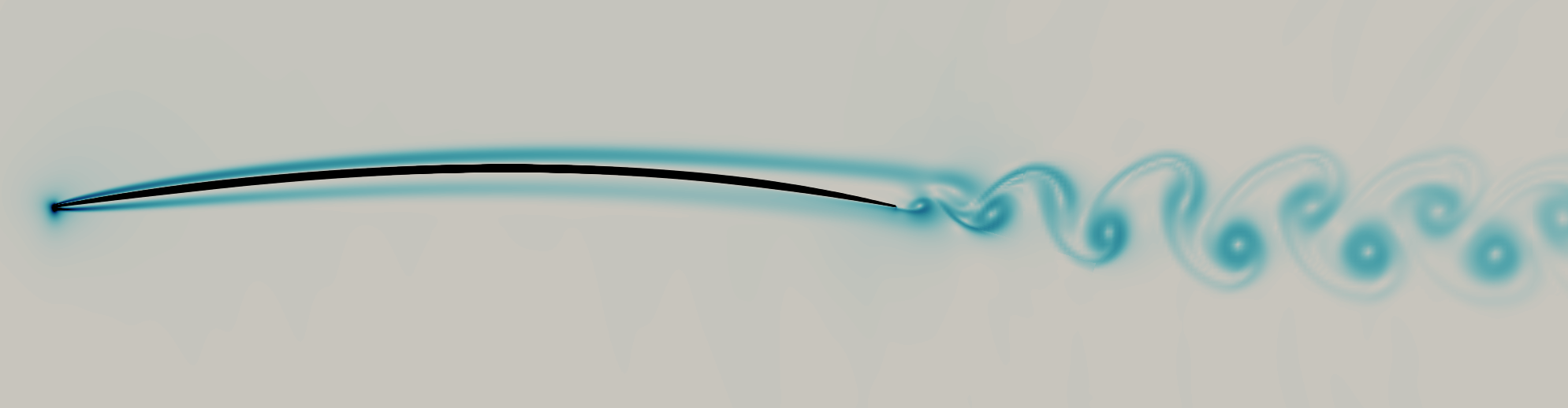}
        \includegraphics[width=\linewidth]{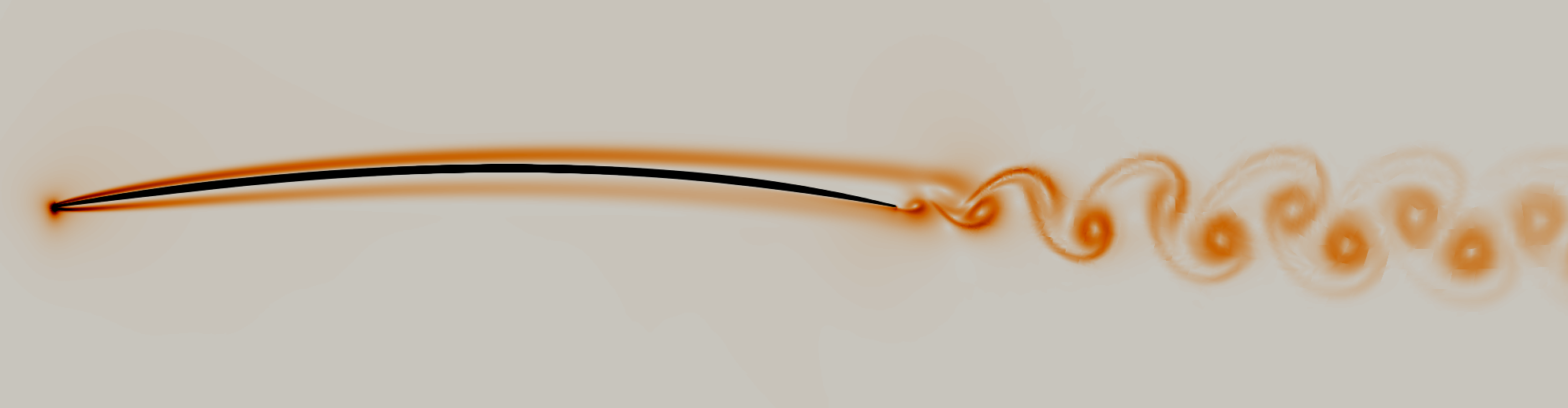}
        \includegraphics[width=\linewidth]{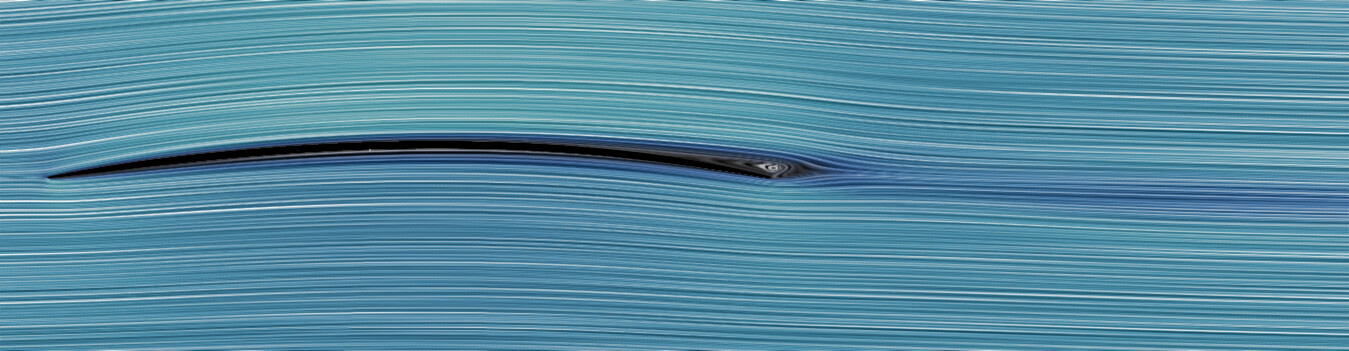}
        \includegraphics[width=\linewidth]{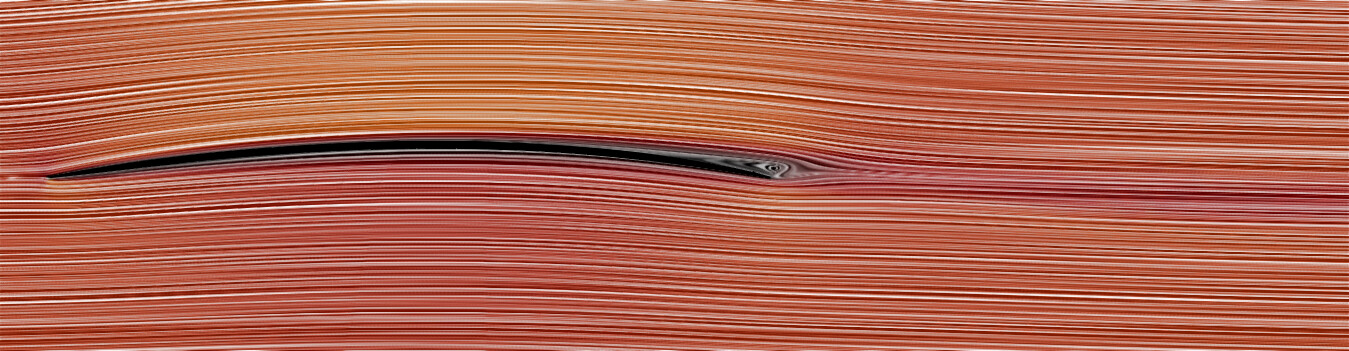}
        \caption{$\alpha = 0^\circ$}
        \end{subfigure}
        \begin{subfigure}[t]{0.46\linewidth}
        \includegraphics[width=\linewidth]{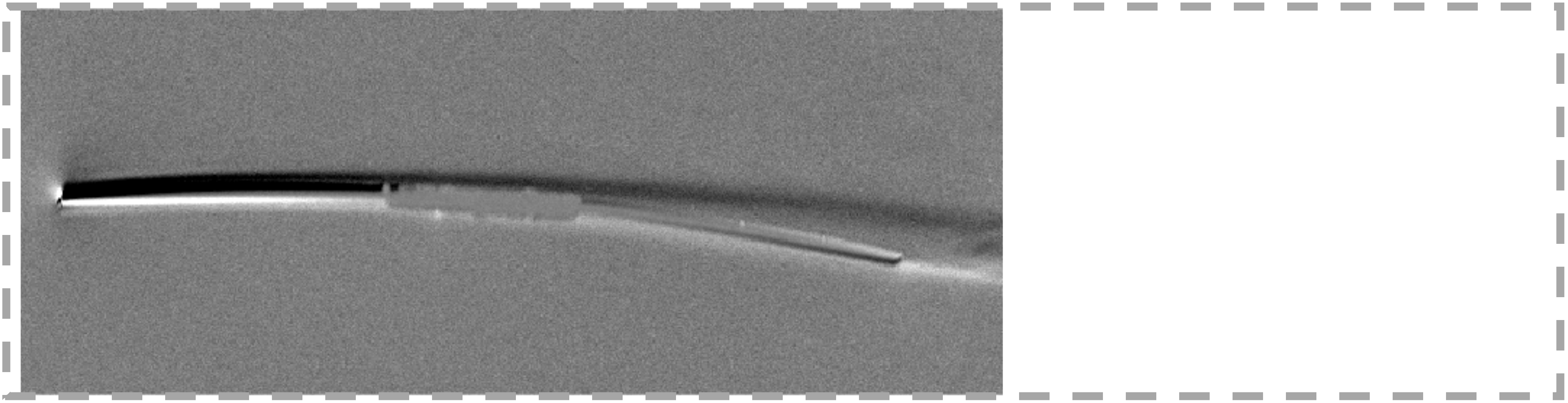}
        \includegraphics[width=\linewidth]{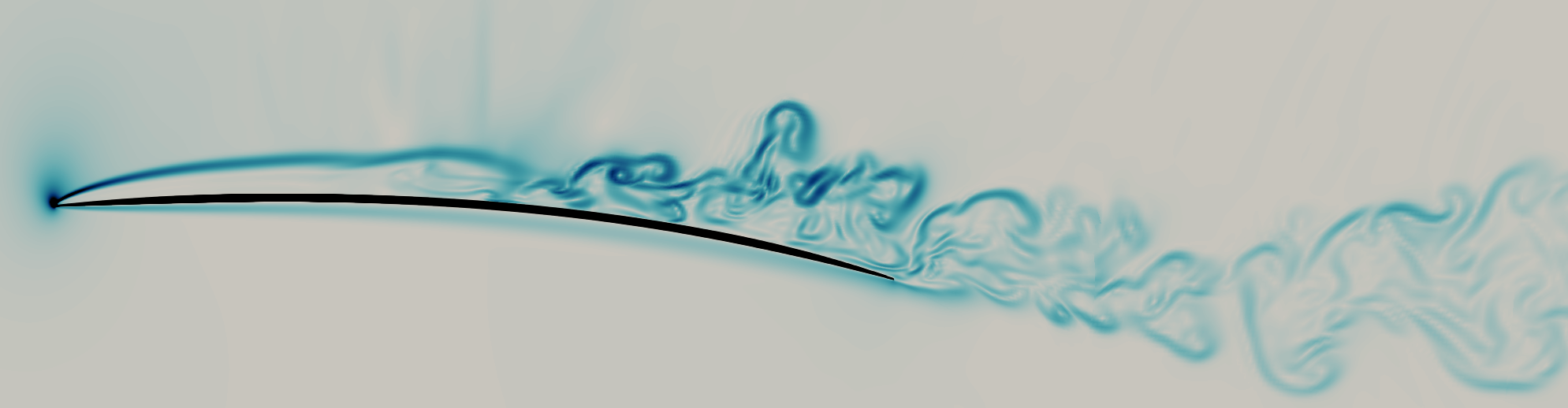}
        \includegraphics[width=\linewidth]{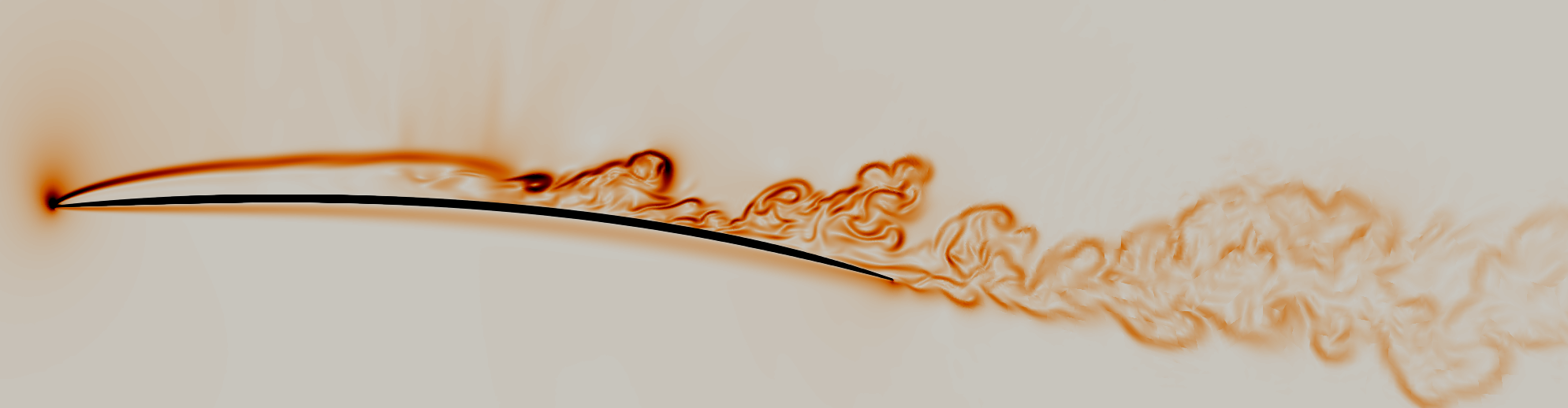}
        \includegraphics[width=\linewidth]{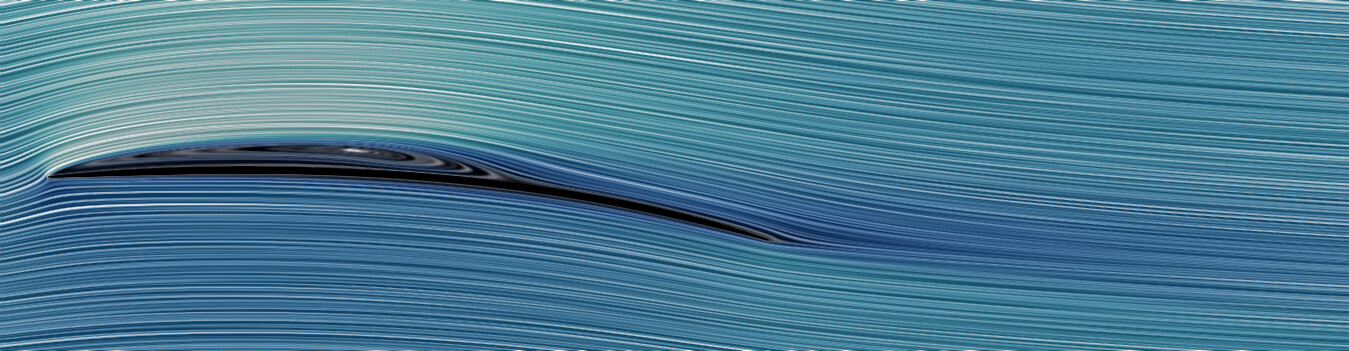}
        \includegraphics[width=\linewidth]{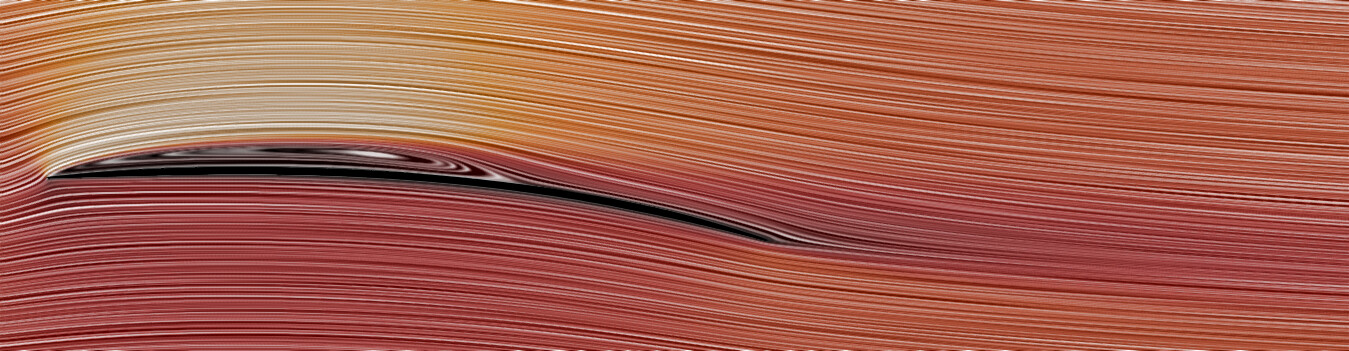}
        \caption{$\alpha = 5^\circ$}
        \end{subfigure}
        \hspace{0.25cm}
        \begin{subfigure}[t]{0.46\linewidth}
        \includegraphics[width=\linewidth]{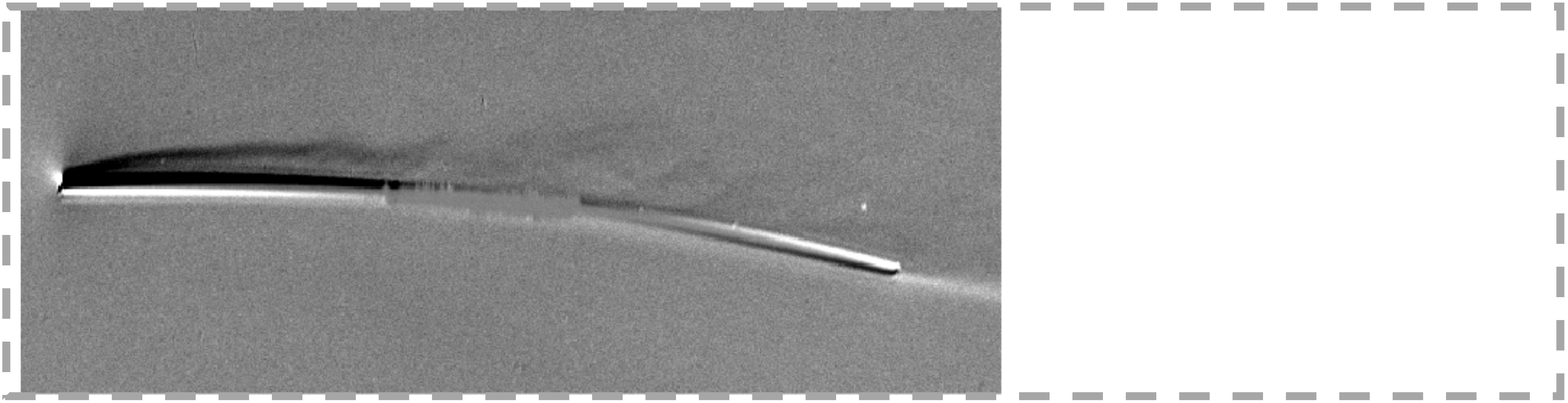}
        \includegraphics[width=\linewidth]{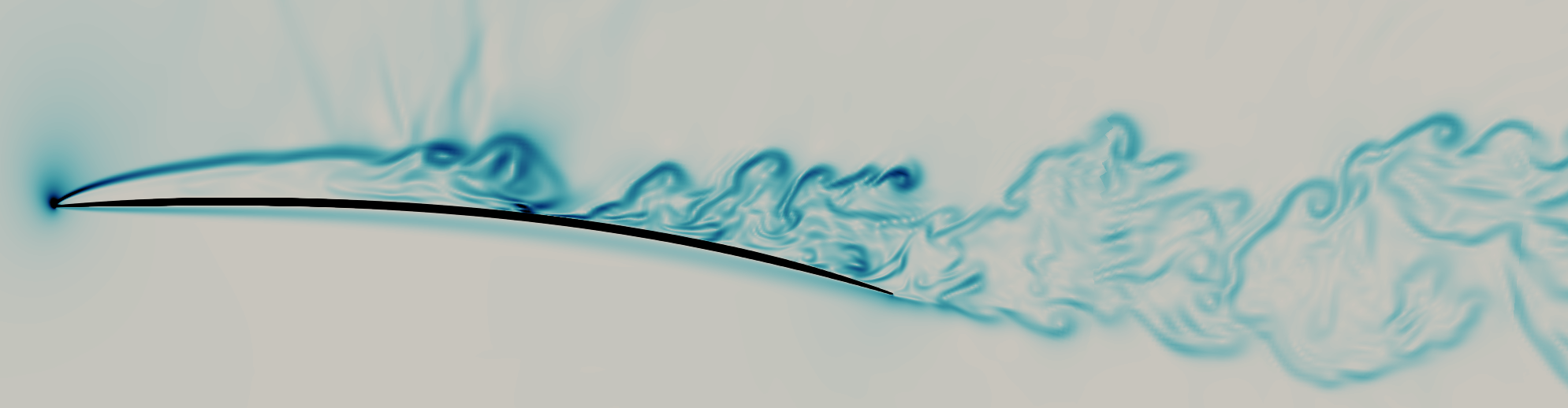}
        \includegraphics[width=\linewidth]{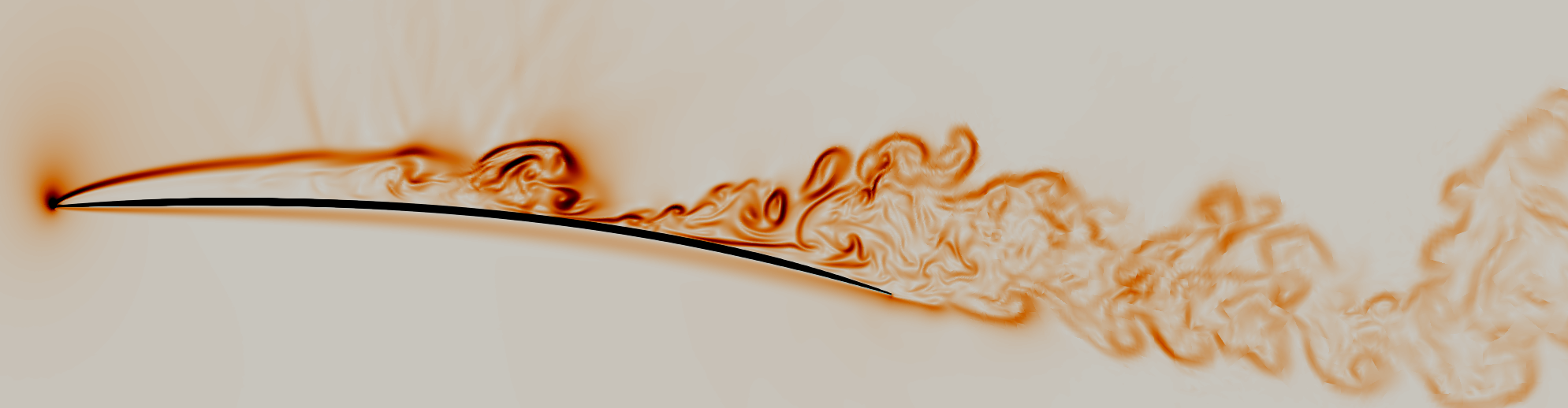}
        \includegraphics[width=\linewidth]{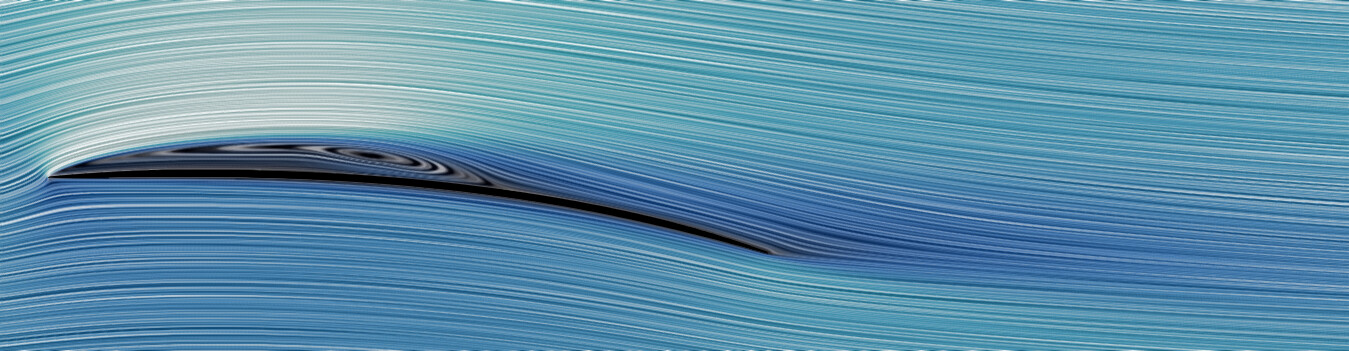}
        \includegraphics[width=\linewidth]{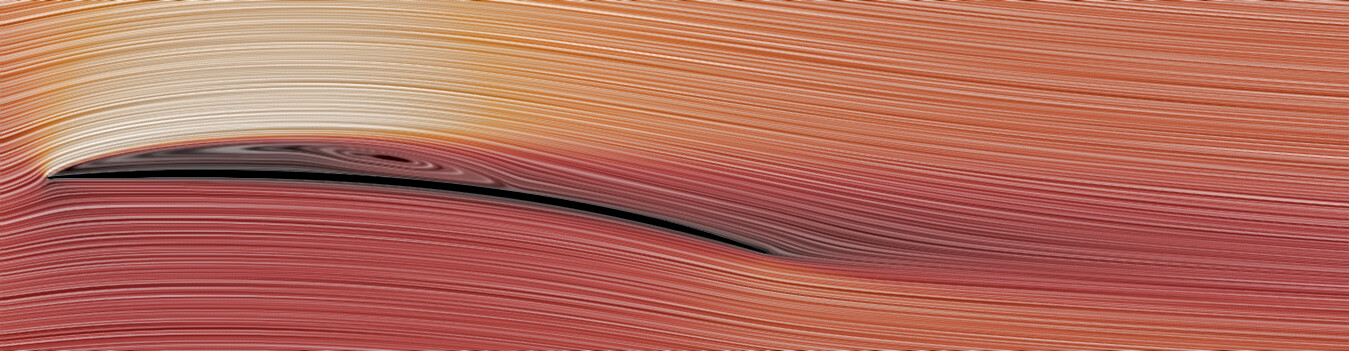}
        \caption{$\alpha = 6^\circ$}
        \end{subfigure}
        \caption{Schlieren images for MWT experiments on roamx-0201-jp-s (top) and images of instantaneous density gradient magnitude $|\boldsymbol{\nabla}\rho|$ (middle) and of time-averaged normalised velocity magnitude $|\mathbf{v}|/v_\infty$ with superimposed LICs (bottom) for 3D-SP OVERFLOW (blue) and 3D-SP PyFR (red) on roamx-0201 for different $\alpha$.}
        \label{fig:schlieren-roamx}
\end{figure}

%% file: Figures-tex/q-criterion.tex
\begin{figure}[h!]
  \centering
        \begin{subfigure}[t]{1\linewidth}
        \centering
        \hspace{1cm}
        \includegraphics[width=0.42\linewidth]{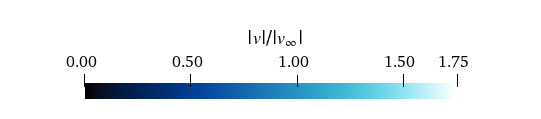}
        \hspace{0.1cm}
        \includegraphics[width=0.42\linewidth]{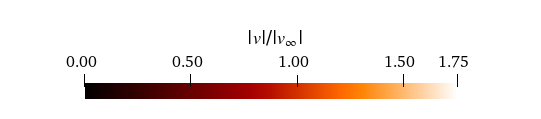}
        \end{subfigure}
        
        \begin{minipage}{0.06\textwidth}
          \centering
          \rotatebox{0}{\small$\alpha = -2^\circ$}
        \end{minipage}
        \begin{minipage}{0.43\textwidth}
          \includegraphics[trim=50 50 0 50,clip,width=\linewidth]{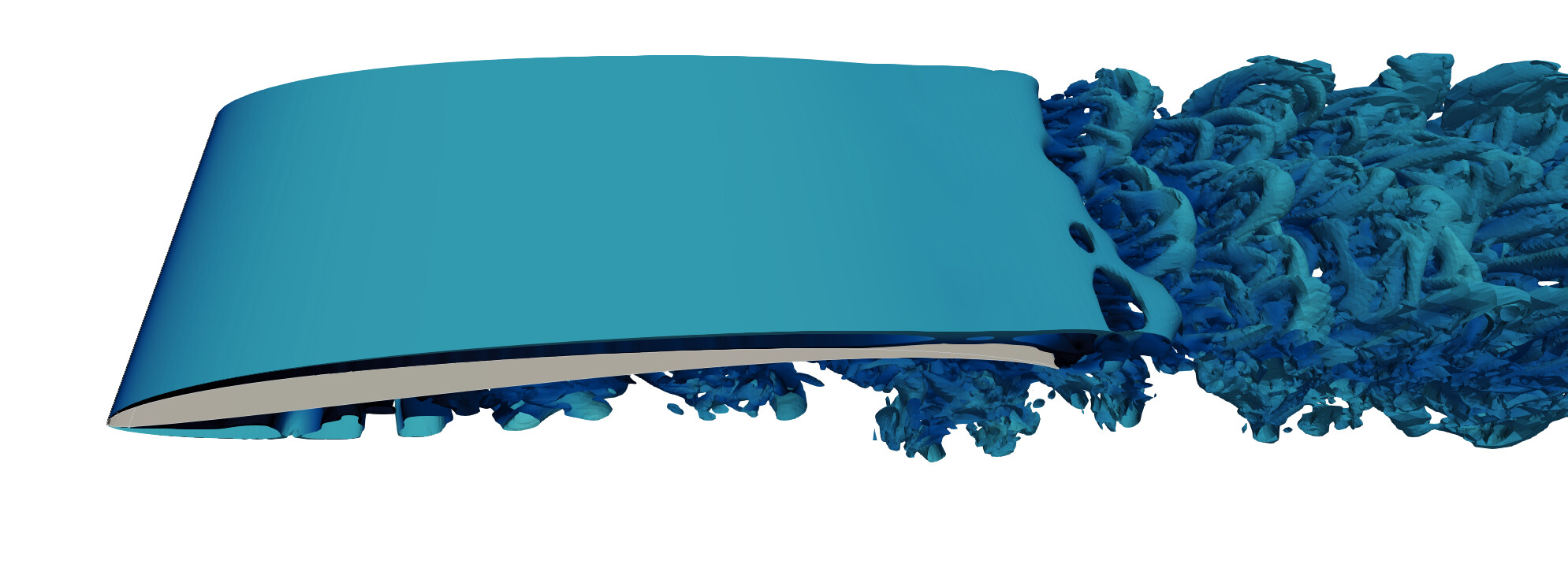}
        \end{minipage}
        \begin{minipage}{0.43\textwidth}
          \includegraphics[trim=50 50 0 50,clip,width=\linewidth]{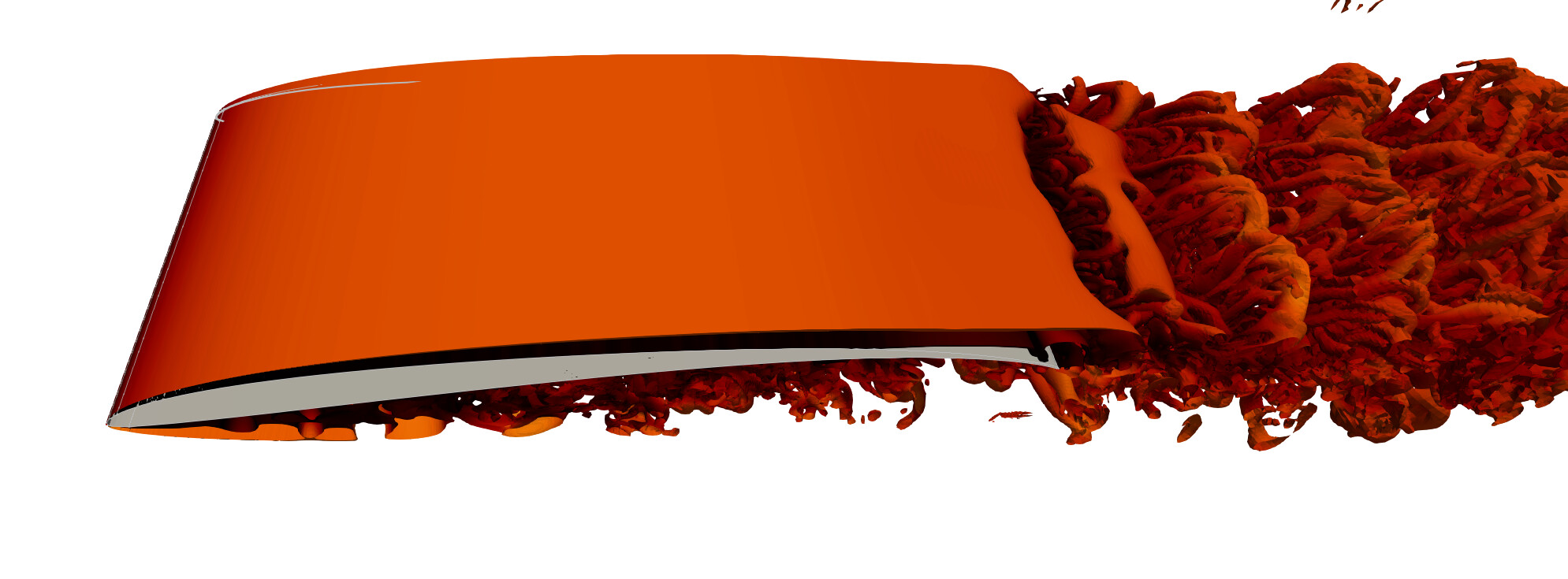}
        \end{minipage}
        
        \begin{minipage}{0.06\textwidth}
          \centering
          \rotatebox{0}{\small$\alpha = 0^\circ$}
        \end{minipage}
        \begin{minipage}{0.43\textwidth}
          \includegraphics[trim=50 50 0 50,clip,width=\linewidth]{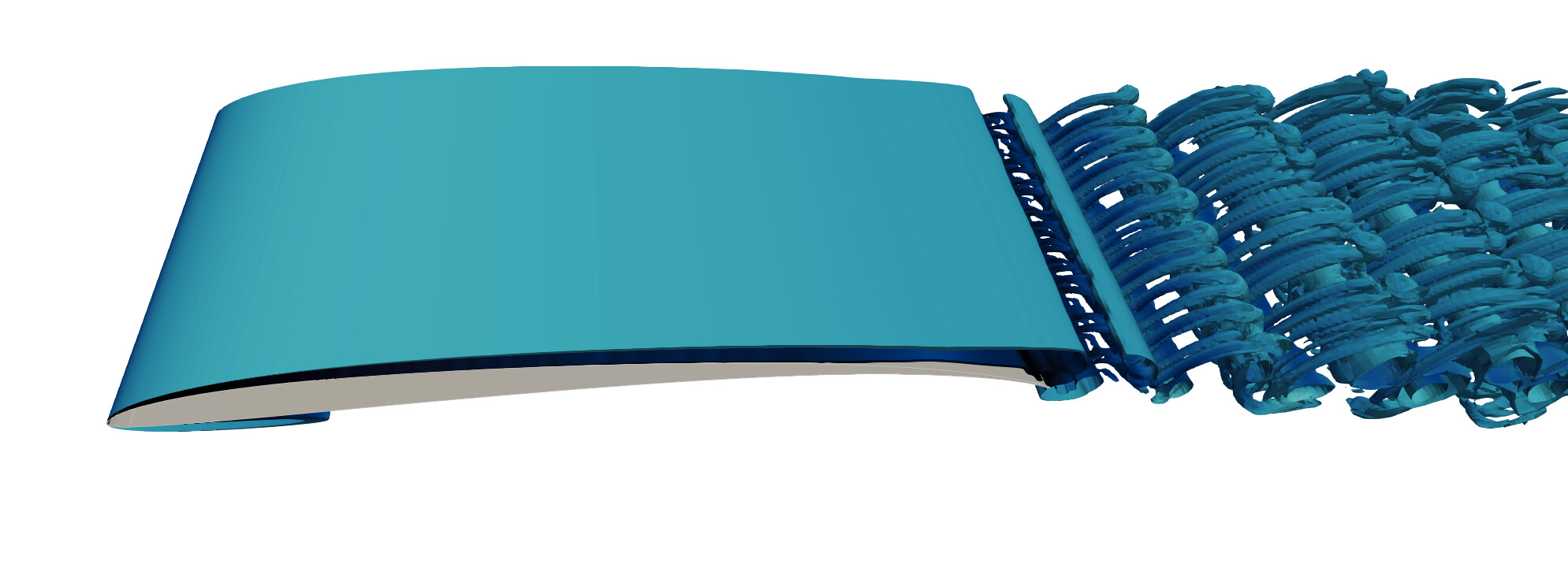}
        \end{minipage}
        \begin{minipage}{0.43\textwidth}
          \includegraphics[trim=50 50 0 50,clip,width=\linewidth]{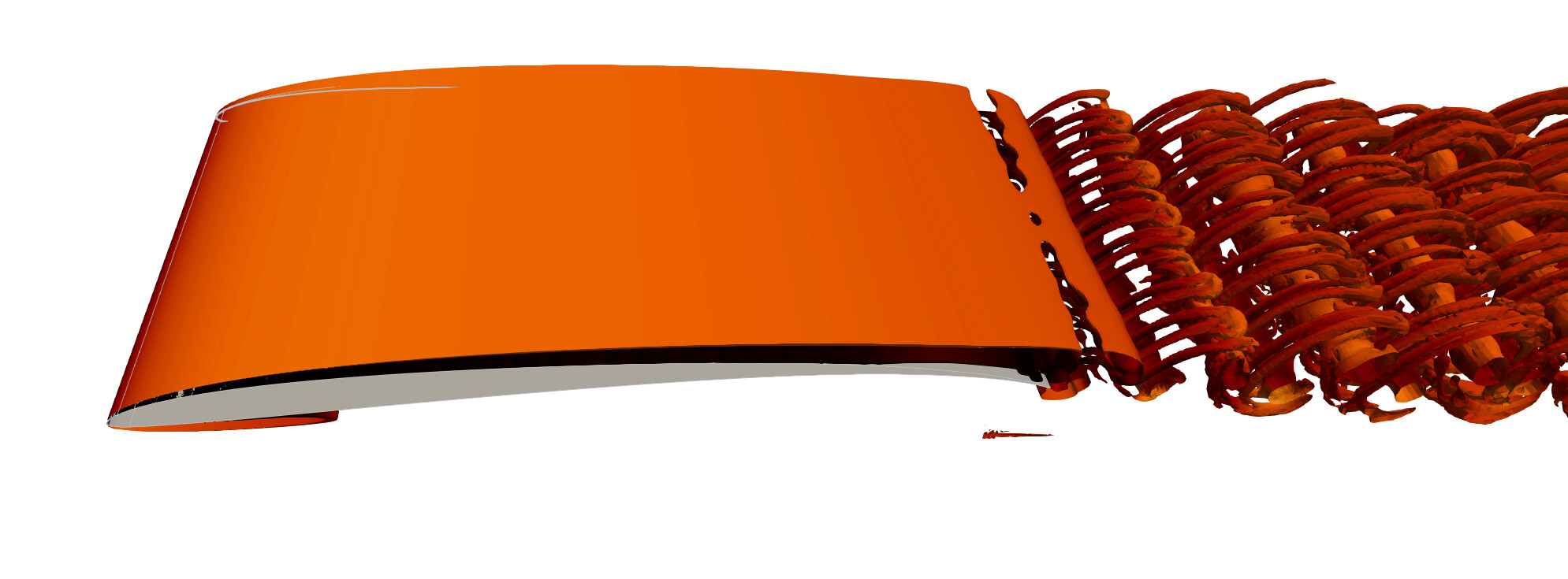}
        \end{minipage}
        
        \begin{minipage}{0.06\textwidth}
          \centering
          \rotatebox{0}{\small$\alpha = 2^\circ$}
        \end{minipage}
        \begin{minipage}{0.43\textwidth}
          \includegraphics[trim=50 50 0 50,clip,width=\linewidth]{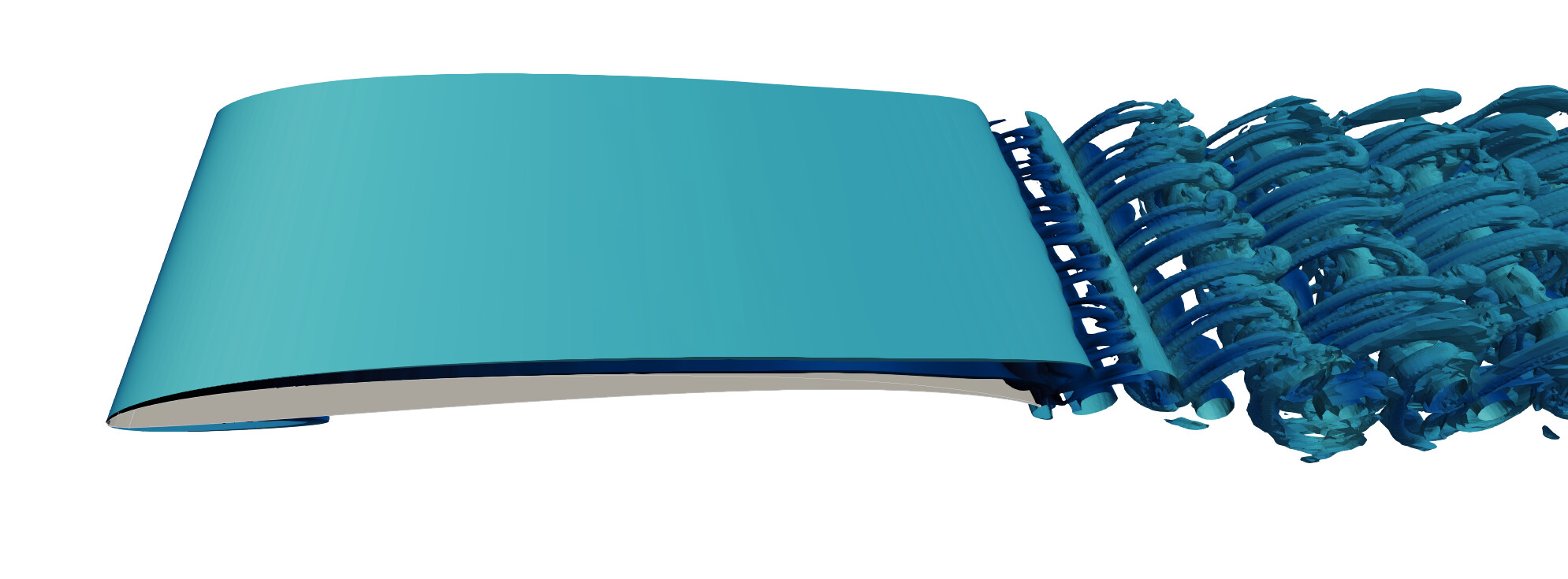}
        \end{minipage}
        \begin{minipage}{0.43\textwidth}
          \includegraphics[trim=50 50 0 50,clip,width=\linewidth]{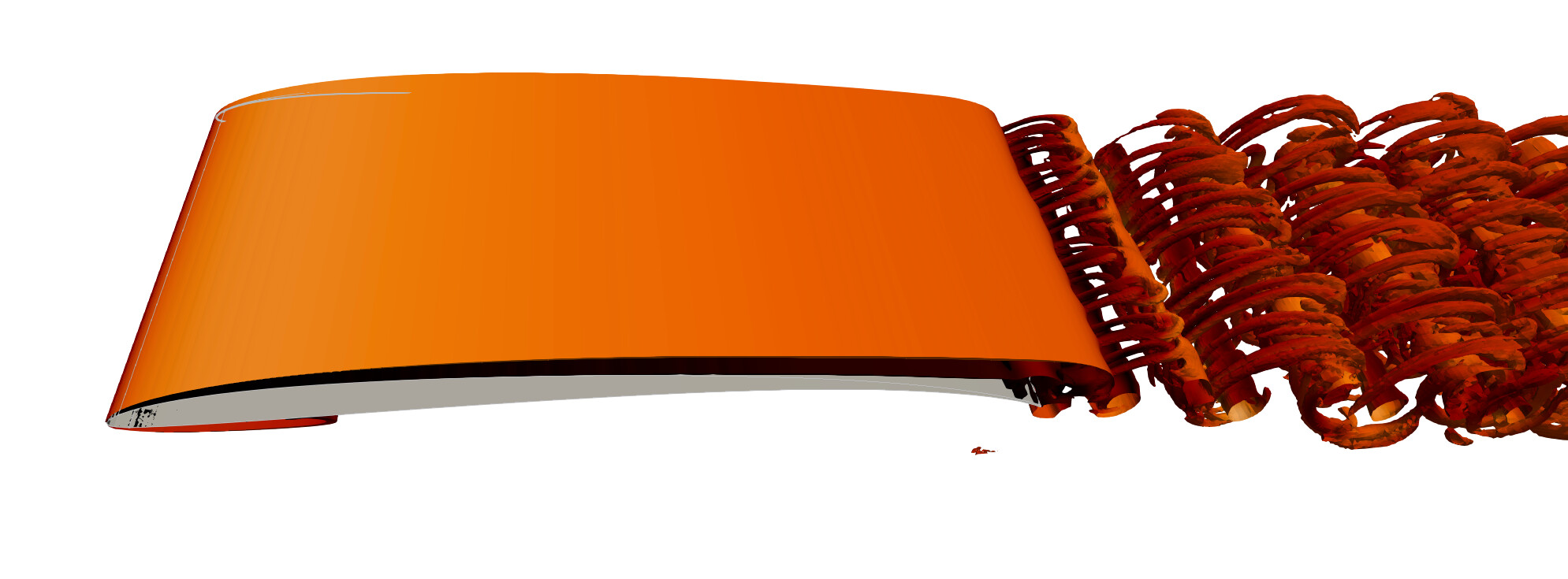}
        \end{minipage}
        
        \begin{minipage}{0.06\textwidth}
          \centering
          \rotatebox{0}{\small$\alpha = 3^\circ$}
        \end{minipage}
        \begin{minipage}{0.43\textwidth}
          \includegraphics[trim=50 50 0 50,clip,width=\linewidth]{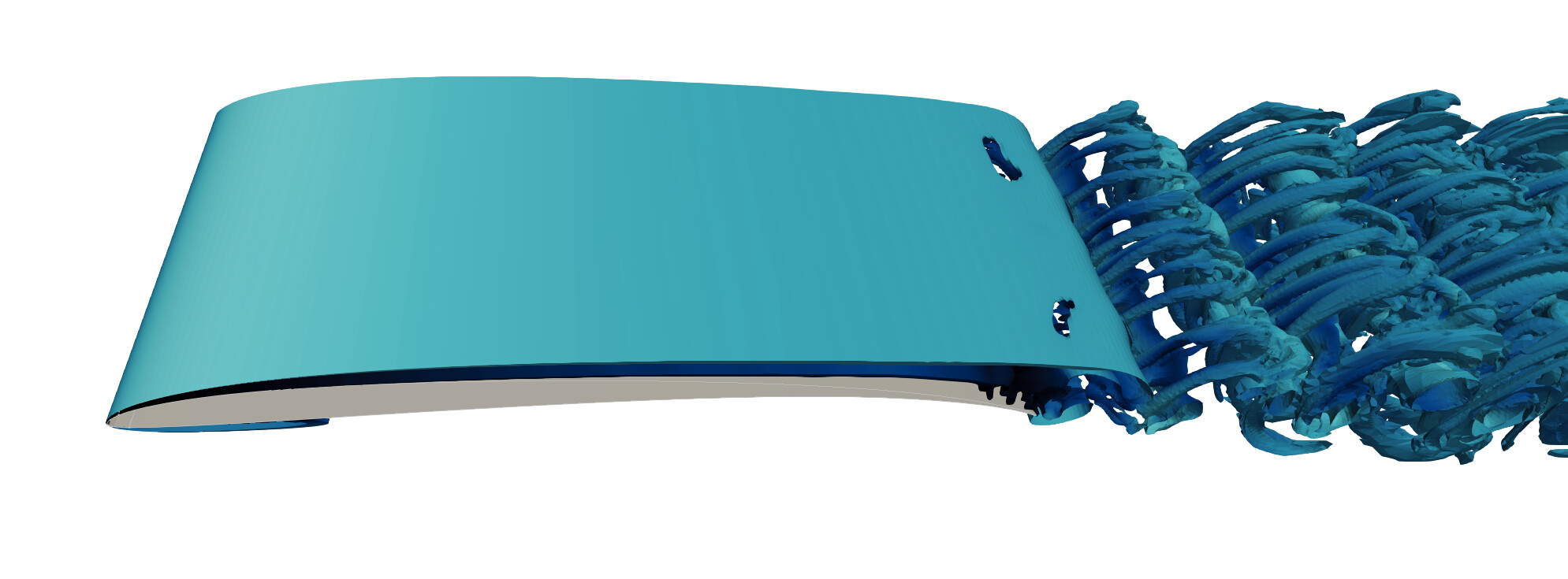}
        \end{minipage}
        \begin{minipage}{0.43\textwidth}
          \includegraphics[trim=50 50 0 50,clip,width=\linewidth]{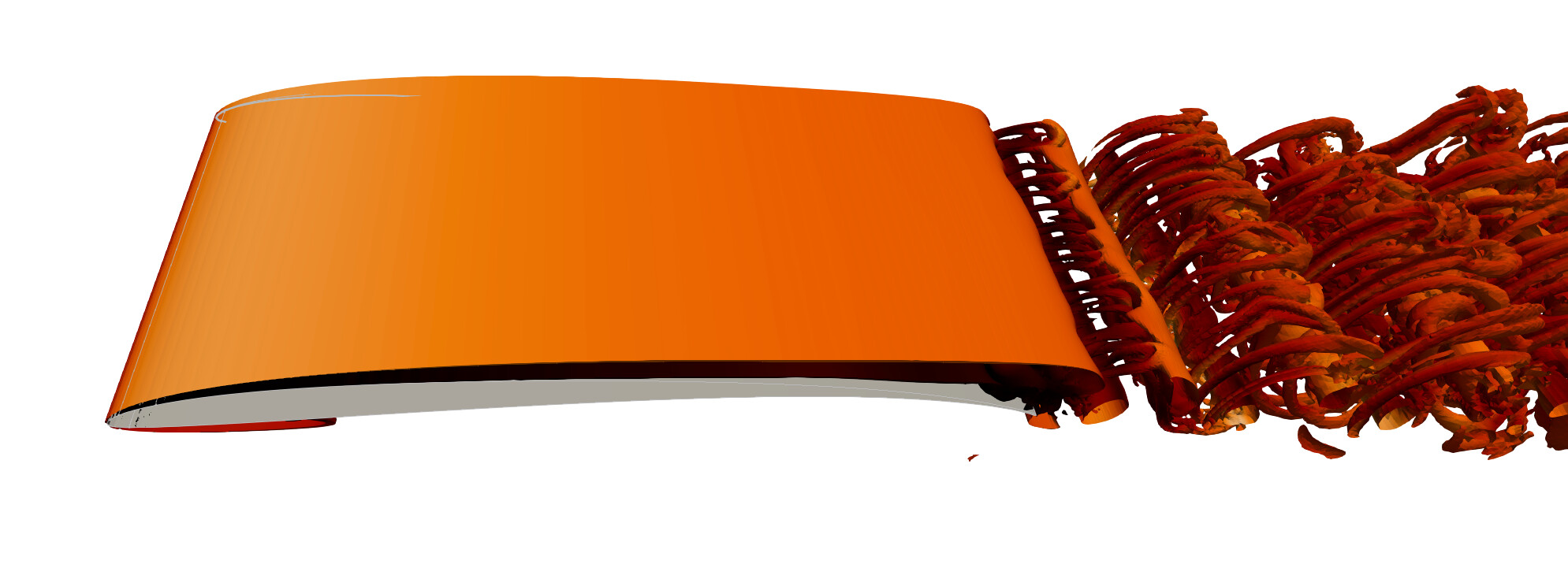}
        \end{minipage}
        
        \begin{minipage}{0.06\textwidth}
          \centering
          \rotatebox{0}{\small$\alpha = 4^\circ$}
        \end{minipage}
        \begin{minipage}{0.43\textwidth}
          \includegraphics[trim=50 50 0 50,clip,width=\linewidth]{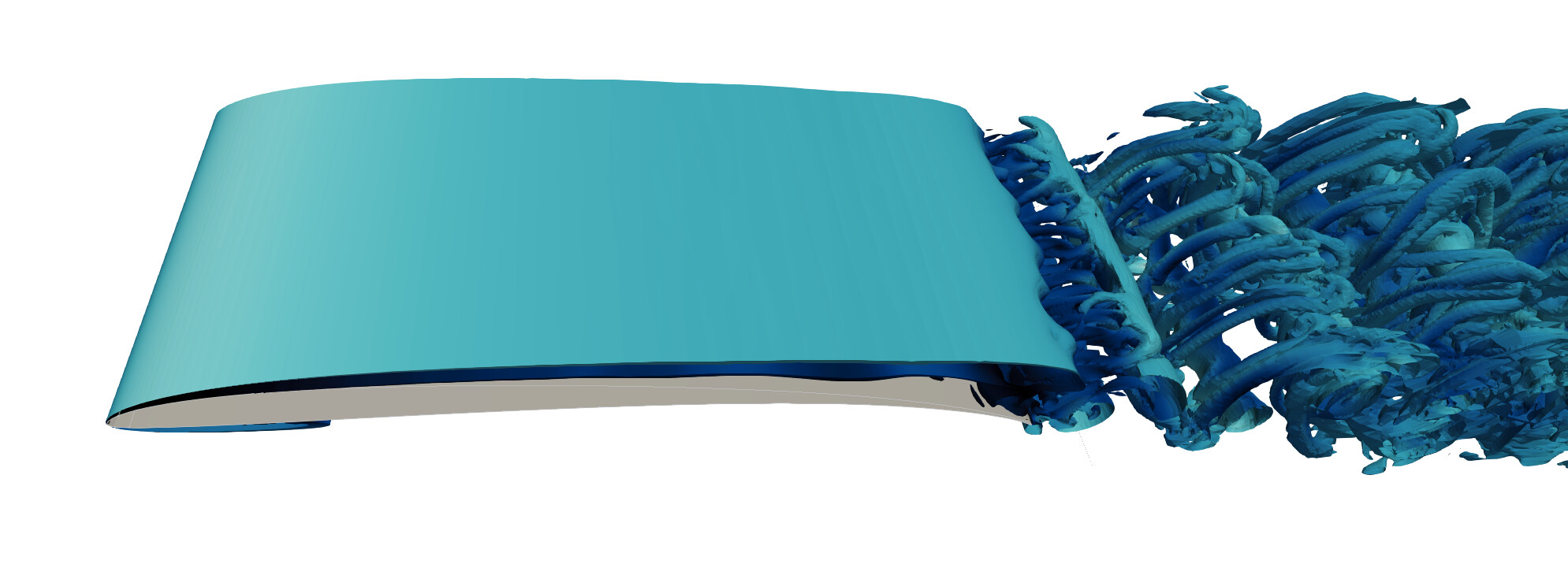}
        \end{minipage}
        \begin{minipage}{0.43\textwidth}
          \includegraphics[trim=50 50 0 50,clip,width=\linewidth]{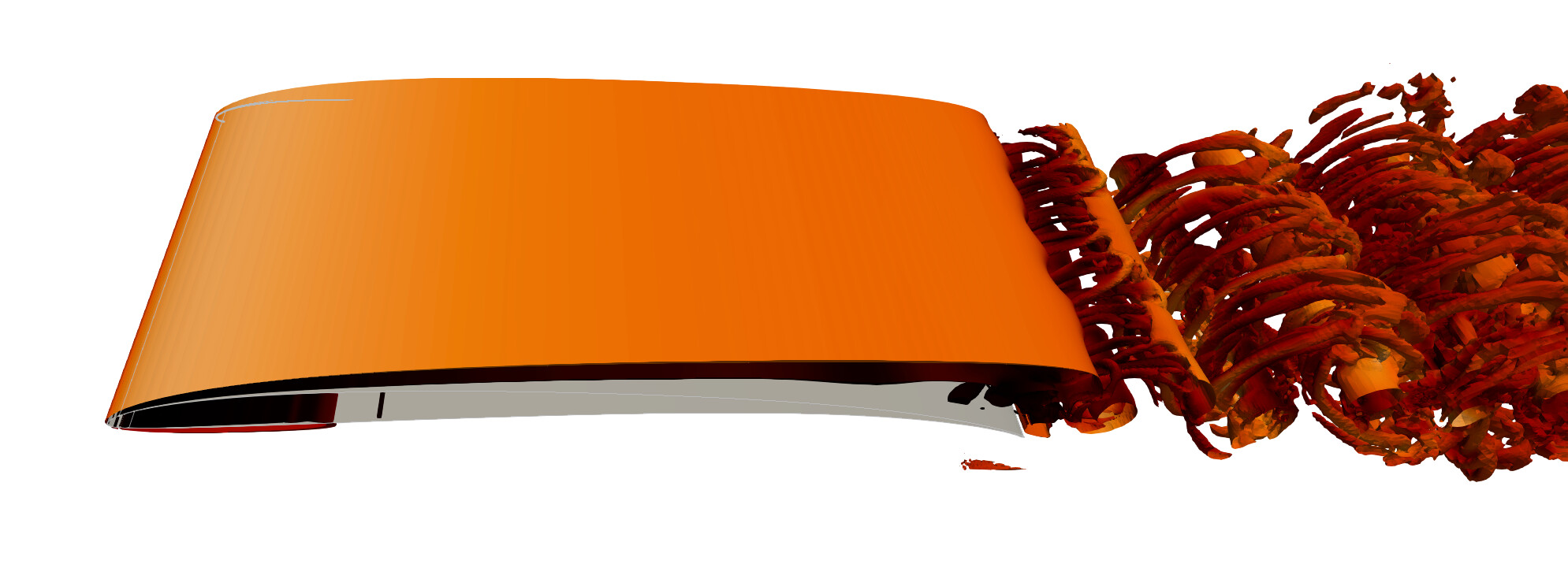}
        \end{minipage}
        
        \begin{minipage}{0.06\textwidth}
          \centering
          \rotatebox{0}{\small$\alpha = 5^\circ$}
        \end{minipage}
        \begin{minipage}{0.43\textwidth}
          \includegraphics[trim=50 50 0 50,clip,width=\linewidth]{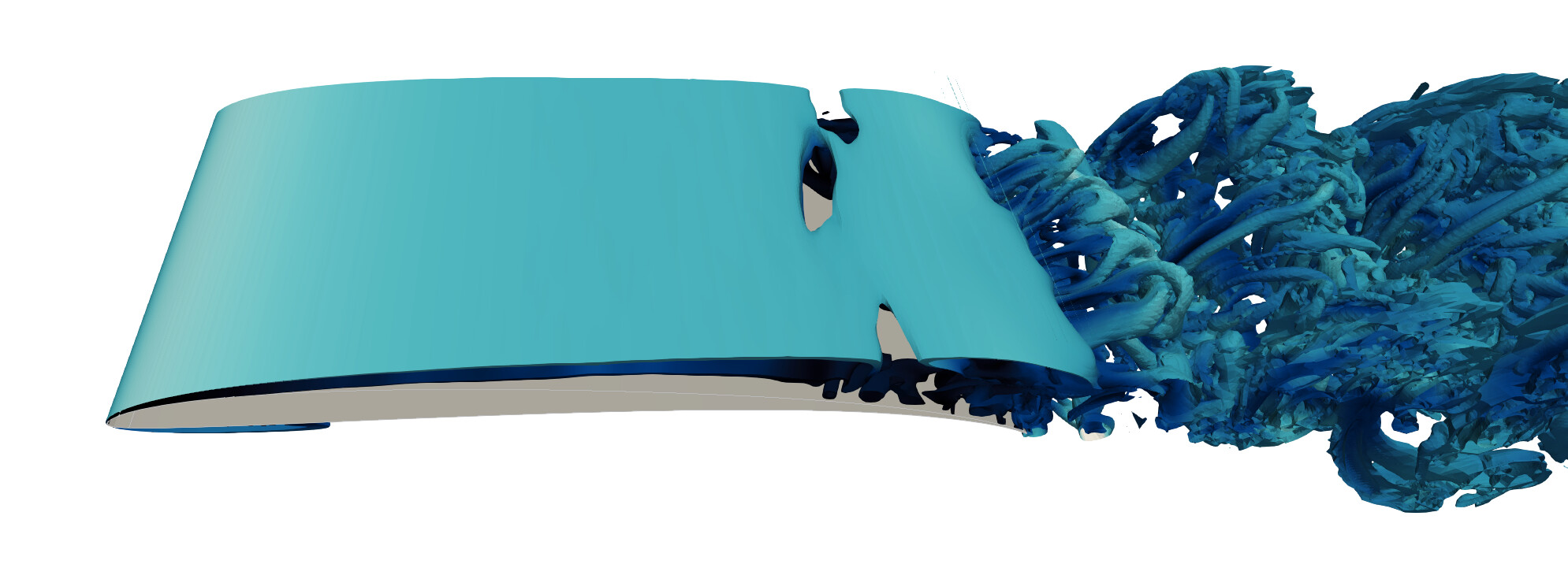}
        \end{minipage}
        \begin{minipage}{0.43\textwidth}
          \includegraphics[trim=50 50 0 50,clip,width=\linewidth]{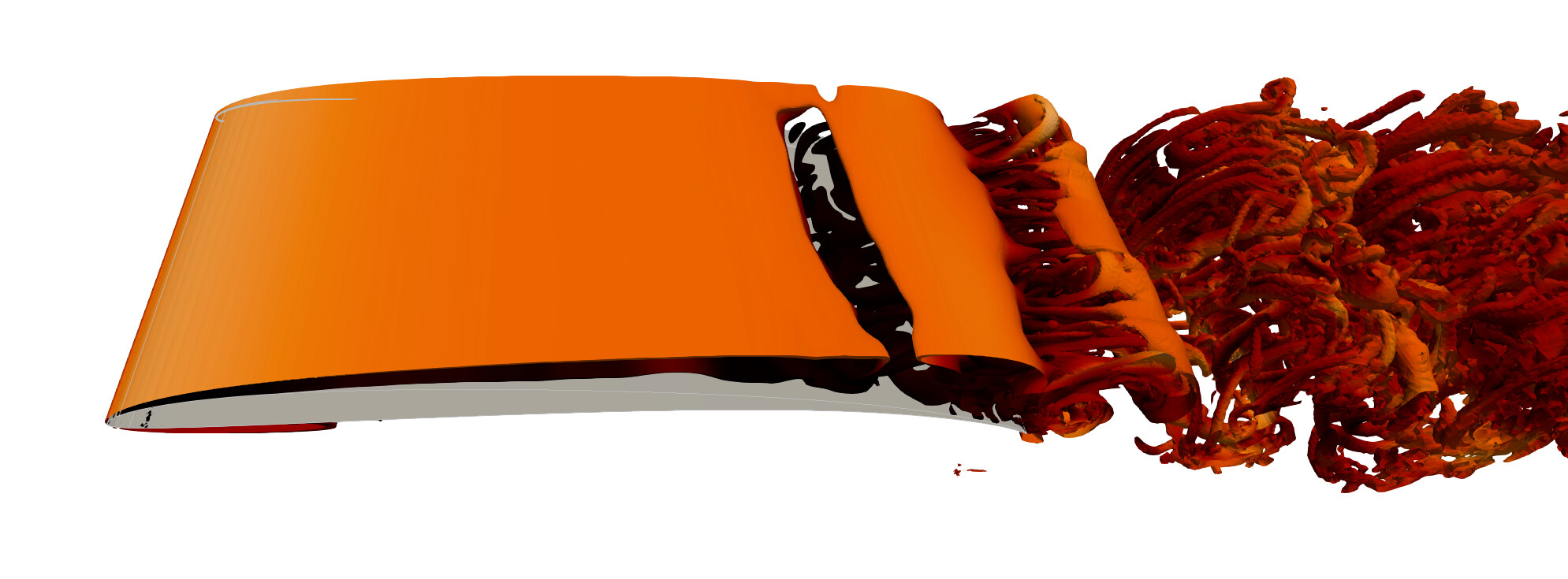}
        \end{minipage}
        
        \begin{minipage}{0.06\textwidth}
          \centering
          \rotatebox{0}{\small$\alpha = 5.5^\circ$}
        \end{minipage}
        \begin{minipage}{0.43\textwidth}
          \includegraphics[trim=50 50 0 50,clip,width=\linewidth]{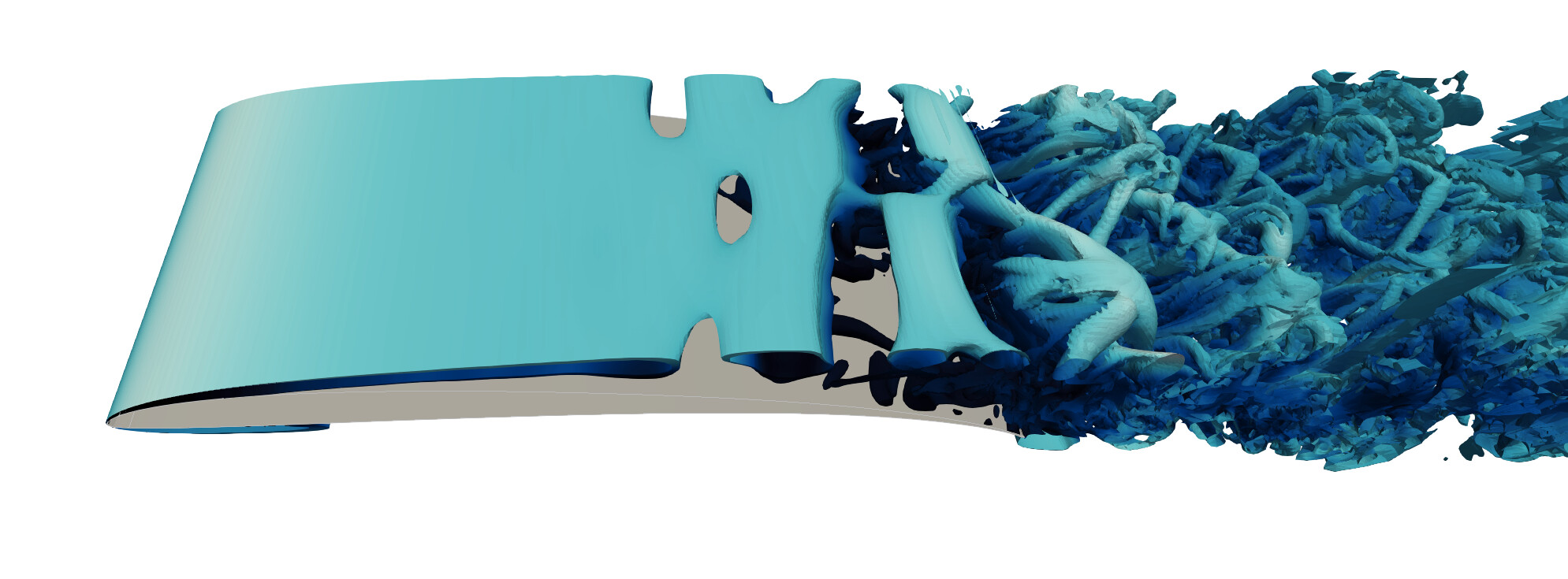}
        \end{minipage}
        \begin{minipage}{0.43\textwidth}
          \includegraphics[trim=50 50 0 50,clip,width=\linewidth]{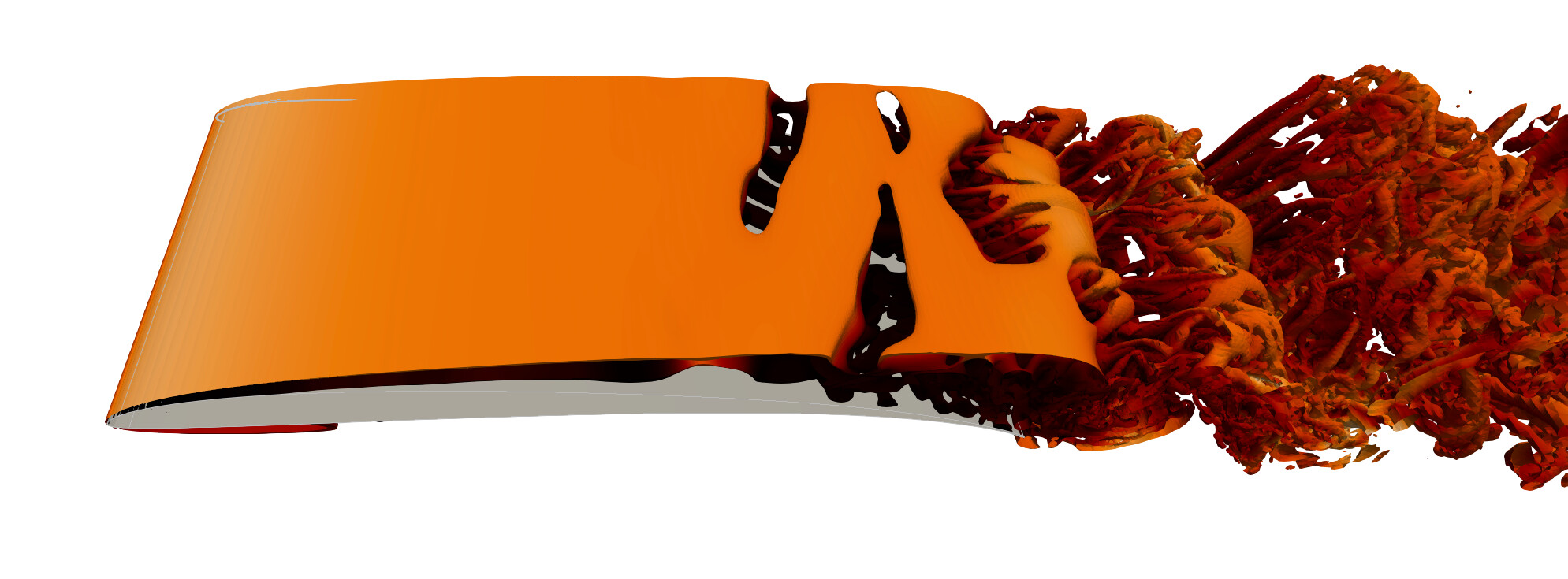}
        \end{minipage}
        
        \begin{minipage}{0.06\textwidth}
          \centering
          \rotatebox{0}{\small$\alpha = 6^\circ$}
        \end{minipage}
        \begin{minipage}{0.43\textwidth}
          \includegraphics[trim=50 50 0 50,clip,width=\linewidth]{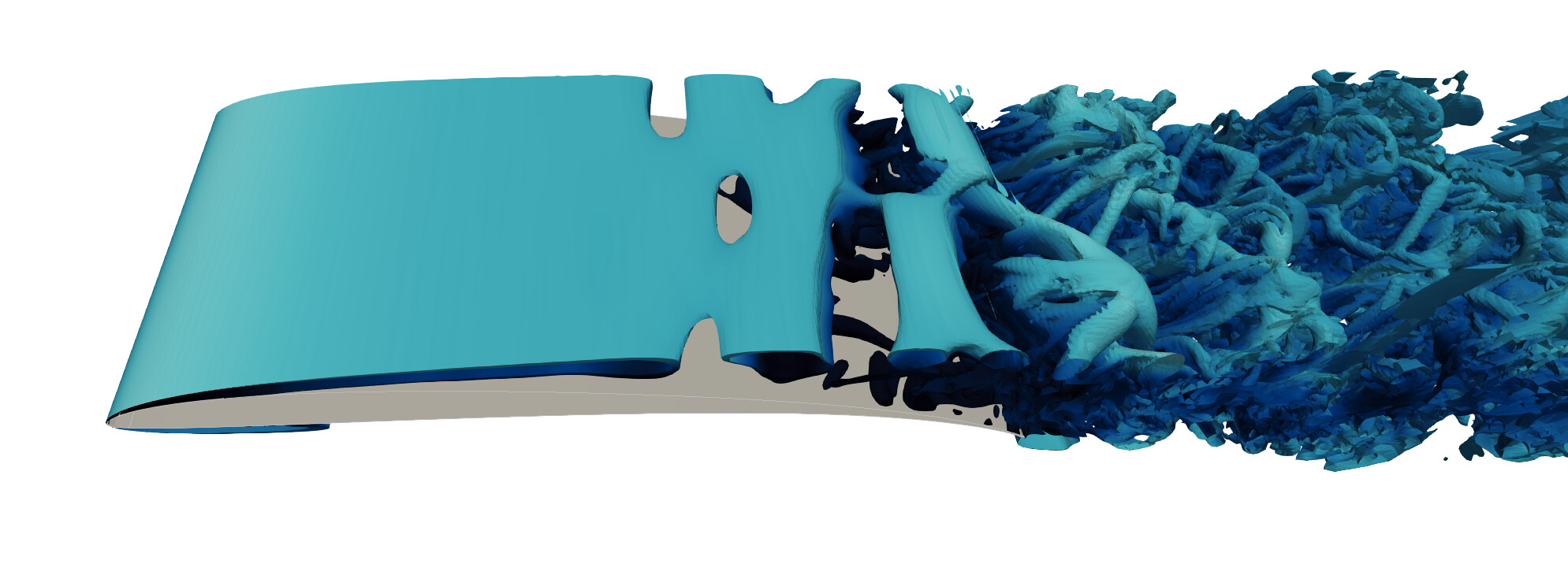}
        \end{minipage}
        \begin{minipage}{0.43\textwidth}
          \includegraphics[trim=50 50 0 50,clip,width=\linewidth]{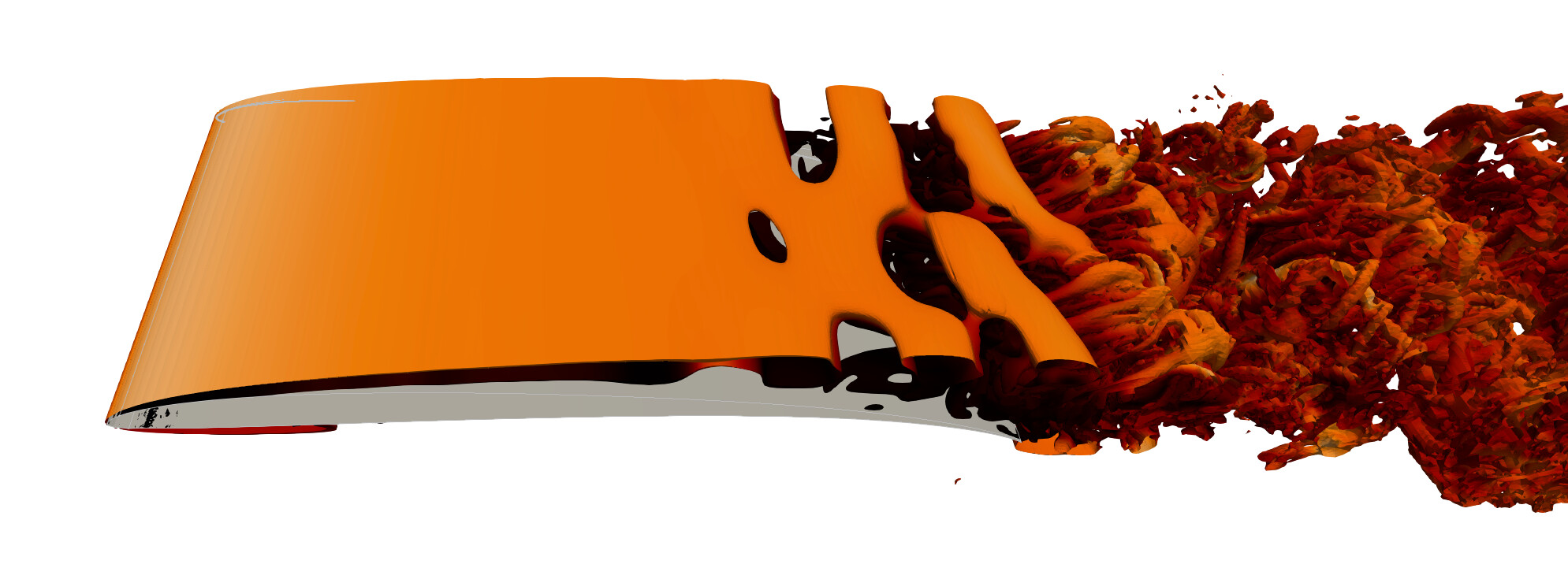}
        \end{minipage}
        \caption{Images of instantaneous Q-criterion isosurfaces $\mathbf{Q = 1}$ coloured by normalised velocity magnitude $|\mathbf{v}|/|\mathbf{v}_\infty|$ obtained with 3D-SP OVERFLOW (left) and 3D-SP PyFR (right) for the CLF5605 airfoil.}
        \label{fig:qcrit-OF-PyFR-q3D-clf}
\end{figure} 

\begin{figure}[h!]
  \centering
        \begin{subfigure}[t]{1\linewidth}
        \centering
        \hspace{1cm}
        \includegraphics[width=0.42\linewidth]{Data/OVERFLOW/Qcriterion/OF_Qcrit_colorbar.png}
        \hspace{0.1cm}
        \includegraphics[width=0.42\linewidth]{Data/PyFR/Qcriterion/PyFR_Qcrit_colorbar.png}
        \end{subfigure}
        
        \begin{minipage}{0.06\textwidth}
          \centering
          \rotatebox{0}{\small$\alpha = -2^\circ$}
        \end{minipage}
        \begin{minipage}{0.43\textwidth}
          \includegraphics[trim=50 50 0 50,clip,width=\linewidth]{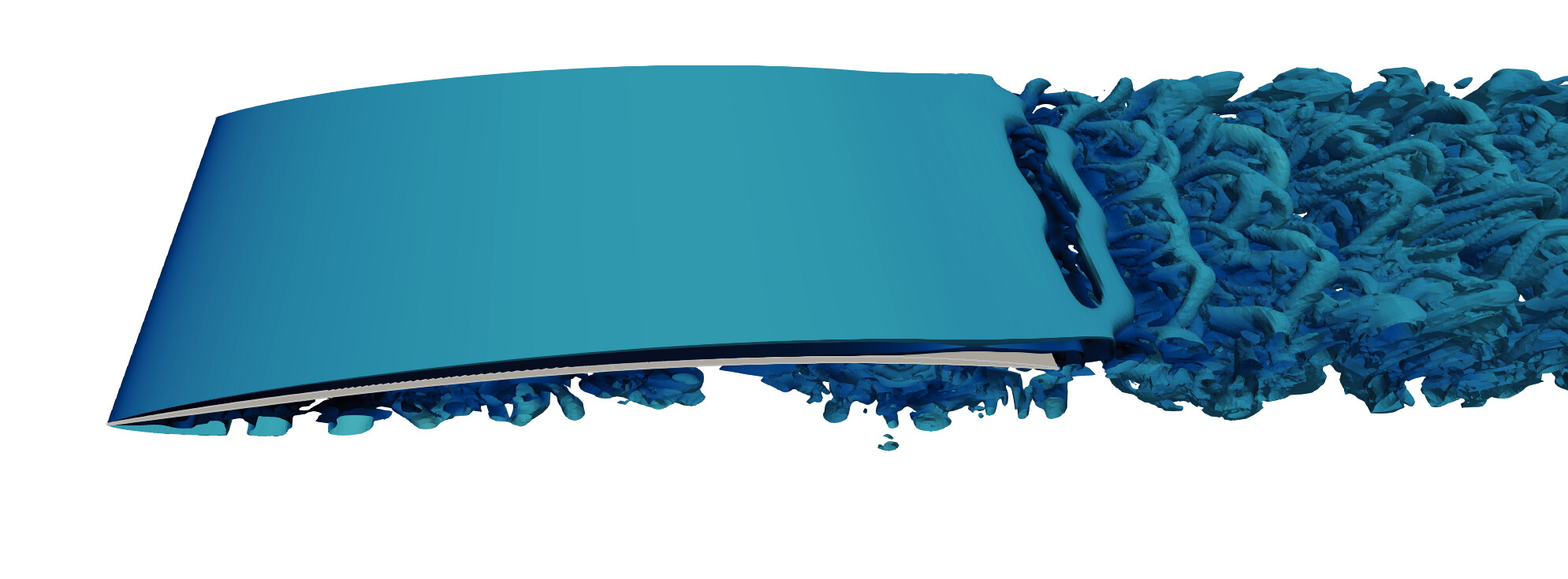}
        \end{minipage}
        \begin{minipage}{0.43\textwidth}
          \includegraphics[trim=50 50 0 50,clip,width=\linewidth]{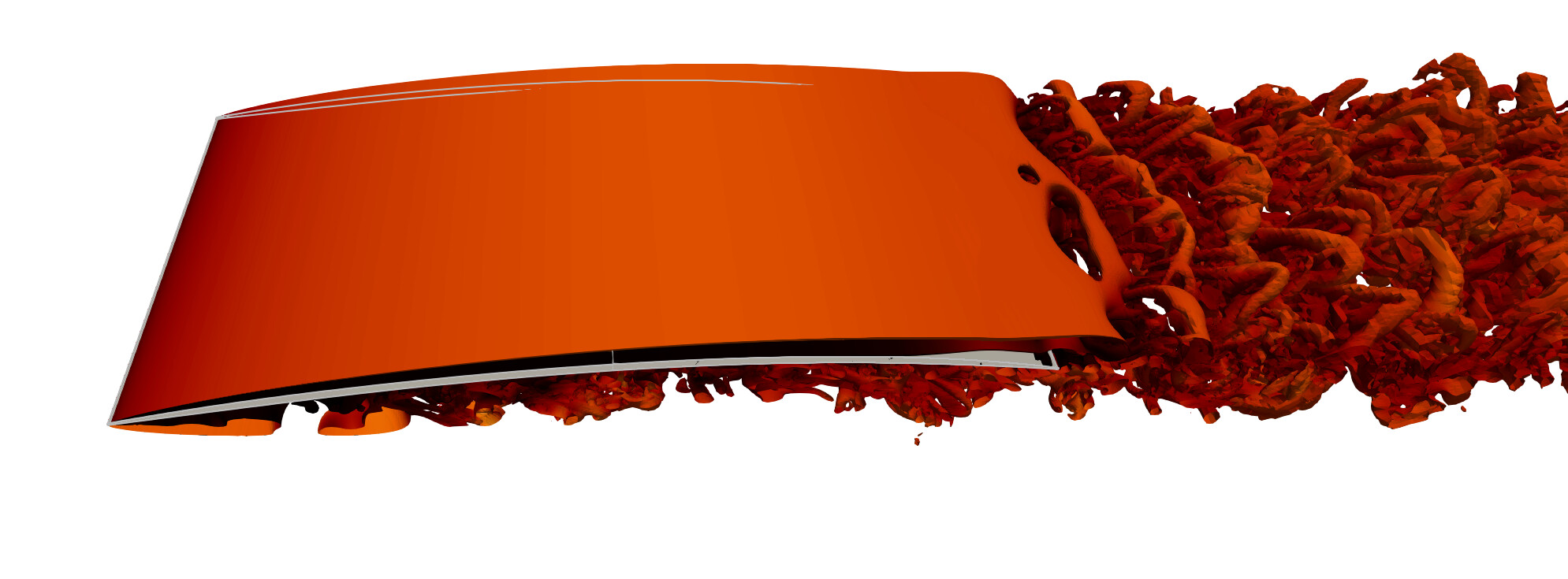}
        \end{minipage}
        
        \begin{minipage}{0.06\textwidth}
          \centering
          \rotatebox{0}{\small$\alpha = 0^\circ$}
        \end{minipage}
        \begin{minipage}{0.43\textwidth}
          \includegraphics[trim=50 50 0 50,clip,width=\linewidth]{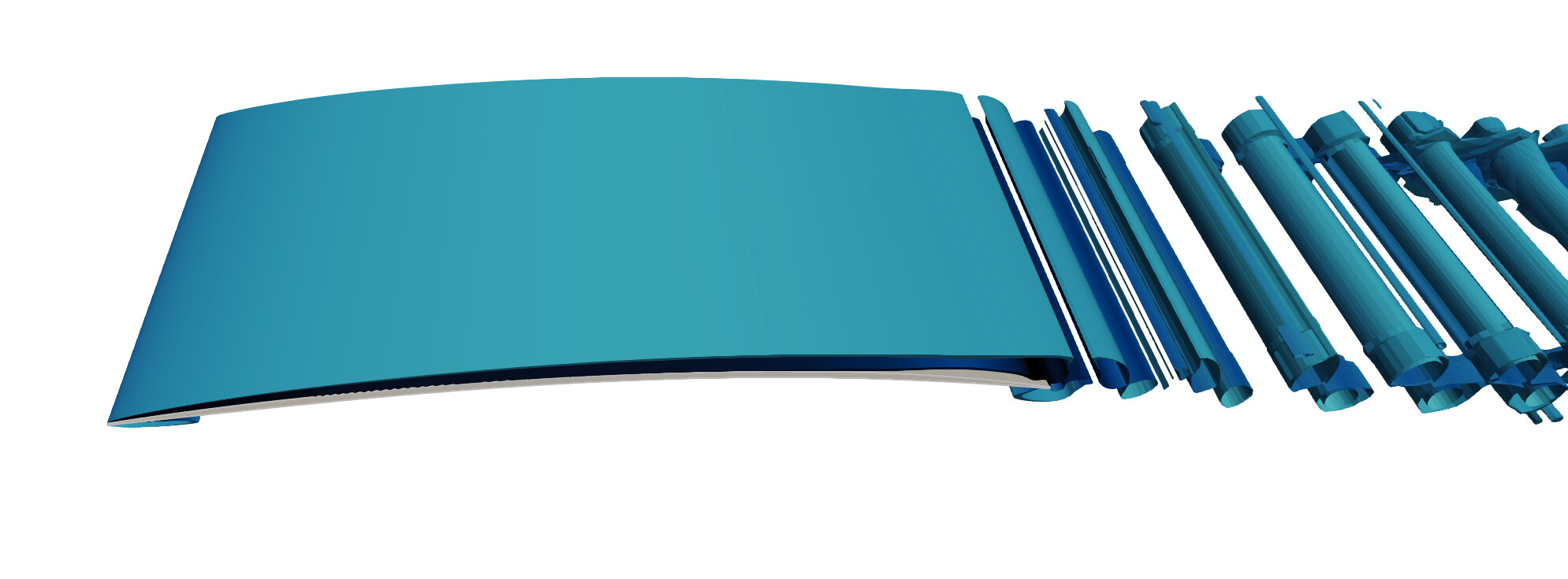}
        \end{minipage}
        \begin{minipage}{0.43\textwidth}
          \includegraphics[trim=50 50 0 50,clip,width=\linewidth]{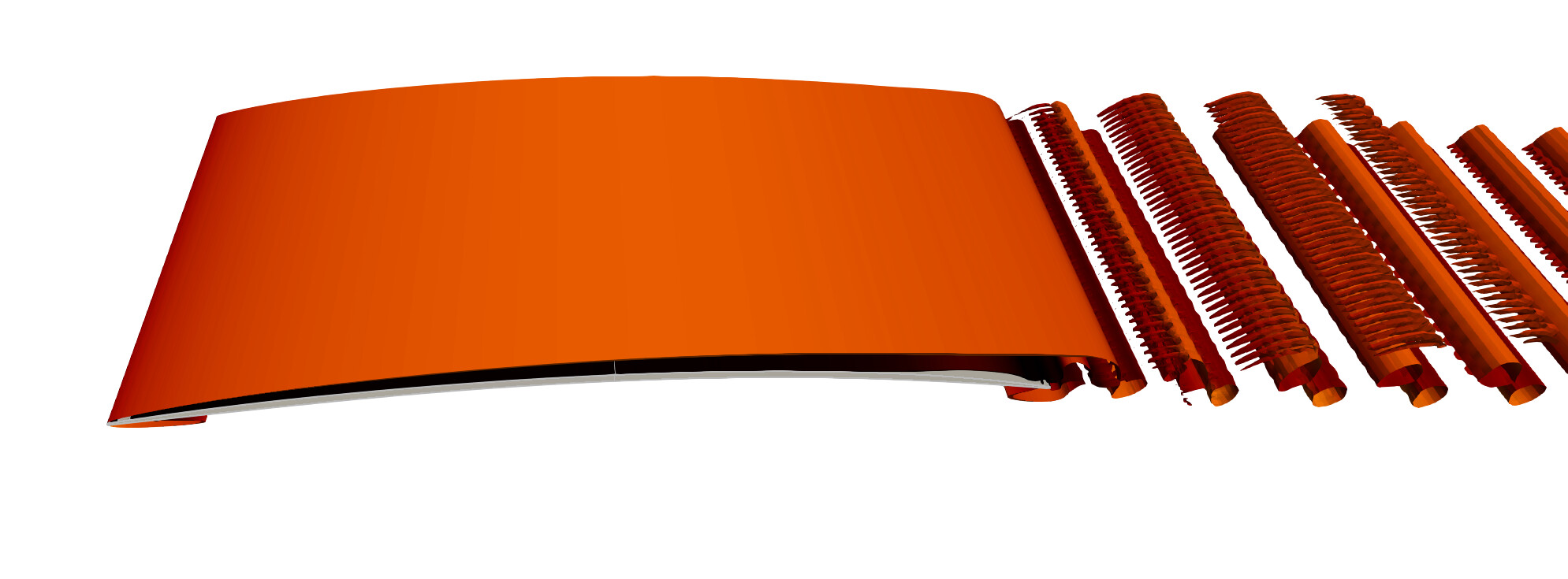}
        \end{minipage}
        
        \begin{minipage}{0.06\textwidth}
          \centering
          \rotatebox{0}{\small$\alpha = 2^\circ$}
        \end{minipage}
        \begin{minipage}{0.43\textwidth}
          \includegraphics[trim=50 50 0 50,clip,width=\linewidth]{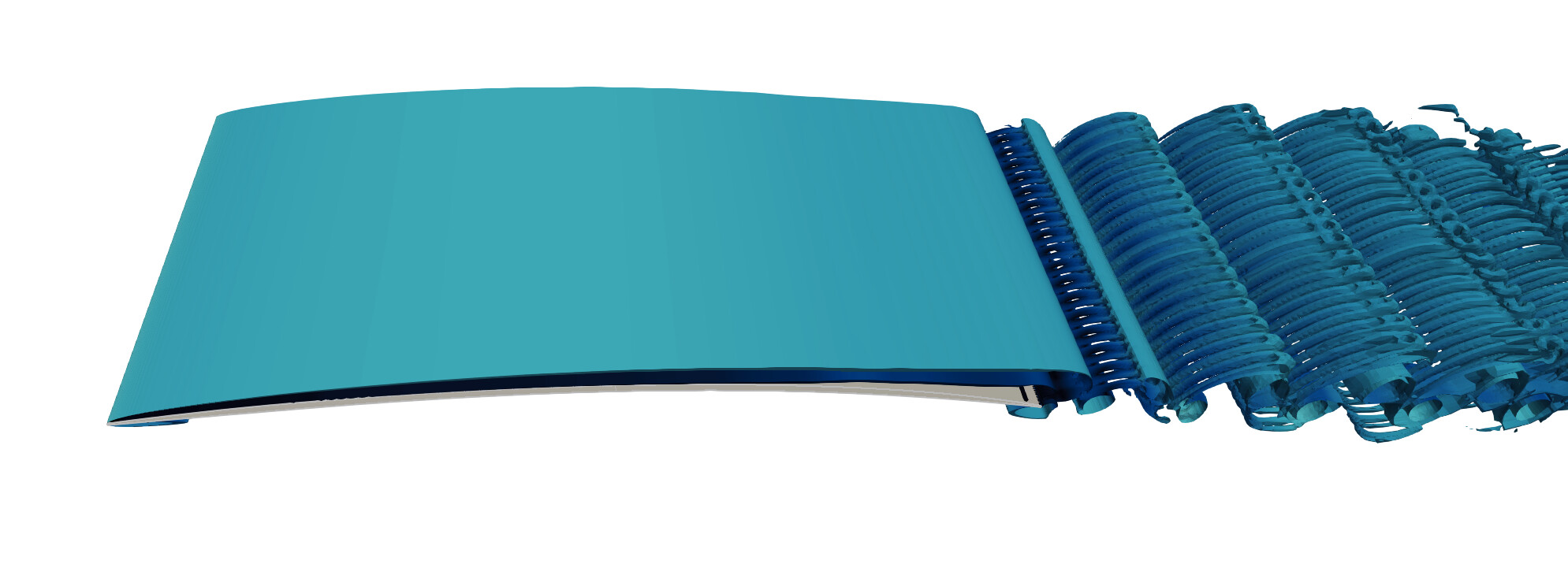}
        \end{minipage}
        \begin{minipage}{0.43\textwidth}
          \includegraphics[trim=50 50 0 50,clip,width=\linewidth]{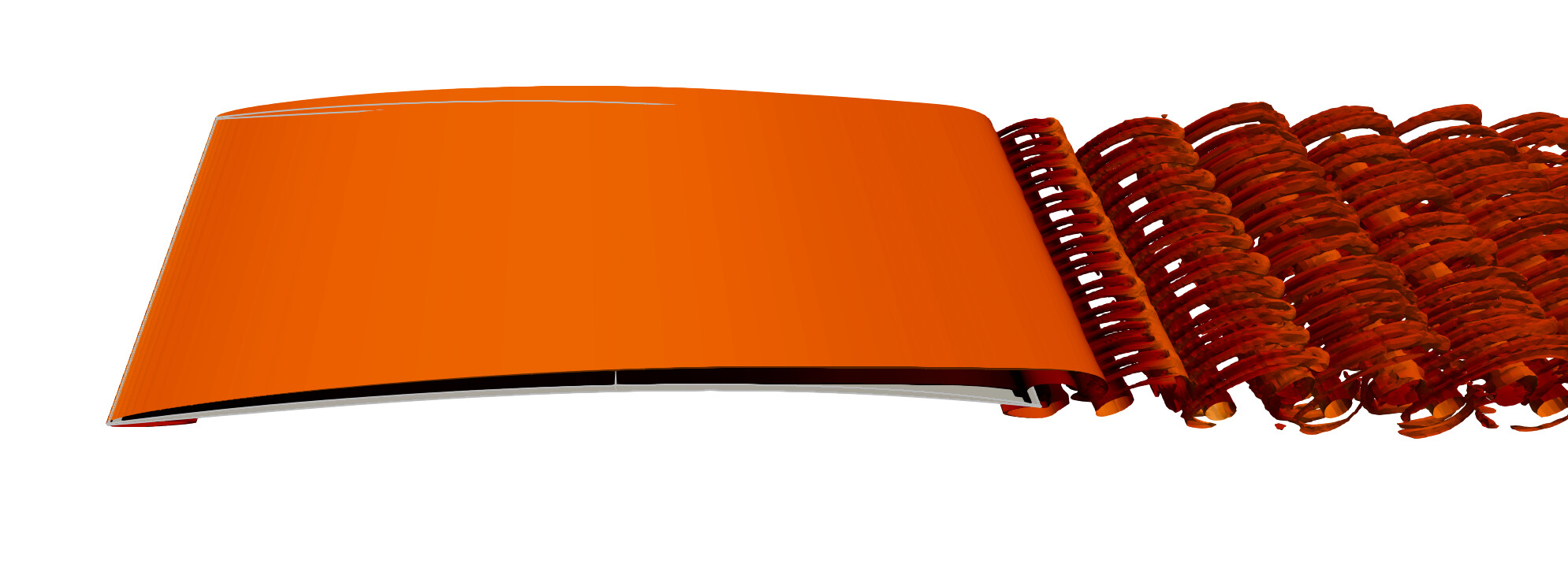}
        \end{minipage}
        
        \begin{minipage}{0.06\textwidth}
          \centering
          \rotatebox{0}{\small$\alpha = 3^\circ$}
        \end{minipage}
        \begin{minipage}{0.43\textwidth}
          \includegraphics[trim=50 50 0 50,clip,width=\linewidth]{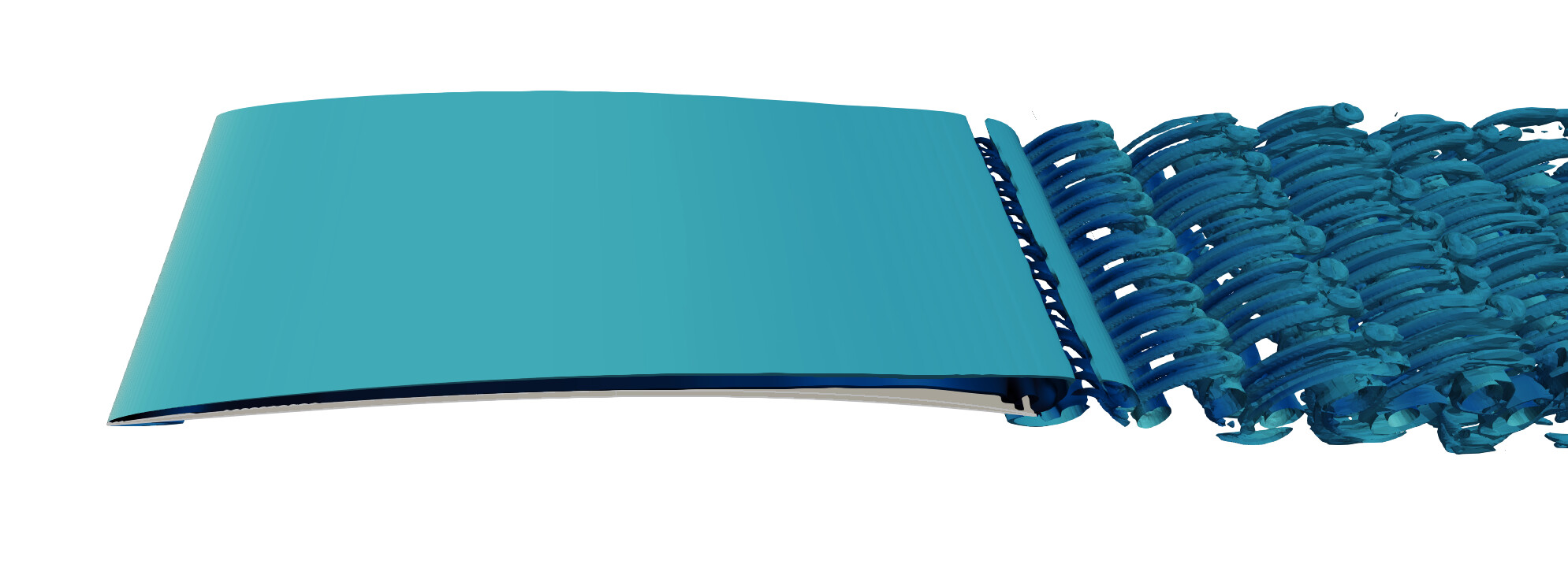}
        \end{minipage}
        \begin{minipage}{0.43\textwidth}
          \includegraphics[trim=50 50 0 50,clip,width=\linewidth]{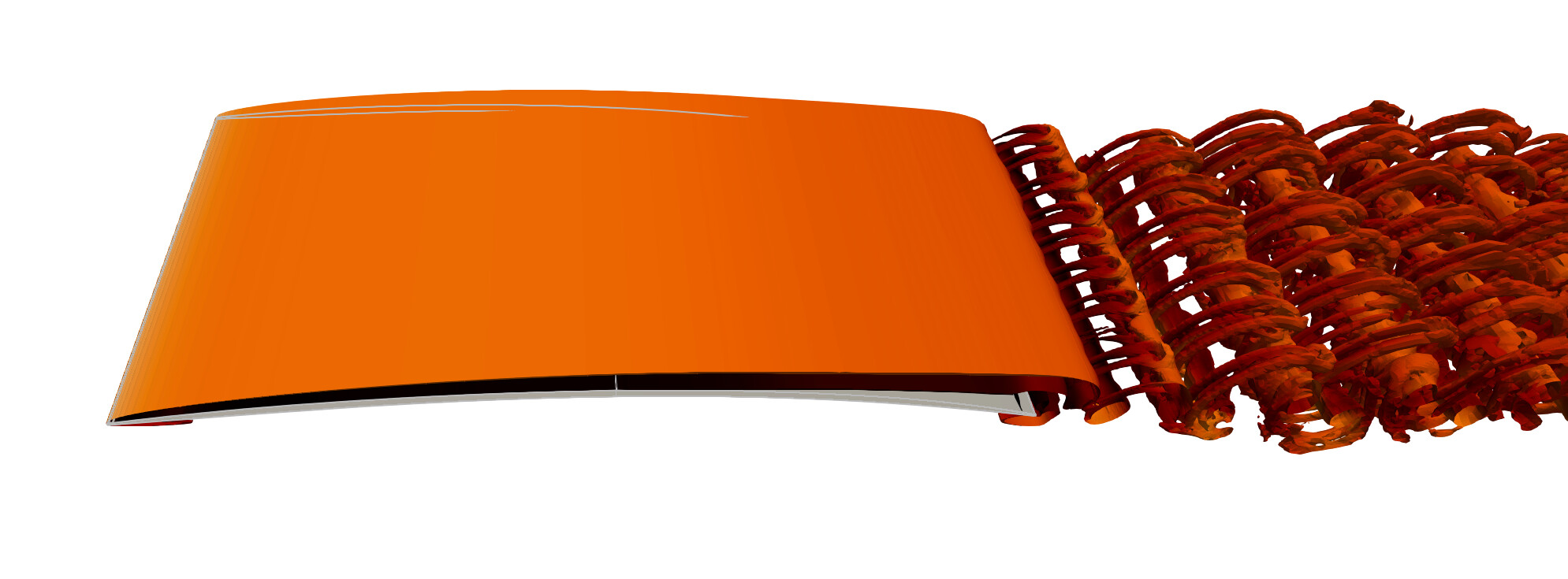}
        \end{minipage}
        
        \begin{minipage}{0.06\textwidth}
          \centering
          \rotatebox{0}{\small$\alpha = 4^\circ$}
        \end{minipage}
        \begin{minipage}{0.43\textwidth}
          \includegraphics[trim=50 50 0 50,clip,width=\linewidth]{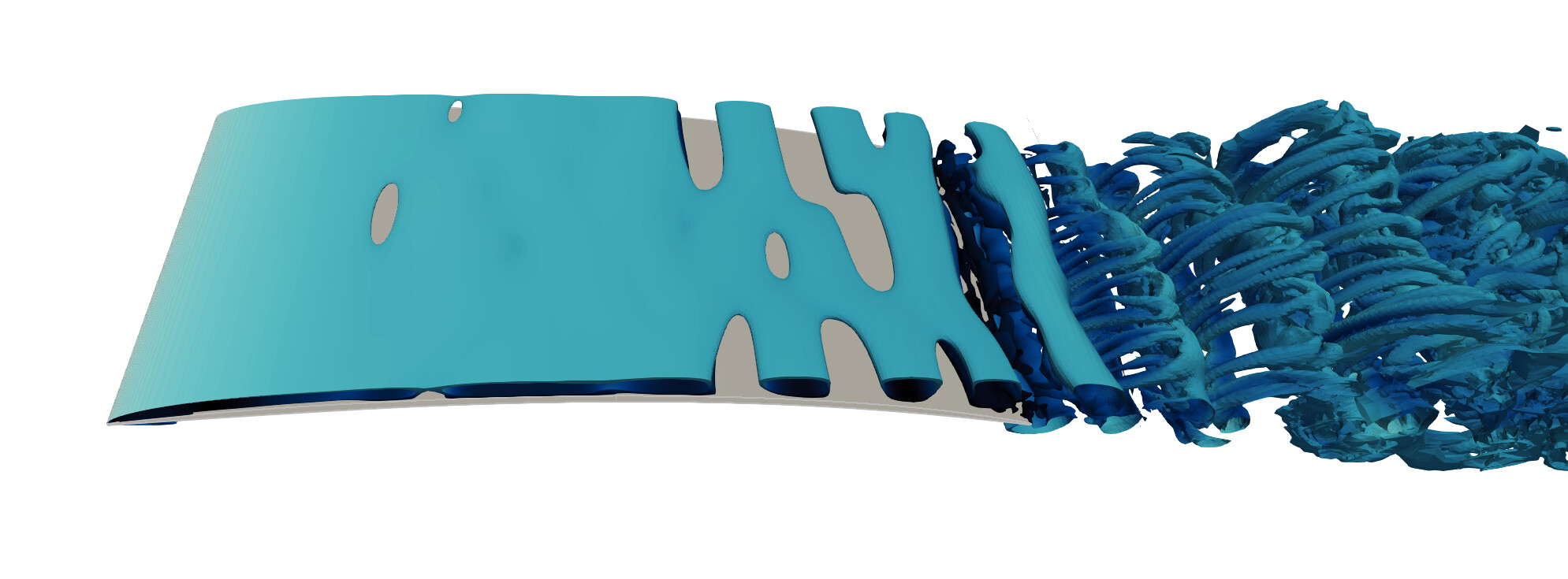}
        \end{minipage}
        \begin{minipage}{0.43\textwidth}
          \includegraphics[trim=50 50 0 50,clip,width=\linewidth]{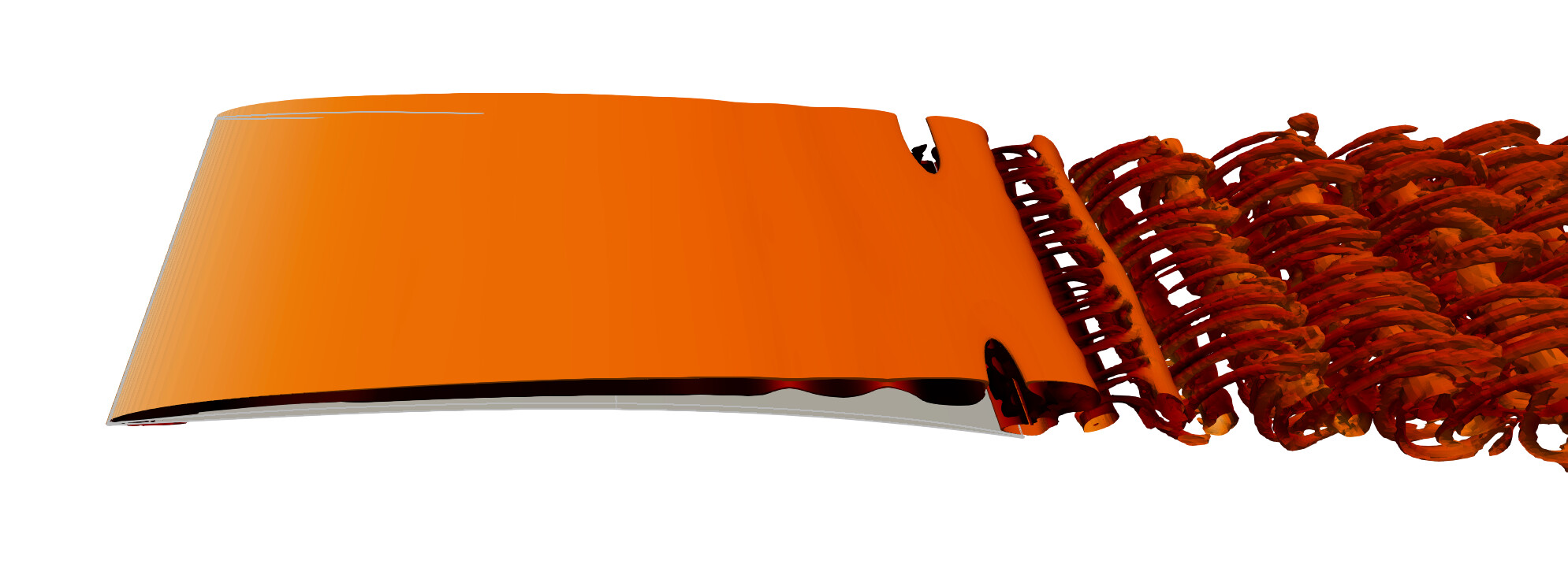}
        \end{minipage}
        
        \begin{minipage}{0.06\textwidth}
          \centering
          \rotatebox{0}{\small$\alpha = 4.5^\circ$}
        \end{minipage}
        \begin{minipage}{0.43\textwidth}
          \includegraphics[trim=50 50 0 50,clip,width=\linewidth]{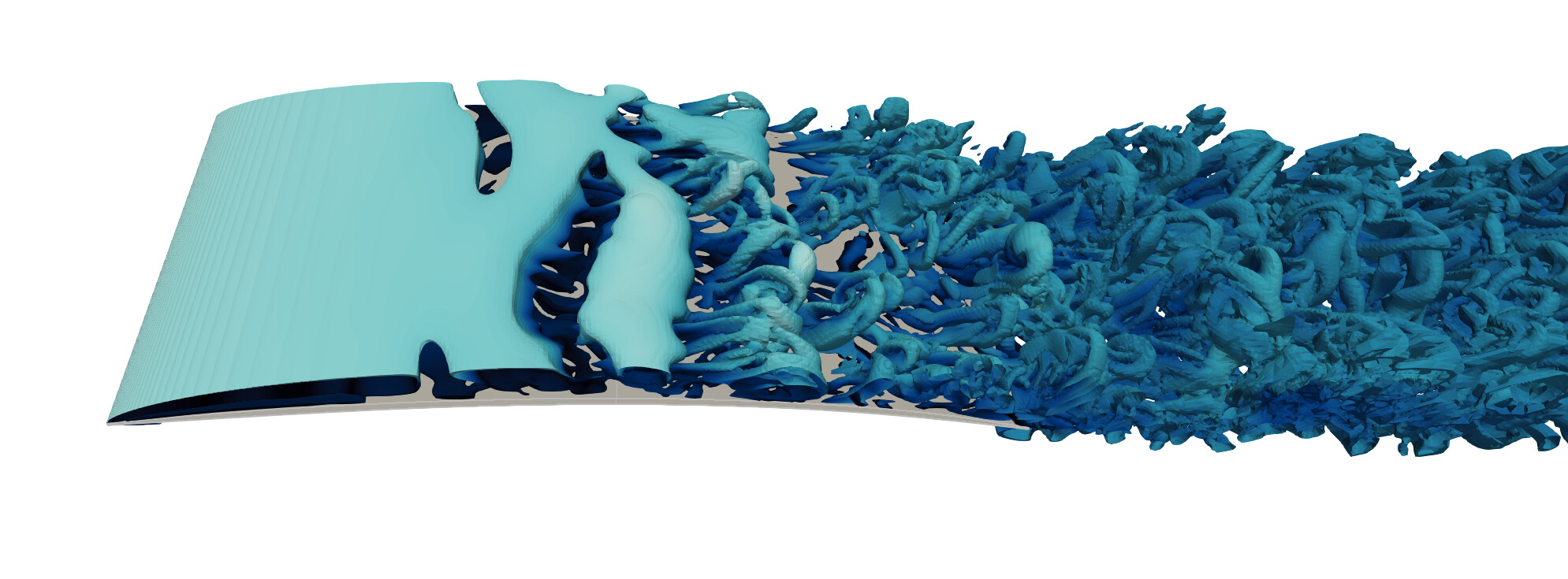}
        \end{minipage}
        \begin{minipage}{0.43\textwidth}
          \includegraphics[trim=50 50 0 50,clip,width=\linewidth]{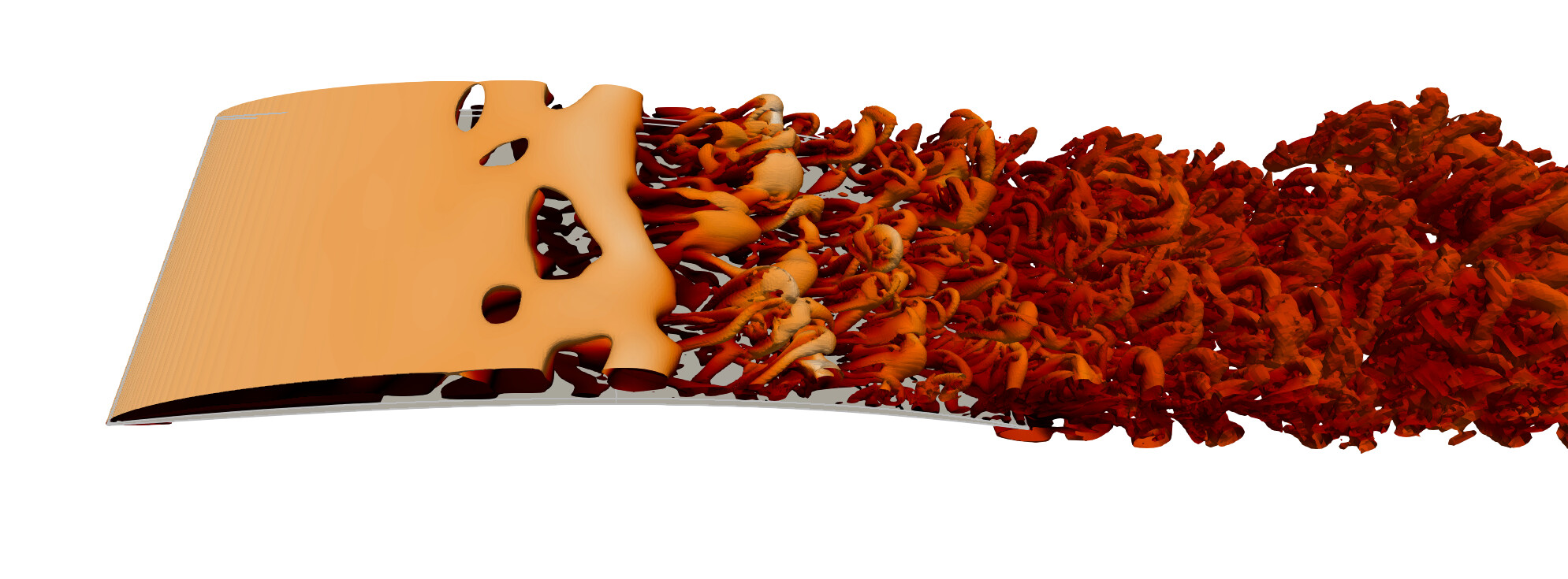}
        \end{minipage}
        
        \begin{minipage}{0.06\textwidth}
          \centering
          \rotatebox{0}{\small$\alpha = 5^\circ$}
        \end{minipage}
        \begin{minipage}{0.43\textwidth}
          \includegraphics[trim=50 50 0 50,clip,width=\linewidth]{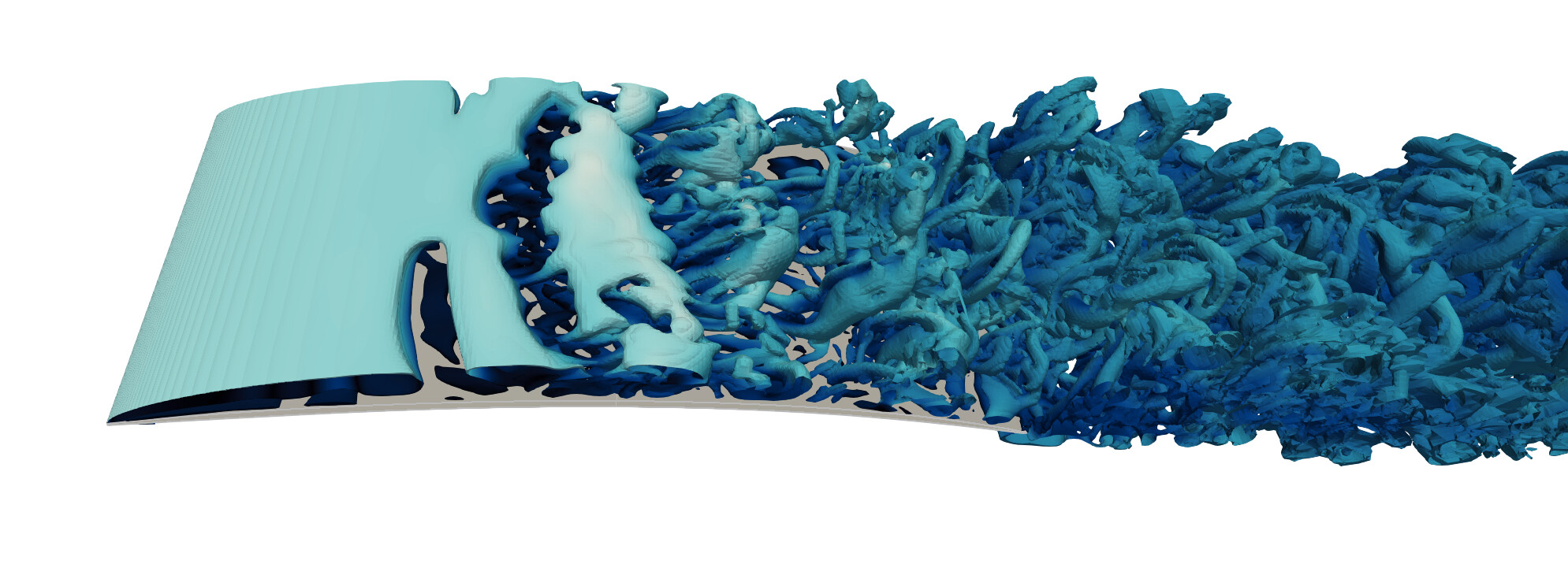}
        \end{minipage}
        \begin{minipage}{0.43\textwidth}
          \includegraphics[trim=50 50 0 50,clip,width=\linewidth]{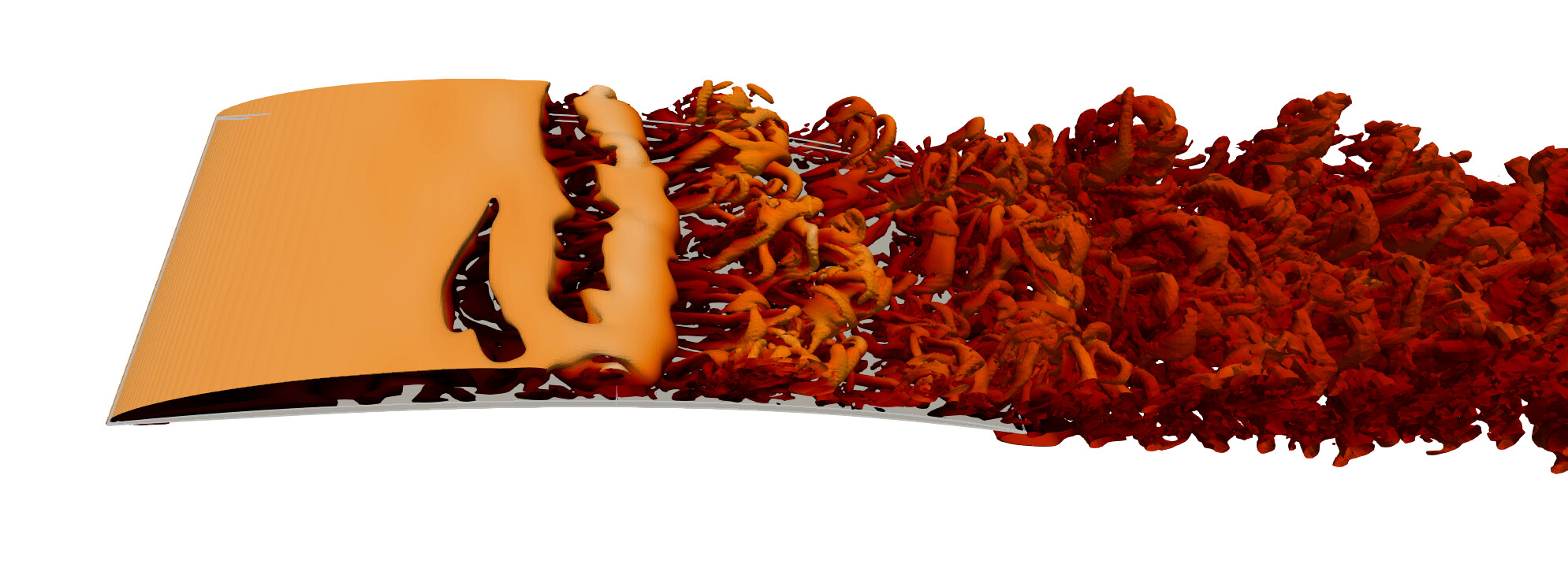}
        \end{minipage}
        
        \begin{minipage}{0.06\textwidth}
          \centering
          \rotatebox{0}{\small$\alpha = 6^\circ$}
        \end{minipage}
        \begin{minipage}{0.43\textwidth}
          \includegraphics[trim=50 50 0 50,clip,width=\linewidth]{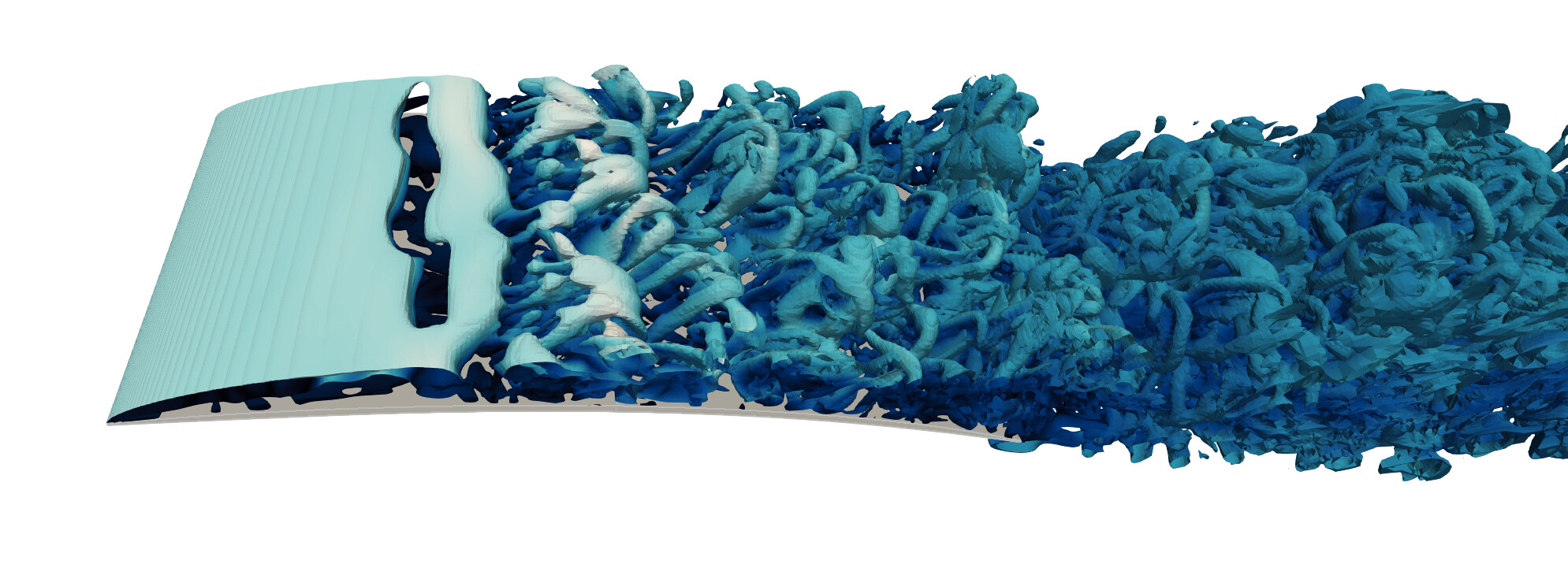}
        \end{minipage}
        \begin{minipage}{0.43\textwidth}
          \includegraphics[trim=50 50 0 50,clip,width=\linewidth]{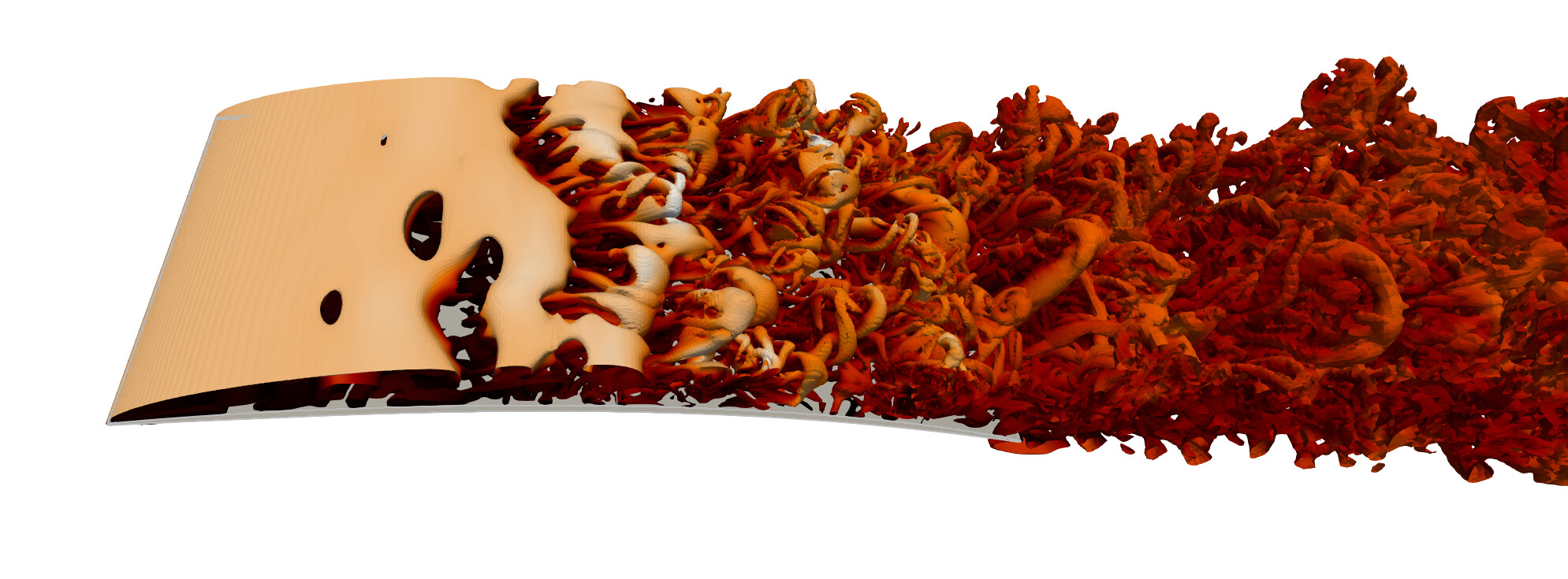}
        \end{minipage}
        \caption{Images of instantaneous Q-criterion isosurfaces $\mathbf{Q = 1}$ coloured by normalised velocity magnitude $|\mathbf{v}|/|\mathbf{v}_\infty|$ obtained with 3D-SP OVERFLOW (left) and 3D-SP PyFR (right) for the roamx-0201 airfoil.}
        \label{fig:qcrit-OF-PyFR-q3D-0201}
\end{figure}

\begin{figure}[h!]
  \centering
        \begin{subfigure}[t]{1\linewidth}
        \centering
        \includegraphics[width=0.5\linewidth]{Data/PyFR/Qcriterion/PyFR_Qcrit_colorbar.png}
        \end{subfigure}
        \begin{subfigure}[t]{0.9\linewidth}
        \includegraphics[width=\linewidth]{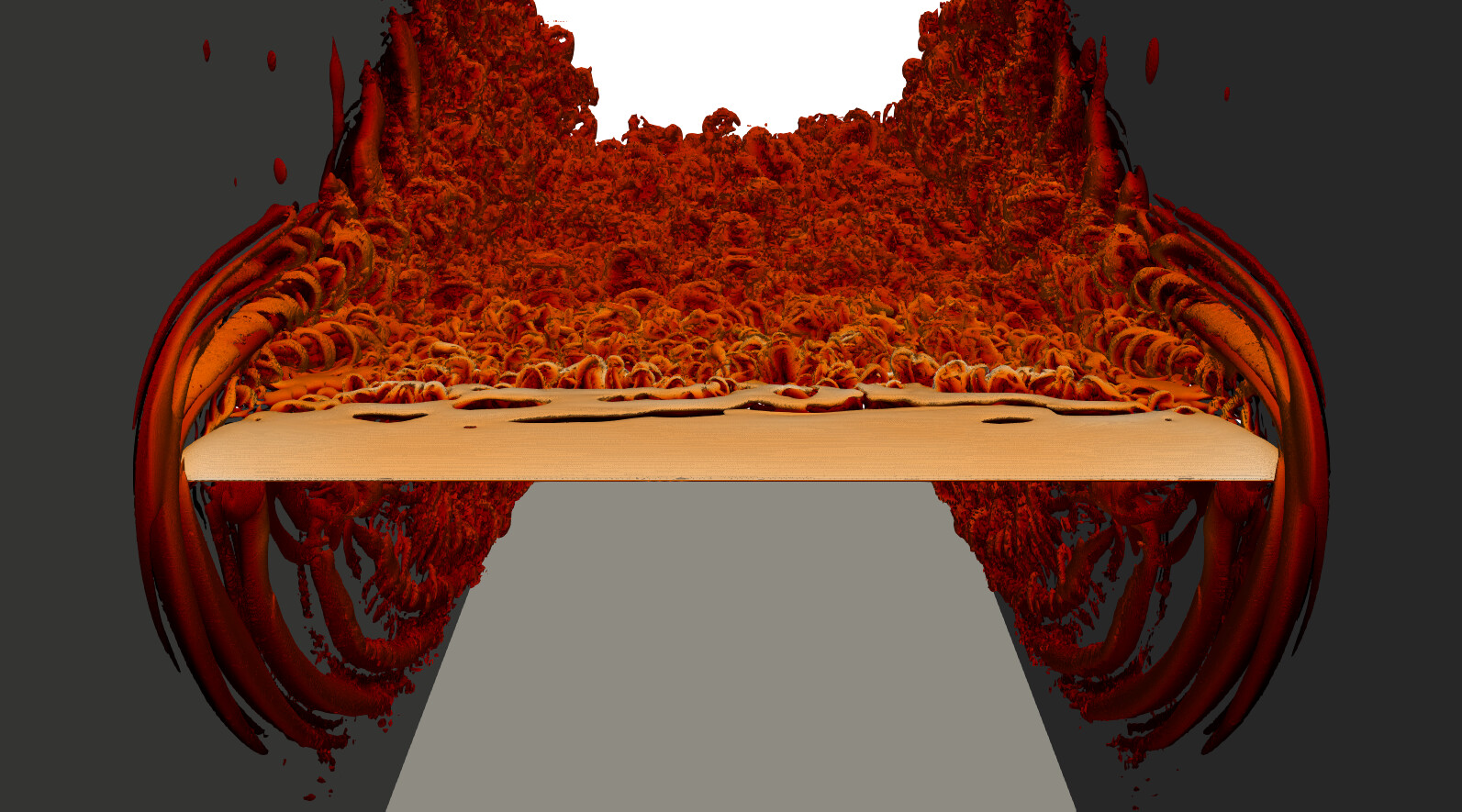}
        \end{subfigure}
        \caption{Instantaneous shapshot of Q-criterion isosurfaces coloured by normalised velocity magnitude $|\mathbf{v}|/|\mathbf{v}_\infty|$ obtained with a VWT PyFR simulation for the roamx-0201 airfoil with $\alpha = 6^\circ$ case.}
        \label{fig:qcrit-VWT-roamx}
        
\end{figure} 

%% file: Figures-tex/time-series.tex
\begin{figure}[h!]
  \centering
        \begin{subfigure}[t]{\textwidth}
        \centering
        \hspace{1.2cm}
        \includegraphics[height=3.7cm]{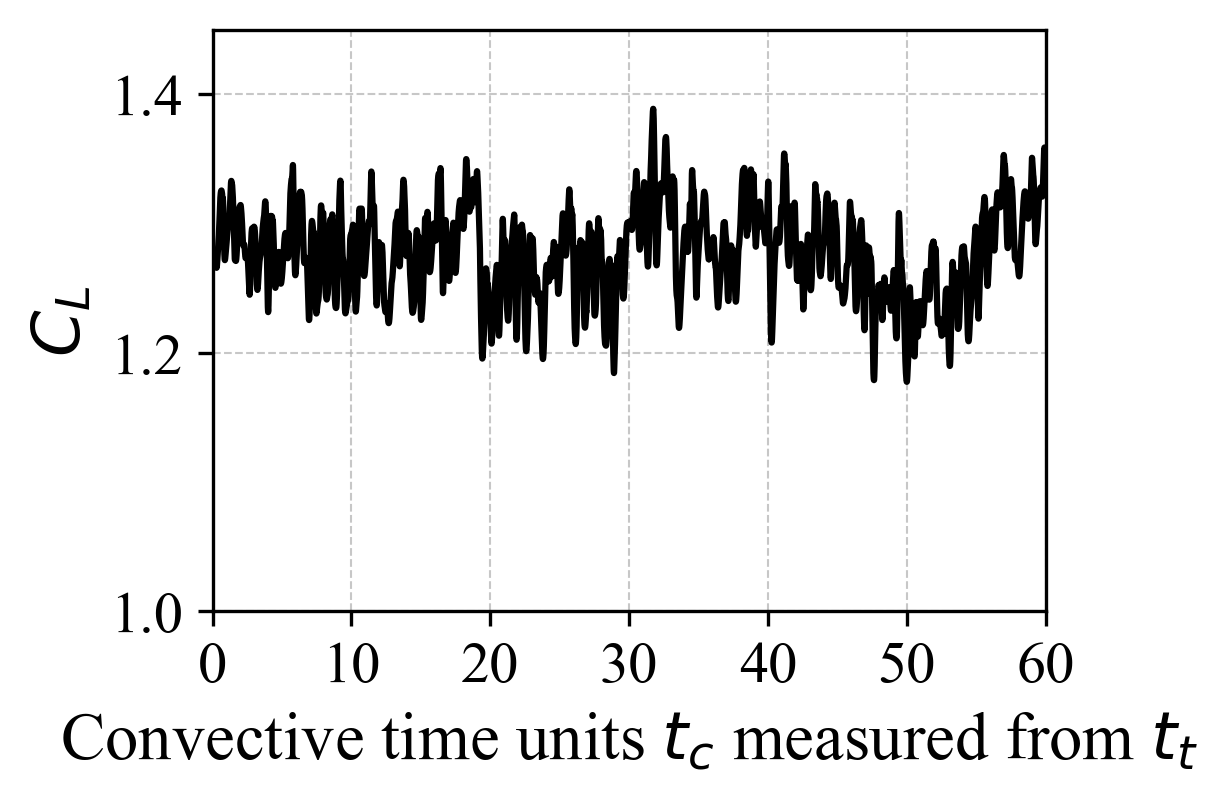}\hspace{0.9cm}
        \raisebox{0.5cm}{\includegraphics[width=0.43\linewidth]{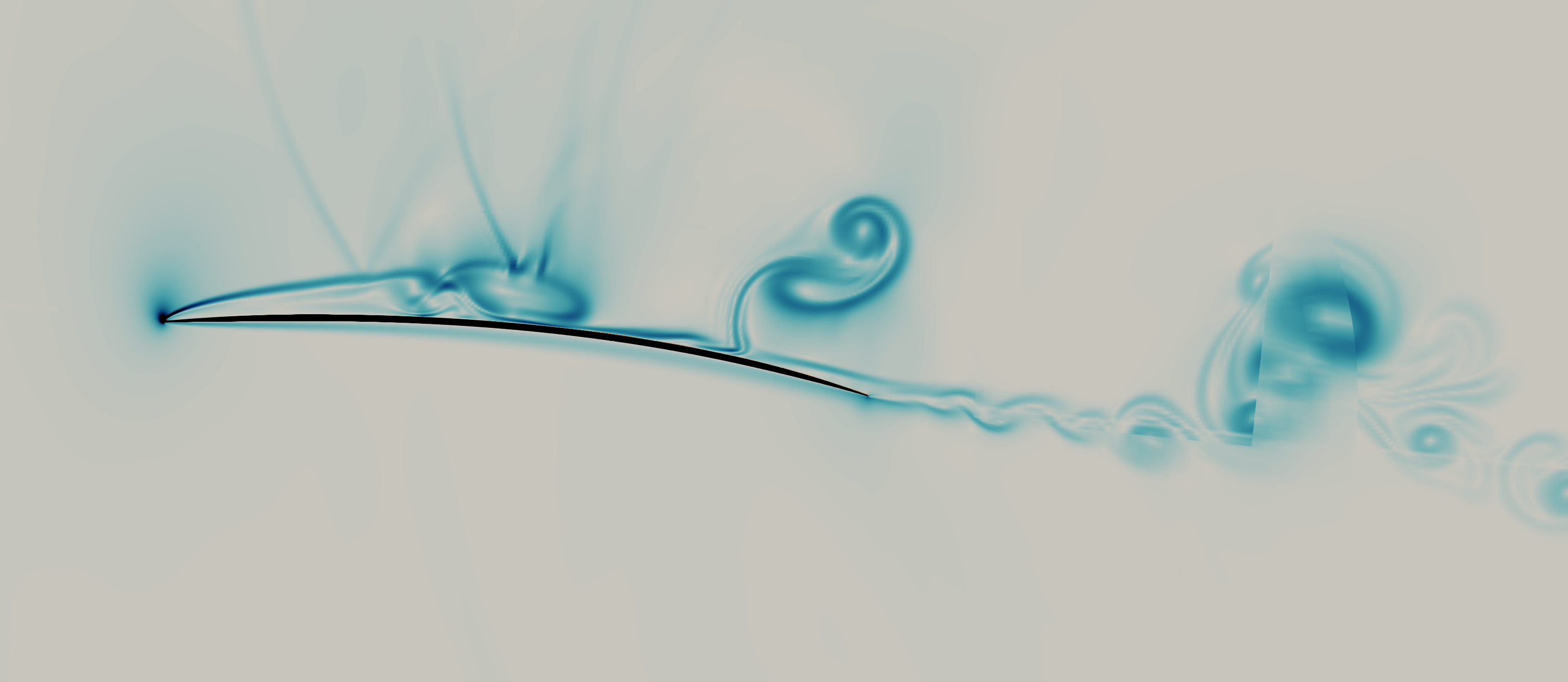}}
        \includegraphics[height=3.7cm]{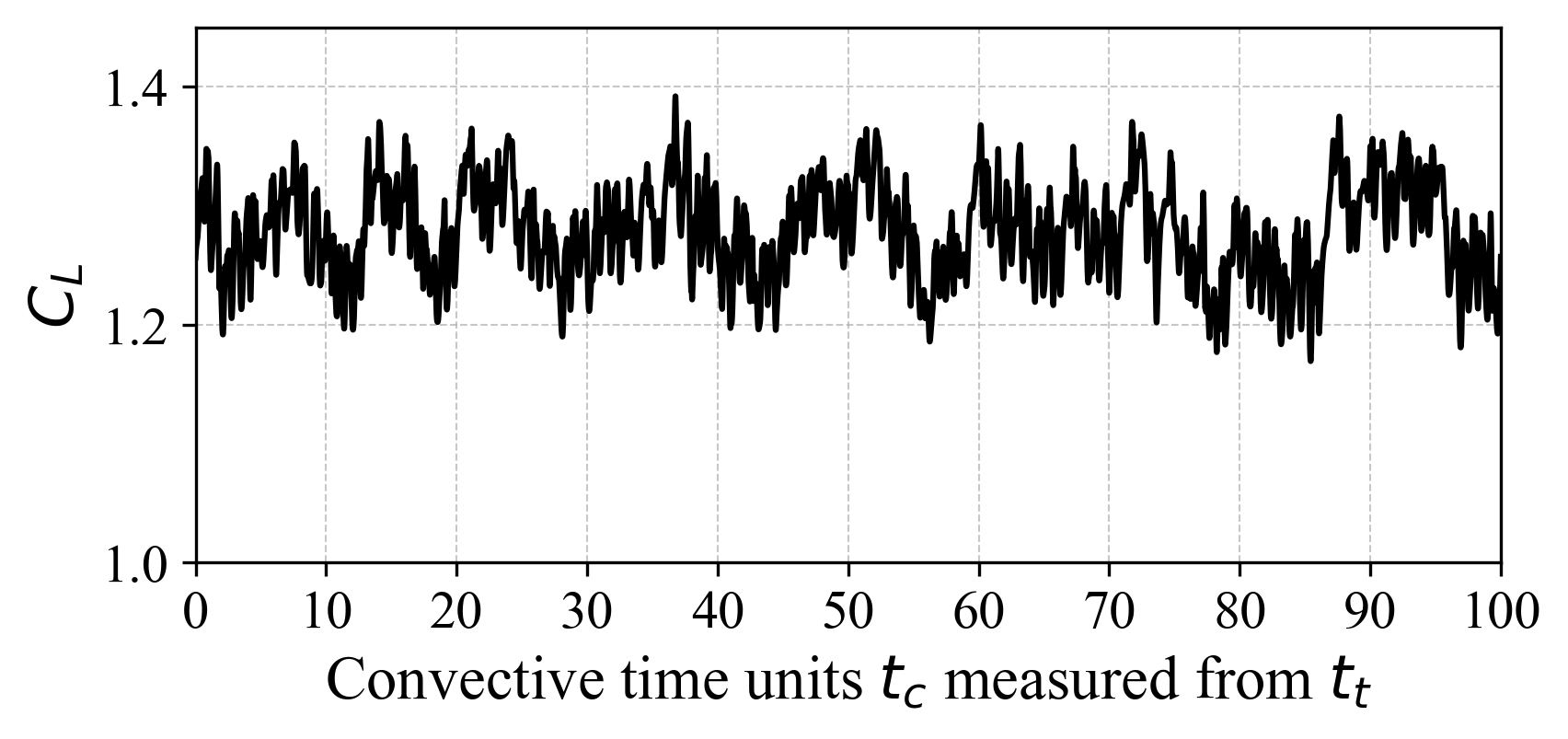}
        \raisebox{0.5cm}{\includegraphics[width=0.43\linewidth]{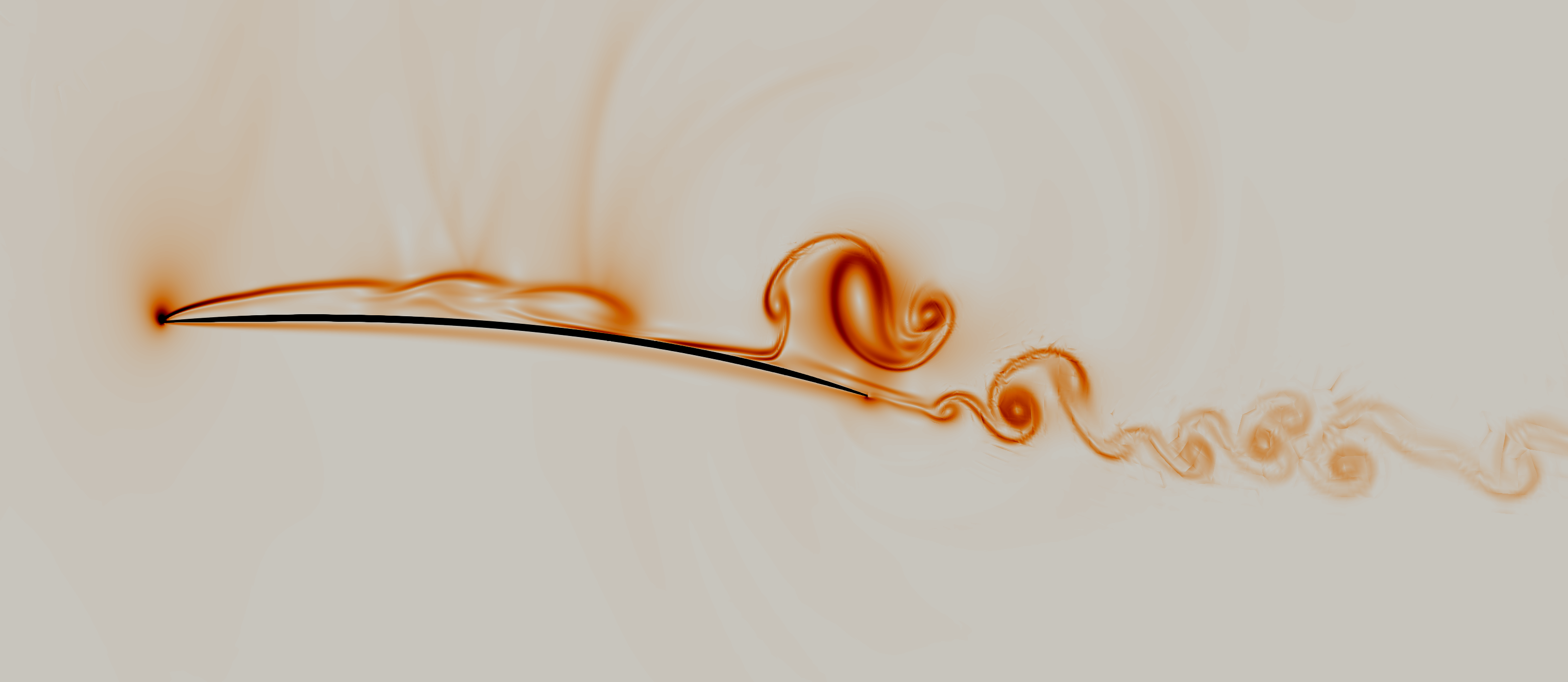}}
        \caption{2D OVERFLOW (top) and 2D PyFR (bottom)}
        \end{subfigure}
        \begin{subfigure}[t]{\textwidth}
        \centering
        \hspace{1.2cm}
        \includegraphics[height=3.7cm]{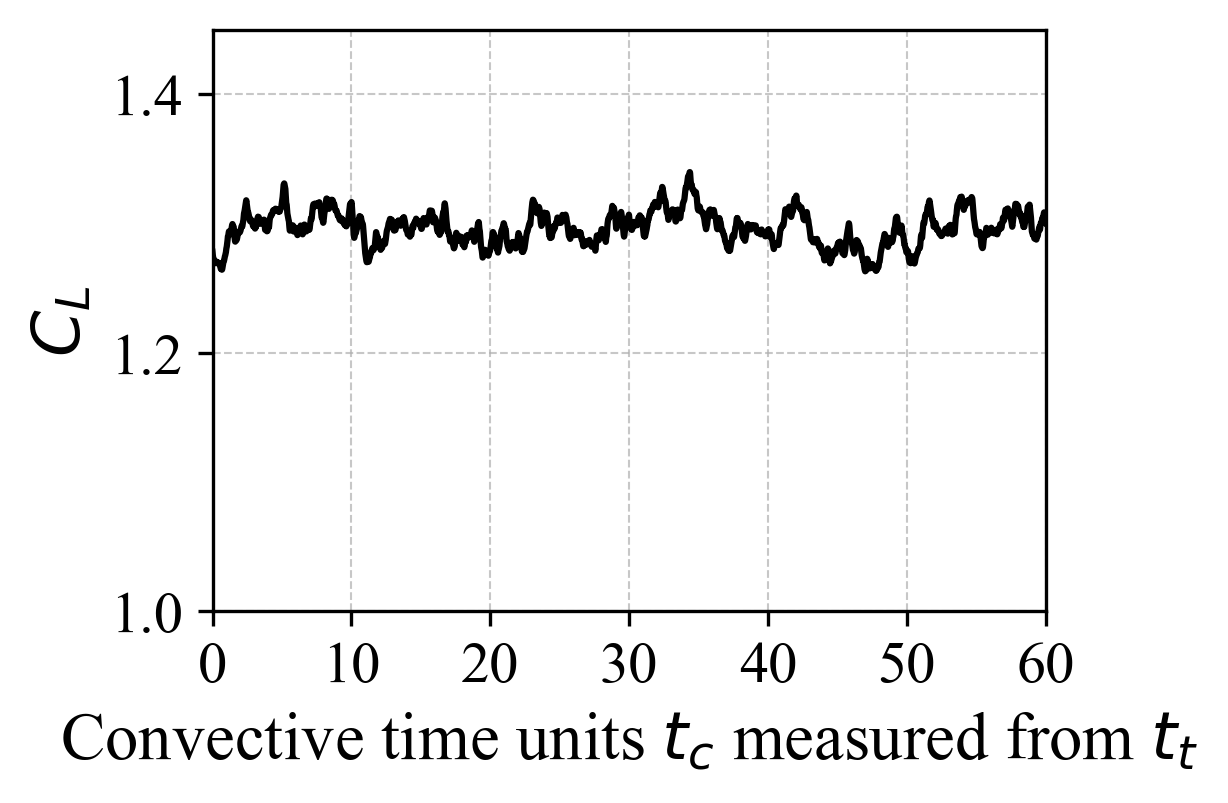}\hspace{0.9cm}
        \raisebox{0.5cm}{\includegraphics[width=0.43\linewidth]{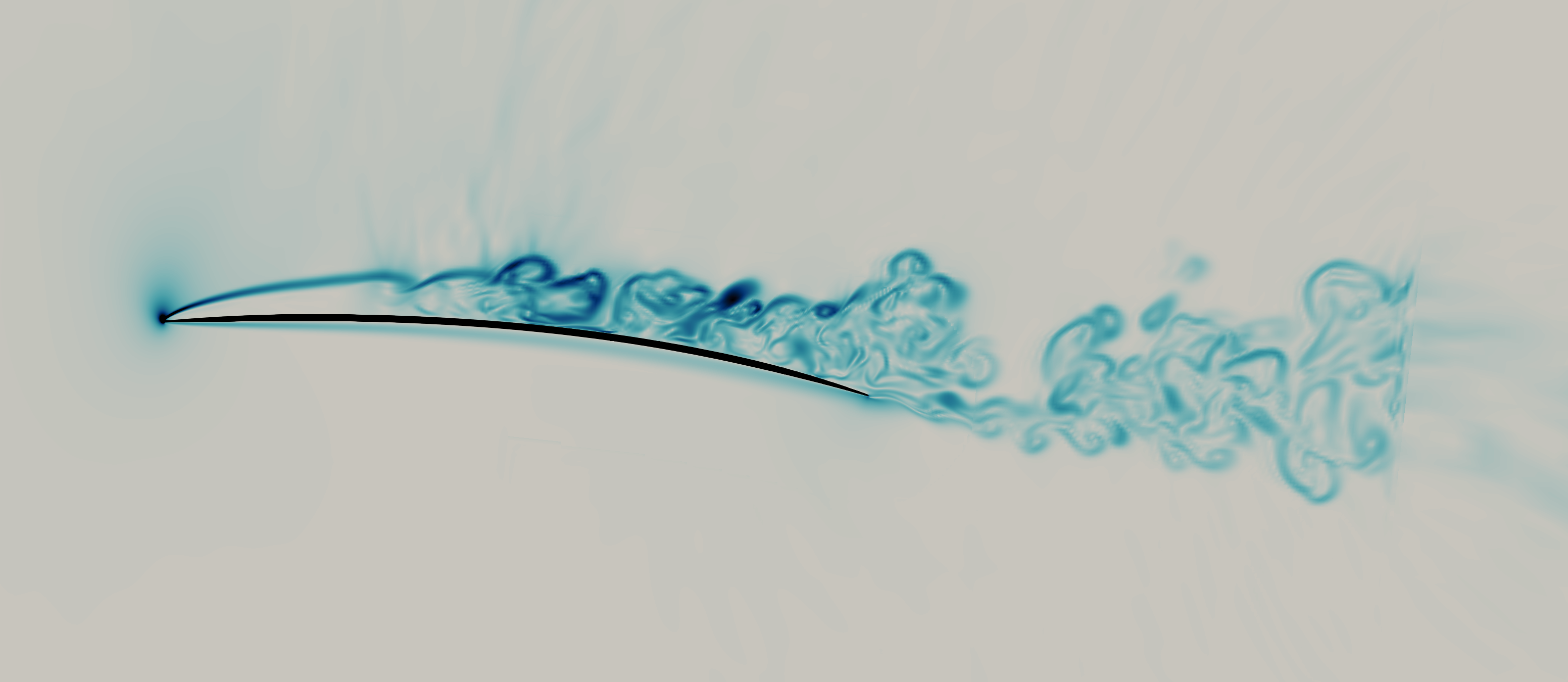}}\\
        
        \hspace{1.2cm}
        \includegraphics[height=3.7cm]{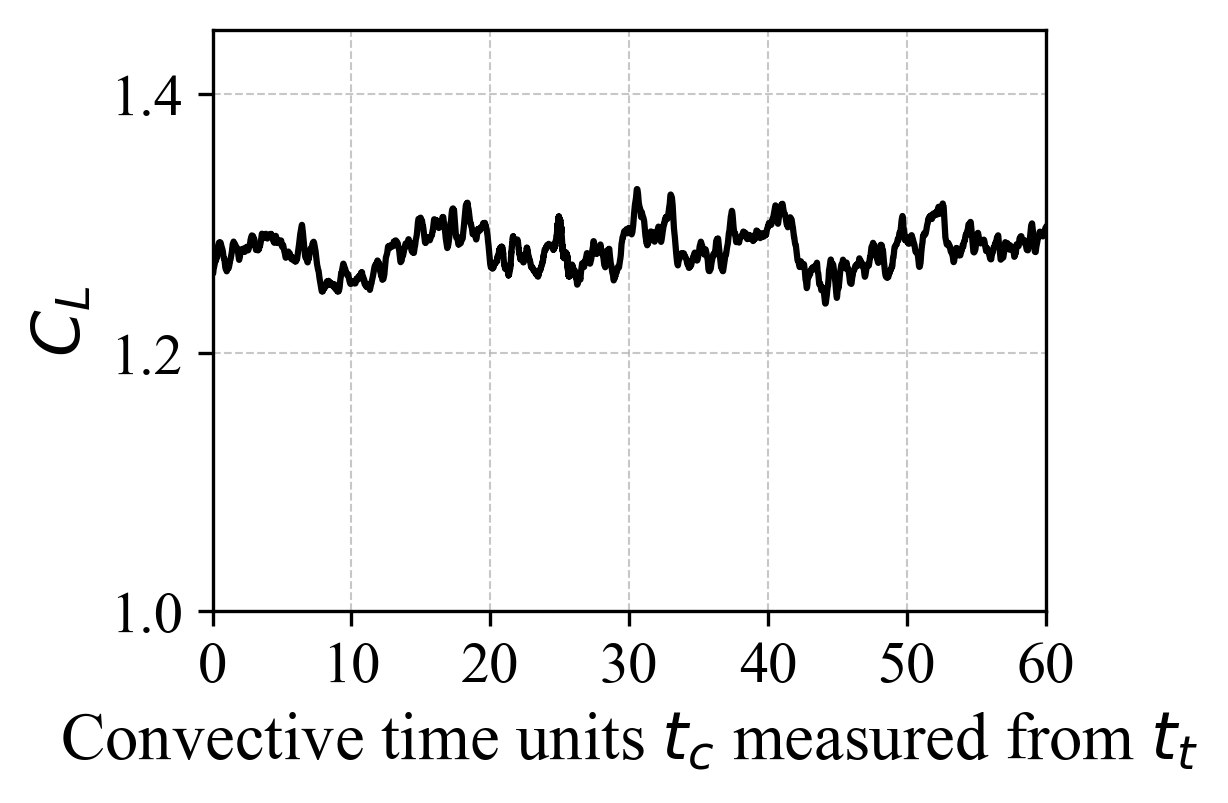}\hspace{0.9cm} 
        \raisebox{0.5cm}{\includegraphics[width=0.43\linewidth]{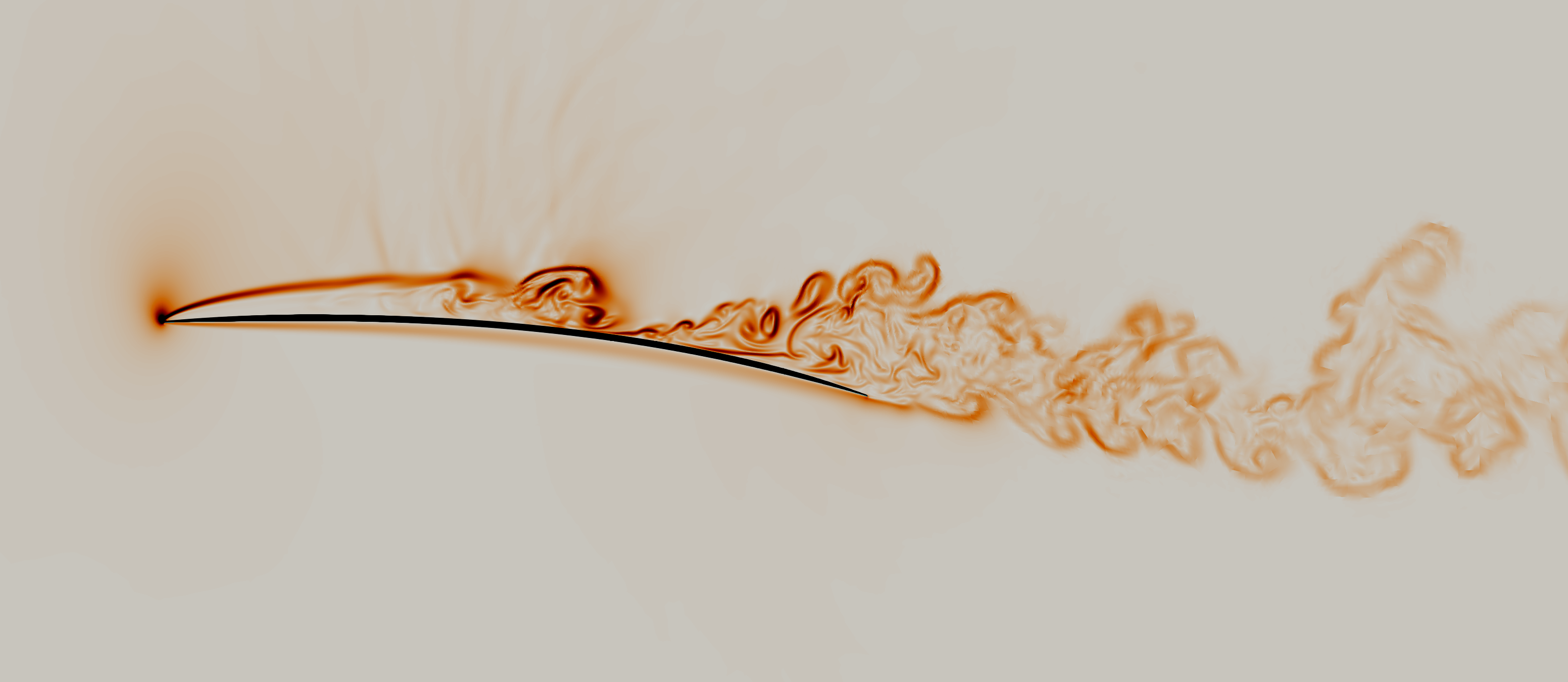}}
        \caption{3D-SP OVERFLOW (top) and 3D-SP PyFR (bottom)}
        \end{subfigure}
        \begin{subfigure}[t]{\textwidth}
        \centering
        \hspace{1.35cm}
        \includegraphics[height=3.7cm]{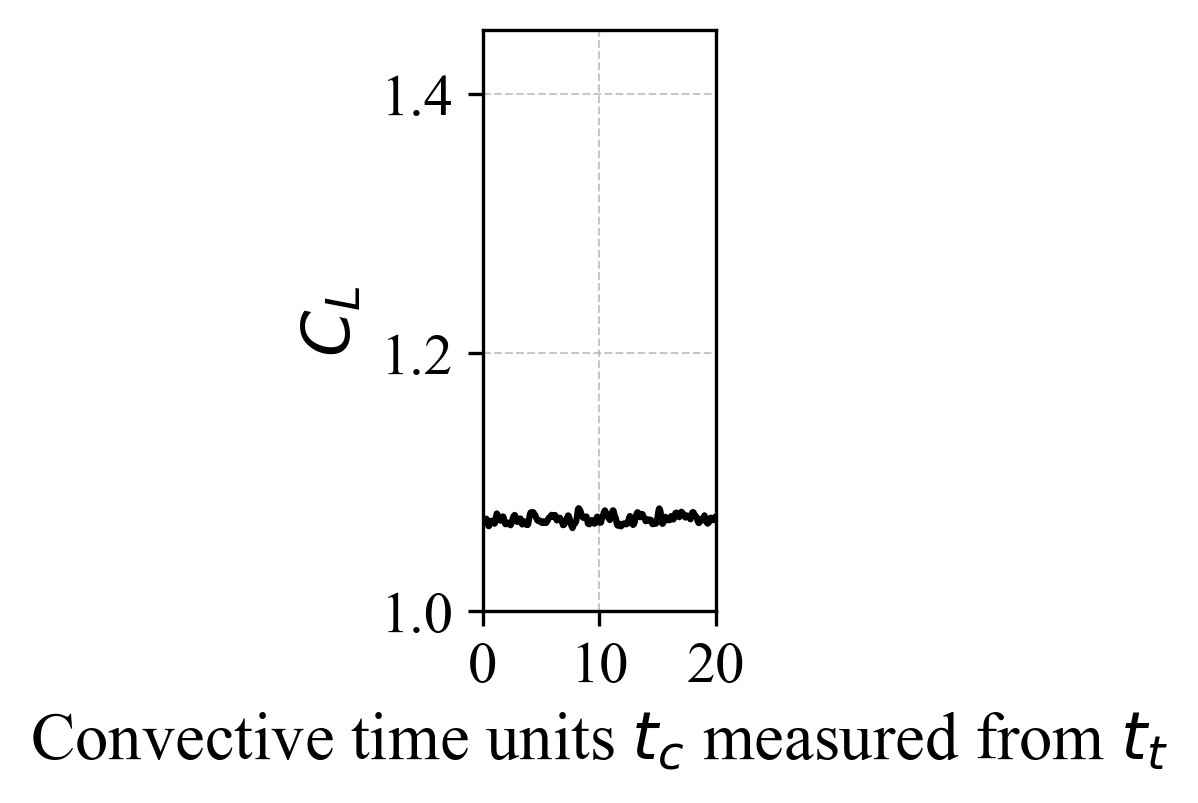} \hspace{0.8cm}
        \raisebox{0.5cm}{\includegraphics[width=0.43\linewidth]{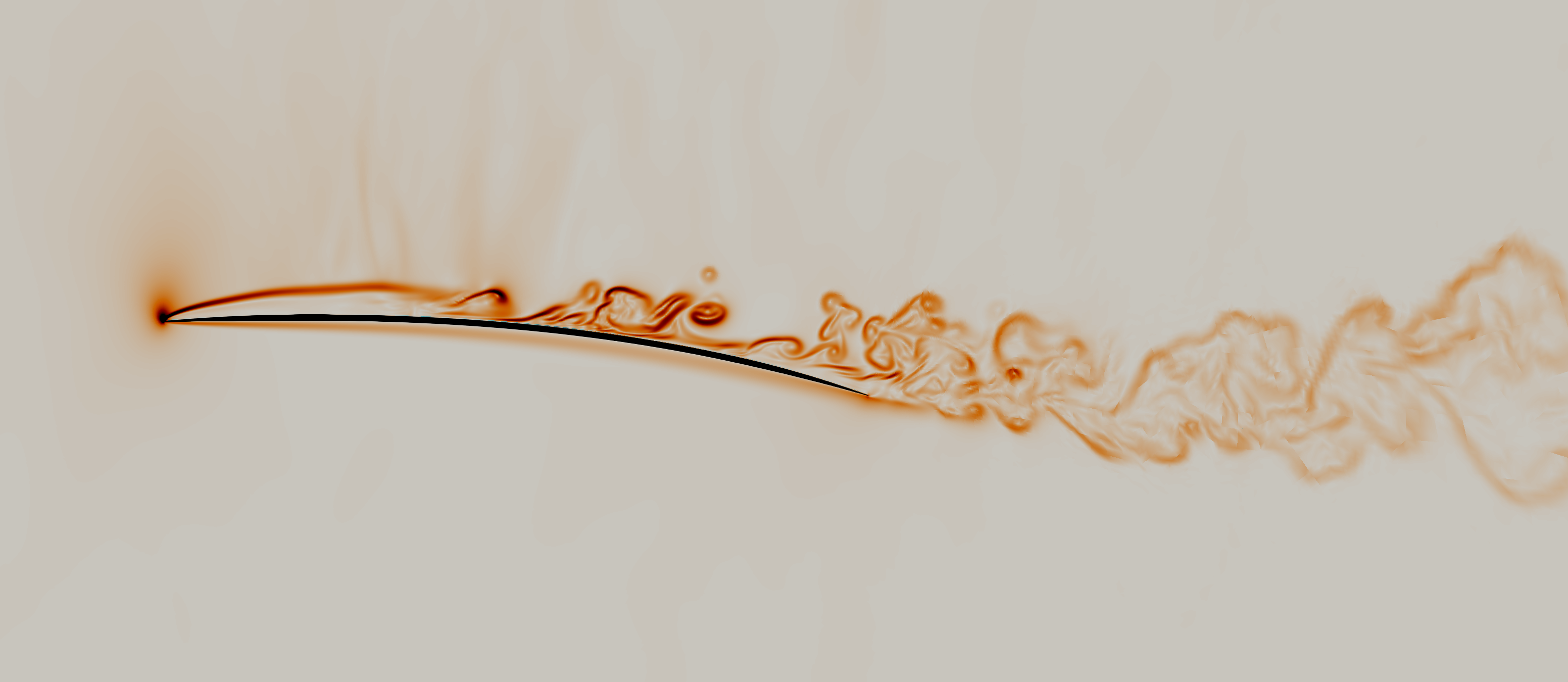}}
        \caption{VWT PyFR}
        \end{subfigure}
        \caption{$C_L$ as a function of time over the Data Extraction Period (left) and images of instantaneous density gradient magnitude $|\boldsymbol{\nabla}\rho|$ (right) for the 2D (a), mid-span 3D-SP (b) and mid-span VWT (c) simulations of the roamx-0201 airfoil. }
        \label{fig:time-series-avg-PyFR}
\end{figure}

%% file: Figures-tex/polars.tex
% Comparison ROAMX vs. CLF

\begin{figure}[!h]
    \centering
    \includegraphics[width=\linewidth]{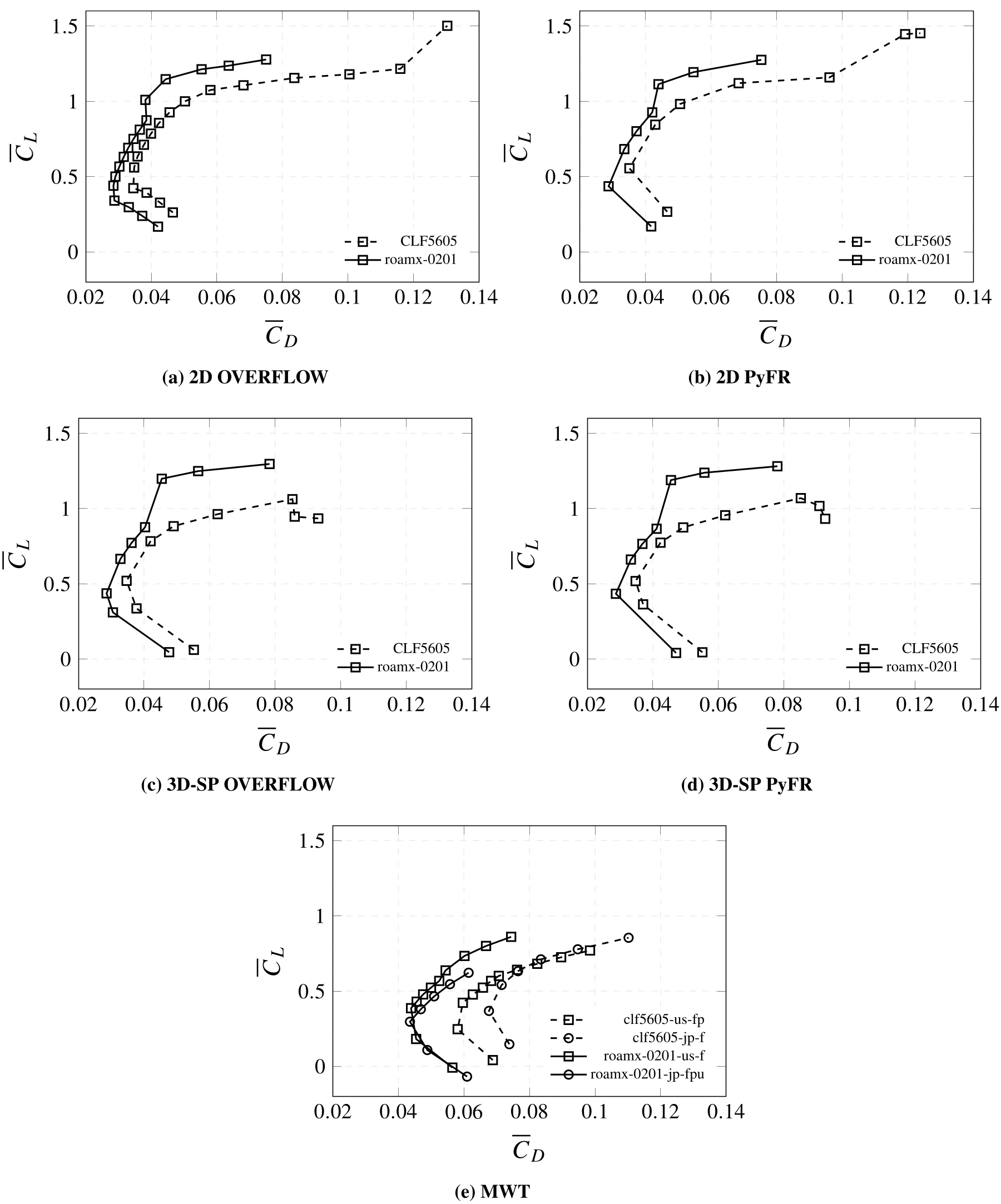}
    \caption{$\overline{C}_L$ vs. $\overline{C}_D$ polars comparing performance of the CLF5605 and the roamx-2021 airfoils for 2D OVERFLOW simulations (a), 2D PyFR simulations (b), 3D-SP OVERFLOW simulations (c), 3D-SP PyFR simulations (d), and MWT experiments (e).}
    \label{fig:Cl-Cd-compare}
\end{figure}

%% file: Figures-tex/cl_over_cd_plots.tex
% Comparison ROAMX vs. CLF

\begin{figure}[!h]
    \centering
    \includegraphics[width=\linewidth]{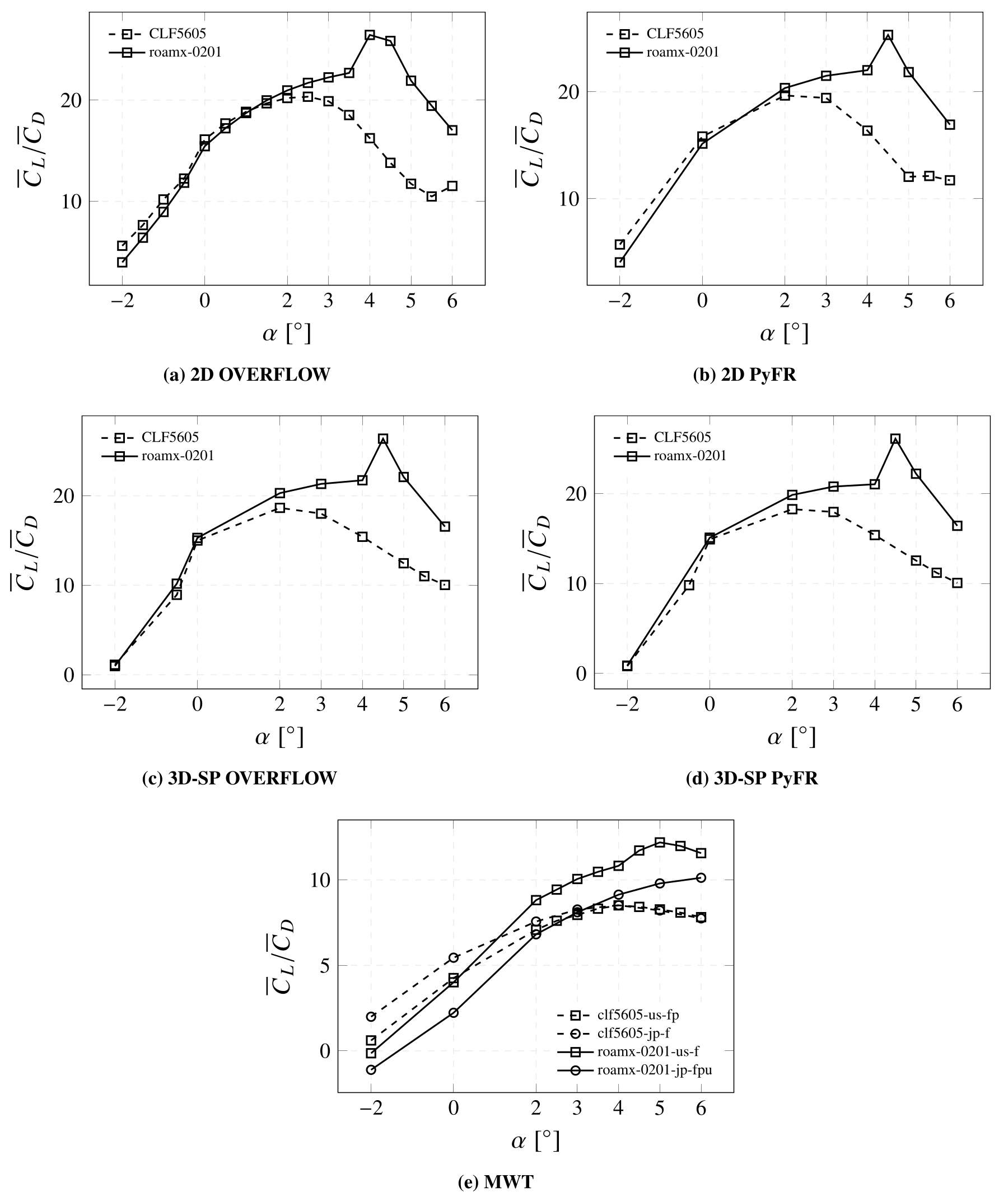}
    \caption{$\overline{C}_L/\overline{C}_D$ as a function of angle of attack $\alpha$ comparing performance of the CLF5605 and the roamx-2021 airfoils for 2D OVERFLOW simulations (a), 2D PyFR simulations (b), 3D-SP OVERFLOW simulations (c), 3D-SP PyFR simulations (d) and MWT experiments (e).}
    \label{fig:E-compare}
\end{figure}

%% file: Figures-tex/Appendix/grid-independence.tex
\begin{figure}[!h]\centering
\includegraphics[width=\linewidth]{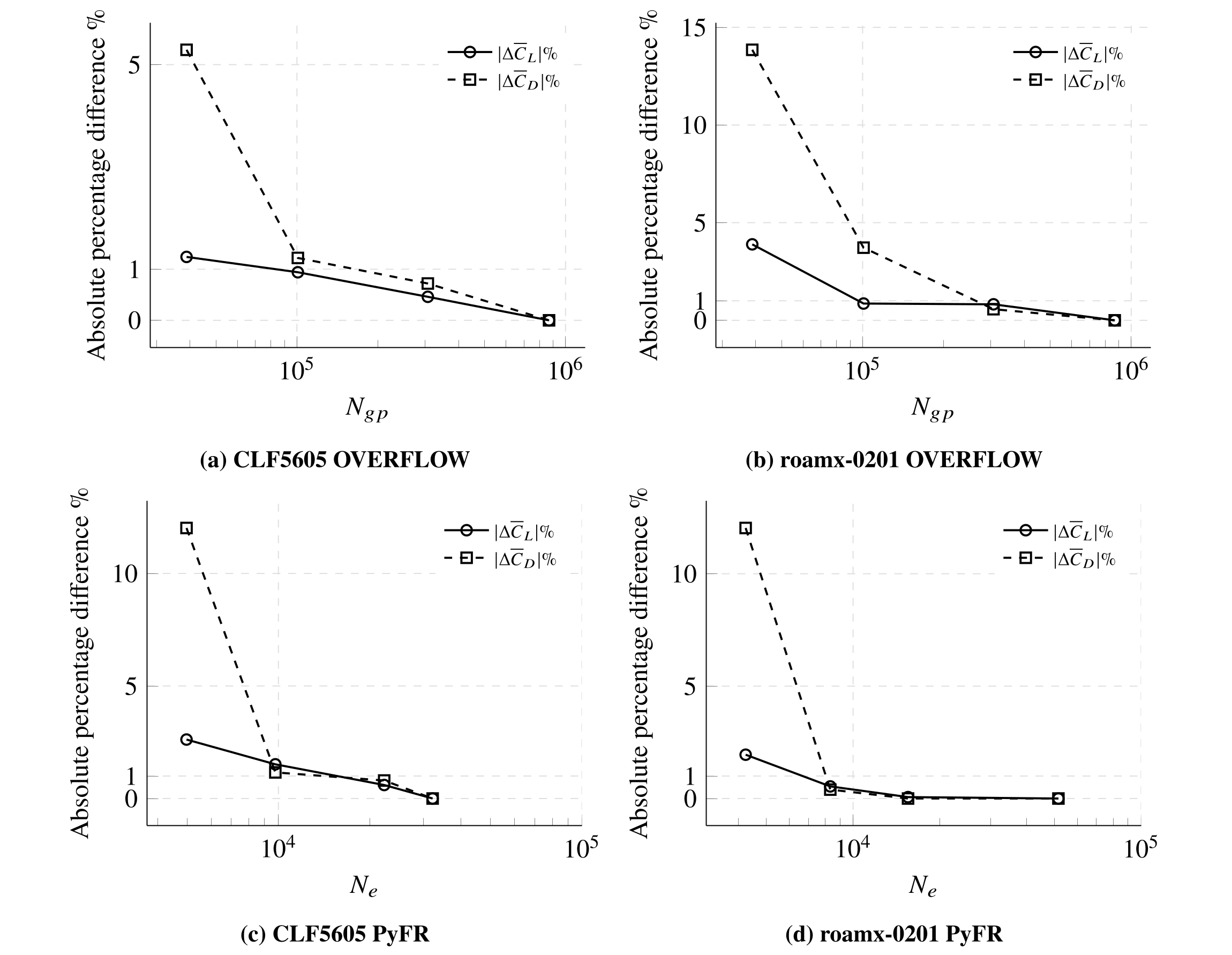}
\caption{Absolute percentage differences in time-averaged lift and drag coefficients with respect to the finest mesh, \( \lvert \Delta \overline{C}_L \rvert \% \) and \( \lvert \Delta \overline{C}_D \rvert \% \), plotted as a function of the number of grid points \( N_{gp} \)  for 2D OVERFLOW and number of elements \( N_e \) for 2D PyFR to assess grid convergence of CLF5605 and roamx-0201 at \( \alpha = 6^\circ \).}
\label{fig:grid-independence}
\end{figure}

%% file: Figures-tex/Appendix/span-independence.tex
\begin{figure}[!h]\centering
\includegraphics[width=\linewidth]{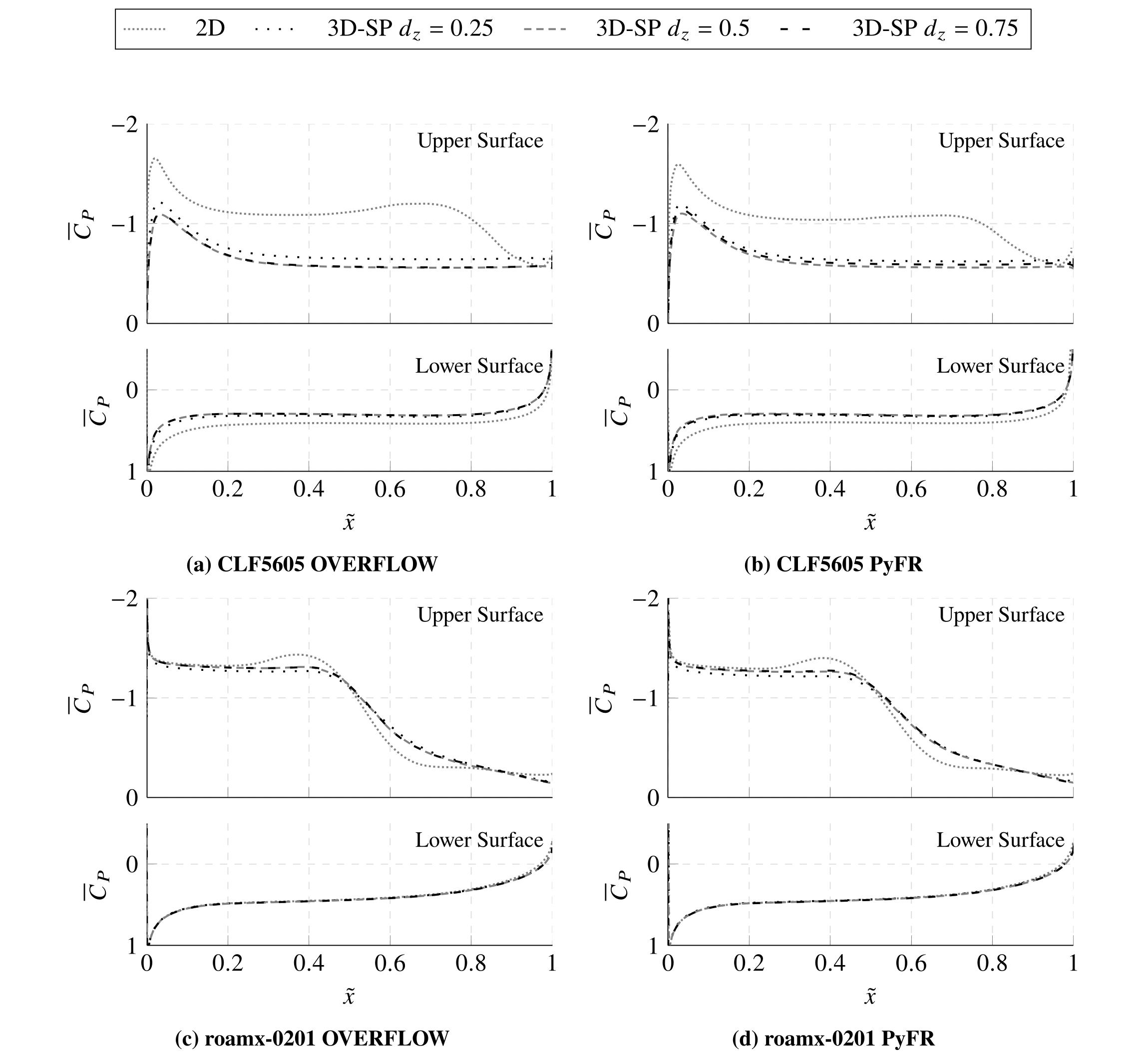}
\caption{Plots of time-averaged pressure coefficient $\overline{C}_P$ as a function of $\tilde{x}$ for different span lengths of the domain $d_z$ to assess domain span length independence for 3D-SP OVERFLOW and PyFR simulations of CLF5605 and roamx-0201 at $\alpha=6^\circ$.}
\label{fig:Cp-span-independence}
\end{figure}

%% file: Figures-tex/Appendix/DNS-resolution.tex
\begin{figure}[!h]\centering
\includegraphics[width=0.98\linewidth]{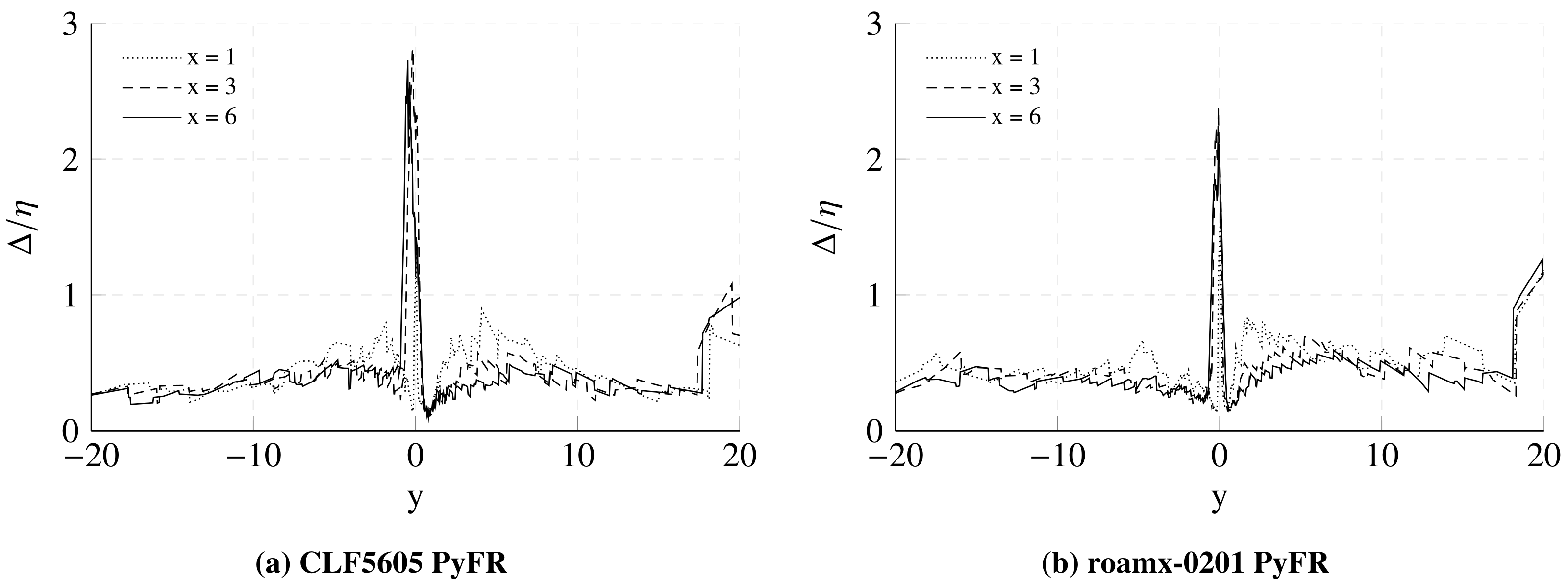}
\caption{Plots of $\boldsymbol{\Delta/\eta}$ in the mid-span plane as a function of $\mathbf{y}$ at $\mathbf{x=1}$, $\mathbf{x=3}$ and $\mathbf{x=6}$ for 3D-SP PyFR.}
\label{fig:DNS-resolution-y}
\end{figure}

\begin{figure}[!h]\centering
\includegraphics[width=0.98\linewidth]{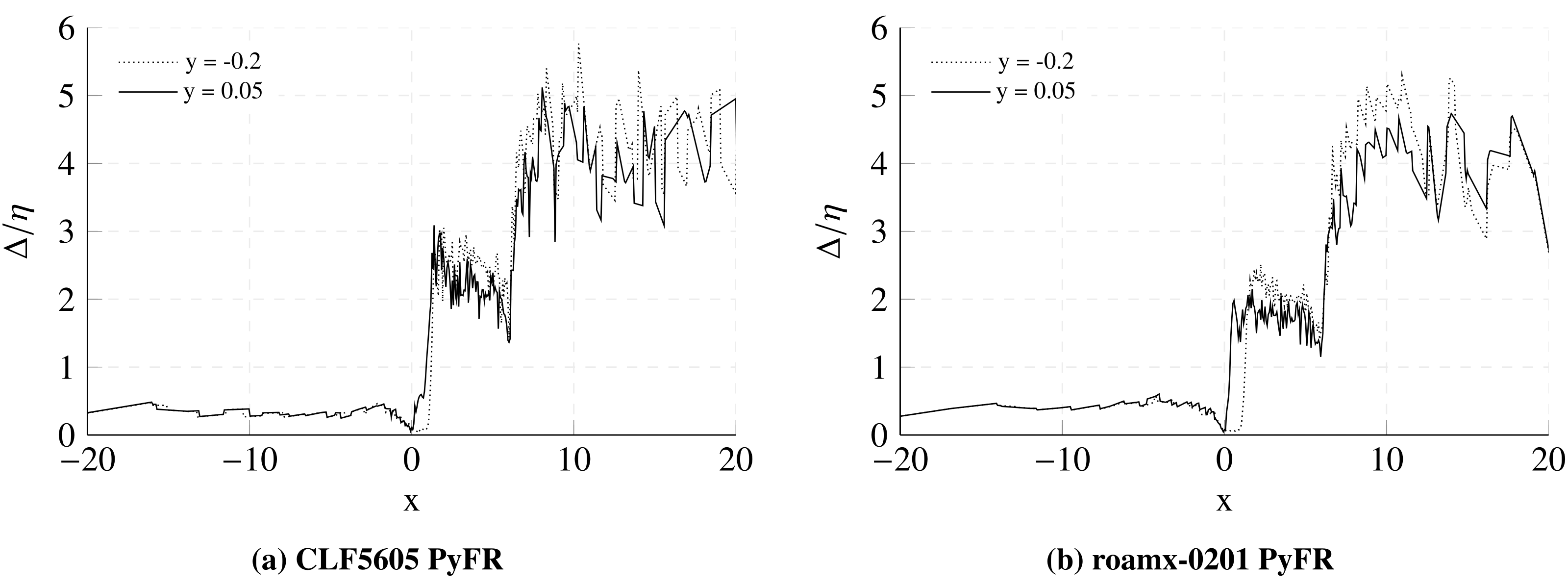}
\caption{Plots of $\boldsymbol{\Delta/\eta}$ in the mid-span plane as a function of $\mathbf{x}$ at $\mathbf{y=0.05}$ and $\mathbf{y=-0.2}$ for 3D-SP PyFR.}
\label{fig:DNS-resolution-x}
\end{figure}

%% file: Figures-tex/Appendix/all-cp.tex
% ---------------------------------------------------------------- clf-5605 Cp all AoA

\begin{figure}[!h]\centering
\includegraphics[width=\linewidth]{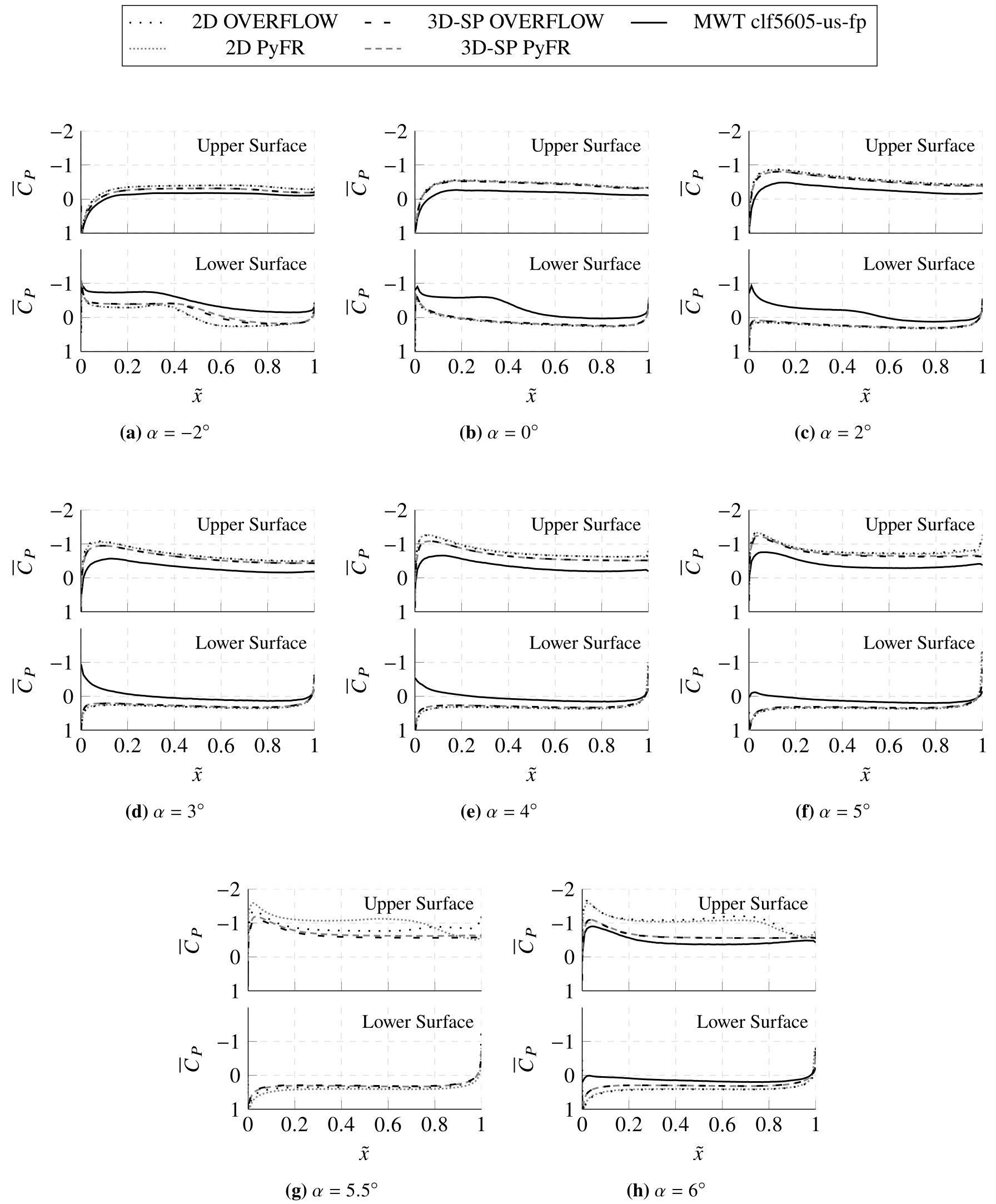}
\caption{Plots of time-averaged pressure coefficient $\overline{C}_P$ as a function of $\tilde{x}$ for different angles of attack $\alpha$ for the CLF5605 airfoil. }
\label{fig:Cp-allAoA-clf-5605}
\end{figure}

% ---------------------------------------------------------------- roamx-0201 Cp all AoA

\begin{figure}[!h]\centering
\includegraphics[width=\linewidth]{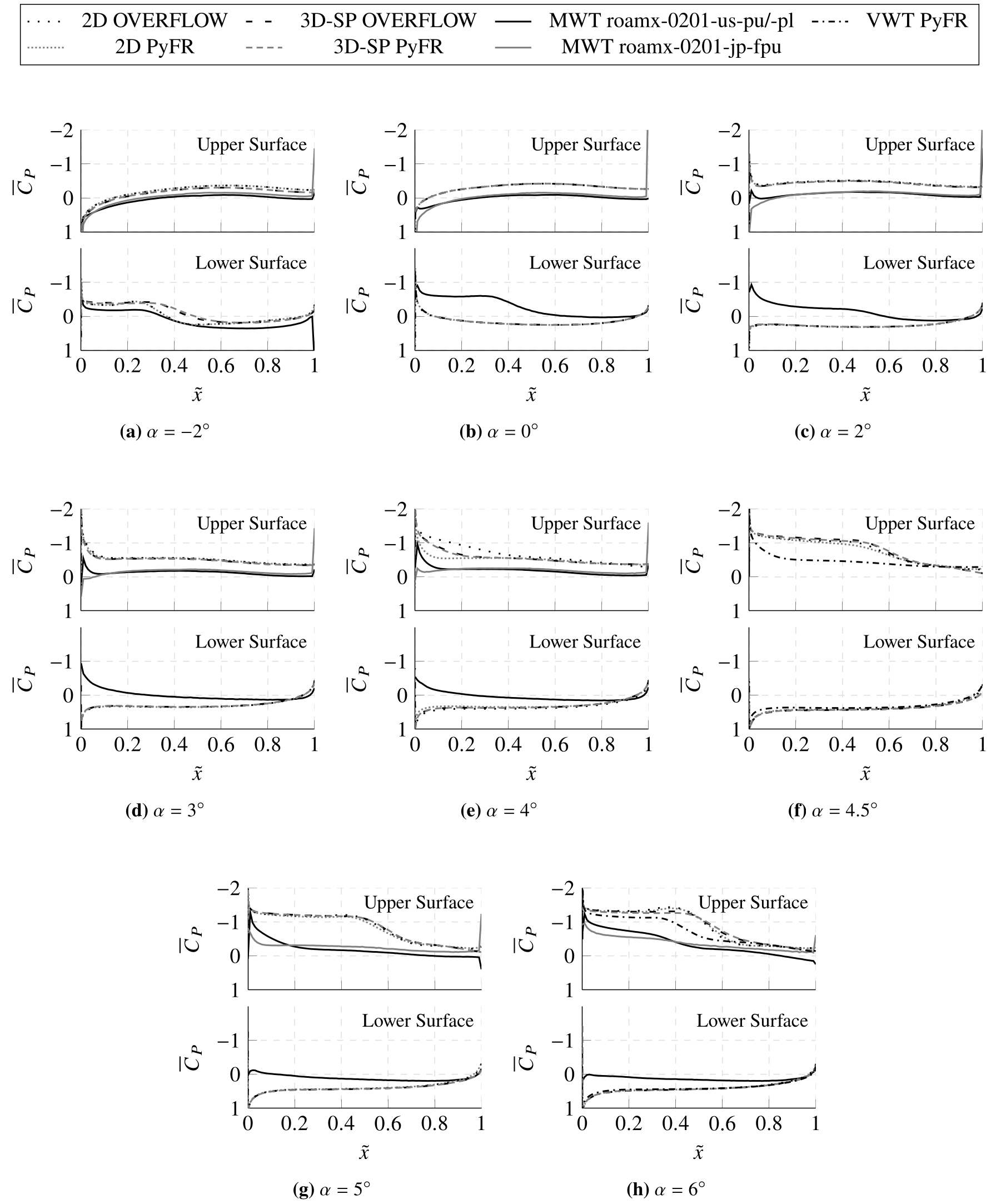}
\caption{Plots of time-averaged pressure coefficient $\overline{C}_P$ as a function of $\tilde{x}$ for different angles of attack $\alpha$ for the roamx-0201 airfoil. }
\label{fig:Cp-allAoA-roamx-0201}
\end{figure}

%% file: Figures-tex/Appendix/all-schlieren.tex
\begin{figure}[p]
  \centering

  % --------- AoA = -2
  \begin{minipage}{\linewidth}
    \centering
    \begin{subfigure}[t]{0.99\linewidth}
      \includegraphics[height=1.6cm]{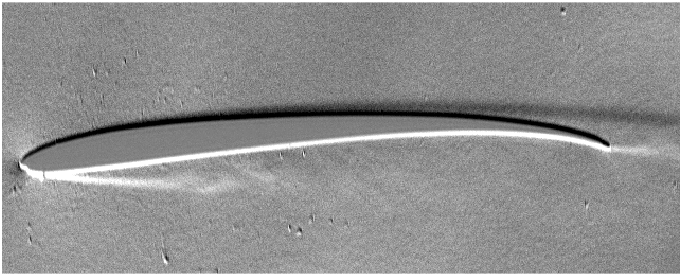}
      \includegraphics[height=1.6cm]{Data/OVERFLOW/Density_gradient/OF-clf5605-q3D--2deg.png}
      \includegraphics[height=1.6cm]{Data/PyFR/Density_gradient/rhograd--2deg-q3D-clf5605.png}
    \end{subfigure}
    \caption*{$\alpha = -2^\circ$}
  \end{minipage}\\[1ex]

  % --------- AoA = 0
  \begin{minipage}{\linewidth}
    \centering
    \begin{subfigure}[t]{0.99\linewidth}
      \includegraphics[height=1.6cm]{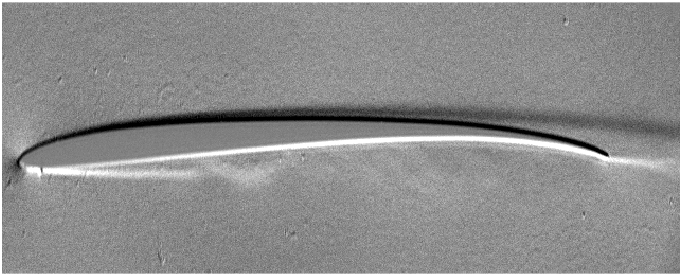}
      \includegraphics[height=1.6cm]{Data/OVERFLOW/Density_gradient/OF-clf5605-q3D-0deg.png}
      \includegraphics[height=1.6cm]{Data/PyFR/Density_gradient/rhograd-0deg-q3D-clf5605.png}
    \end{subfigure}
    \caption*{$\alpha = 0^\circ$}
  \end{minipage}\\[1ex]

  % --------- AoA = 2
  \begin{minipage}{\linewidth}
    \centering
    \begin{subfigure}[t]{0.99\linewidth}
      \includegraphics[height=1.6cm]{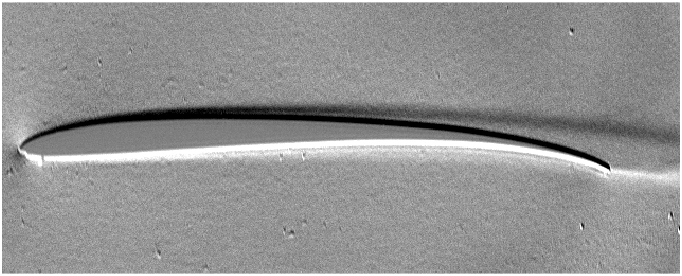}
      \includegraphics[height=1.6cm]{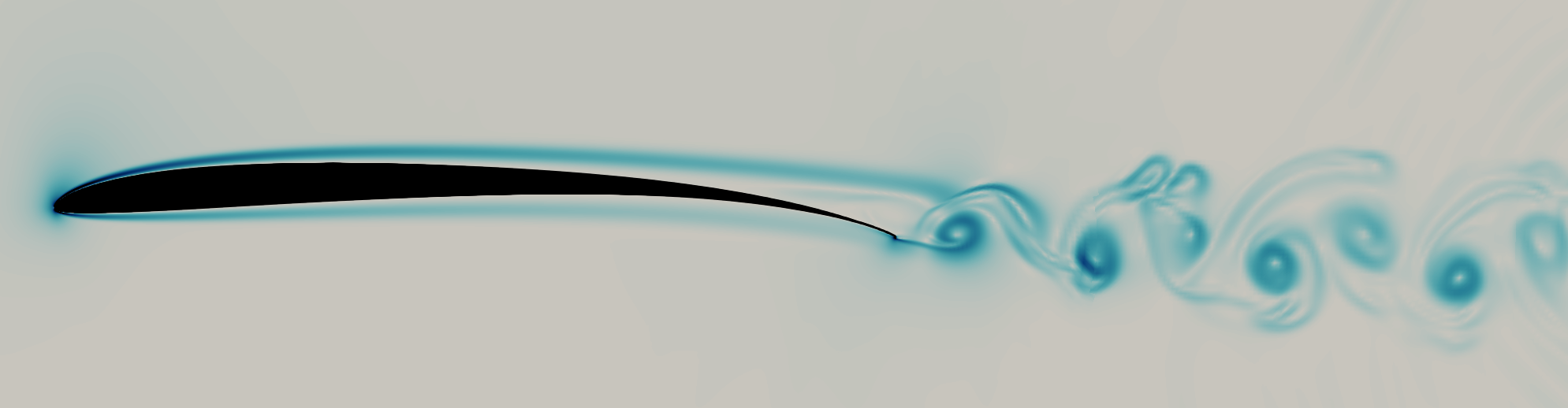}
      \includegraphics[height=1.6cm]{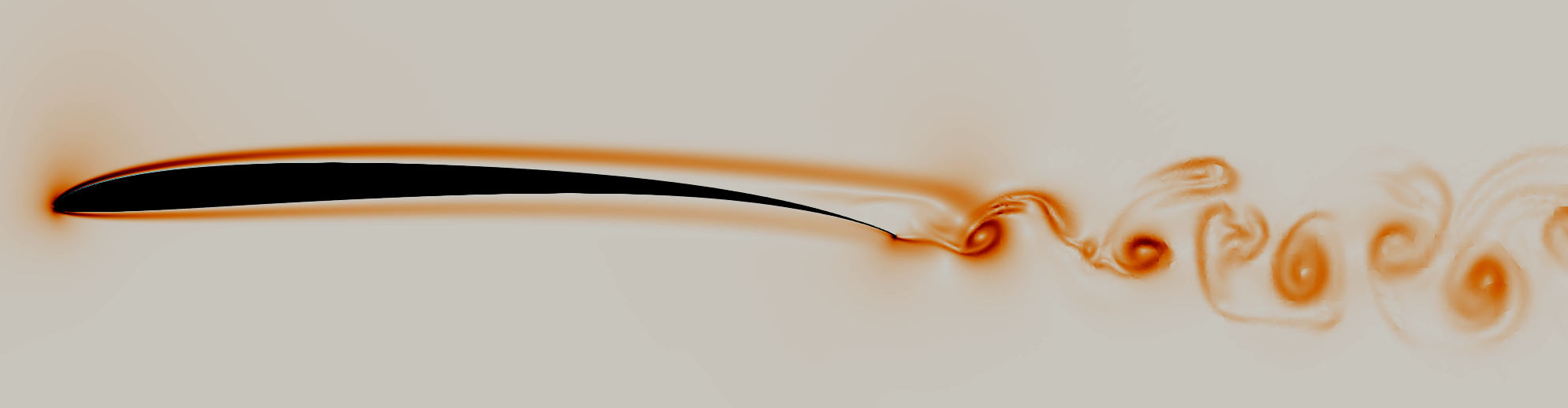}
    \end{subfigure}
    \caption*{$\alpha = 2^\circ$}
  \end{minipage}\\[1ex]

  % --------- AoA = 3
  \begin{minipage}{\linewidth}
    \centering
    \begin{subfigure}[t]{0.99\linewidth}
      \includegraphics[height=1.6cm]{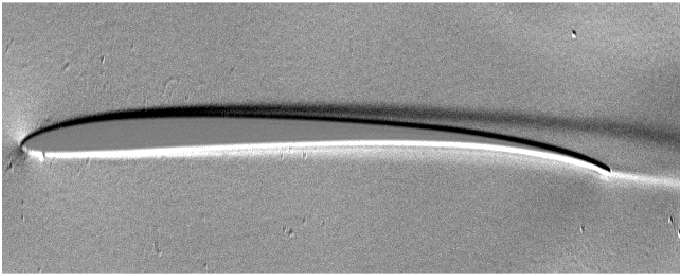}
      \includegraphics[height=1.6cm]{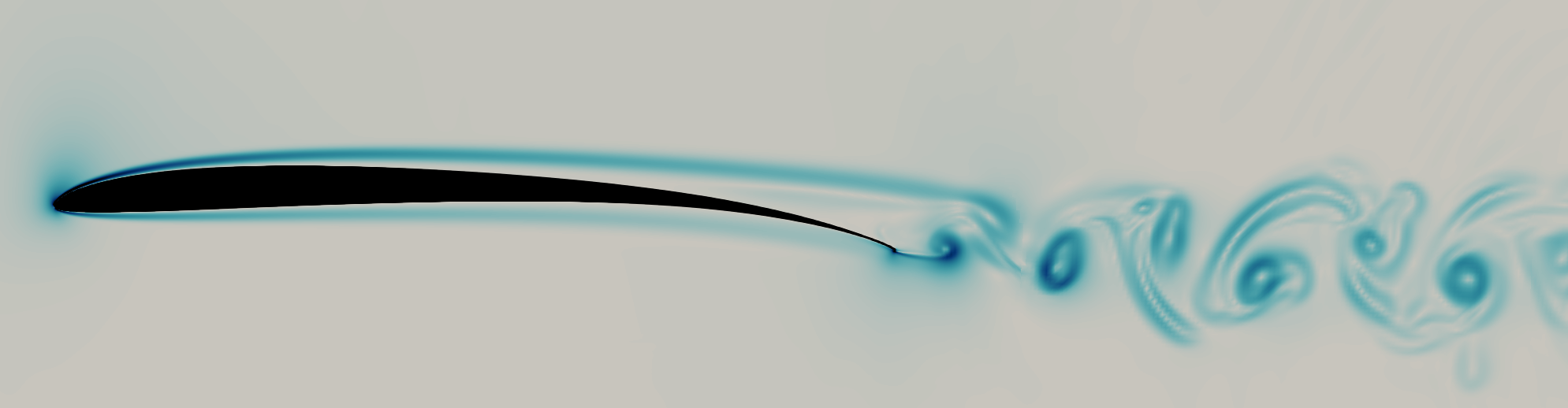}
      \includegraphics[height=1.6cm]{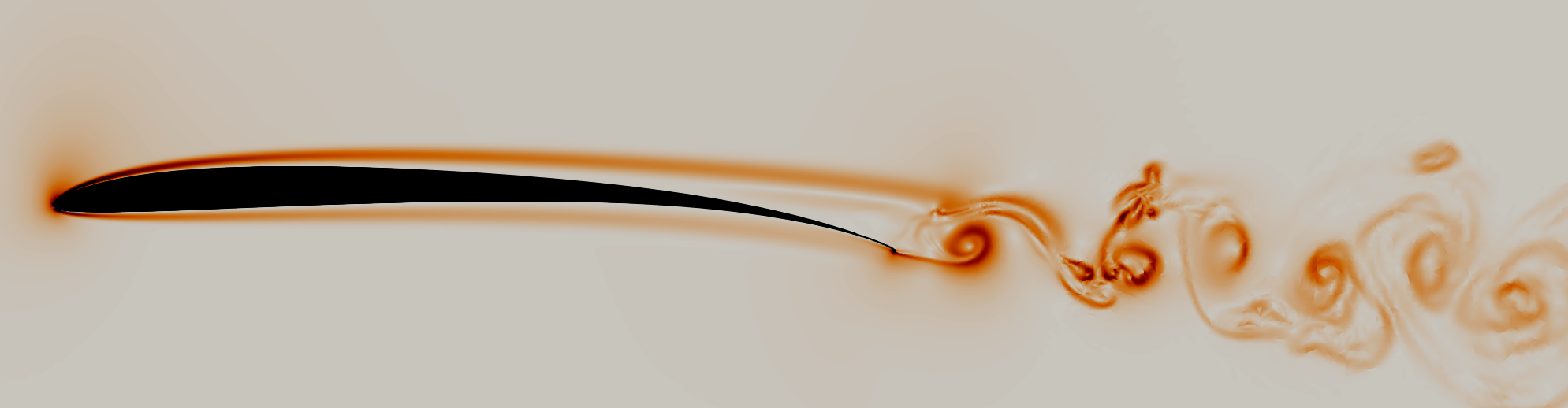}
    \end{subfigure}
    \caption*{$\alpha = 3^\circ$}
  \end{minipage}\\[1ex]

  % --------- AoA = 4
  \begin{minipage}{\linewidth}
    \centering
    \begin{subfigure}[t]{0.99\linewidth}
      \includegraphics[height=1.6cm]{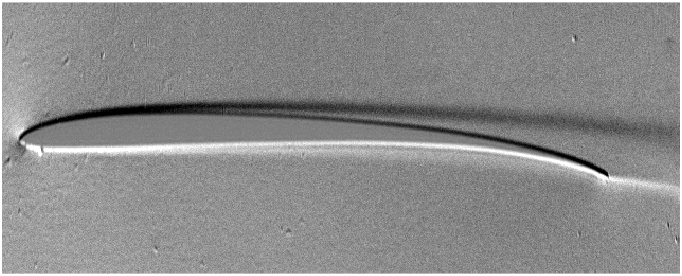}
      \includegraphics[height=1.6cm]{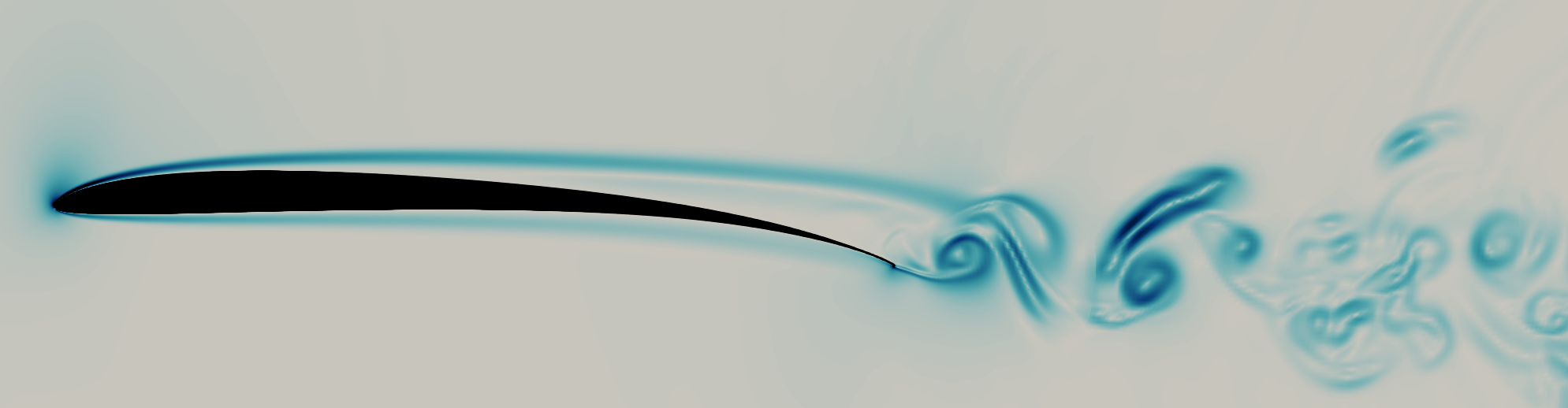}
      \includegraphics[height=1.6cm]{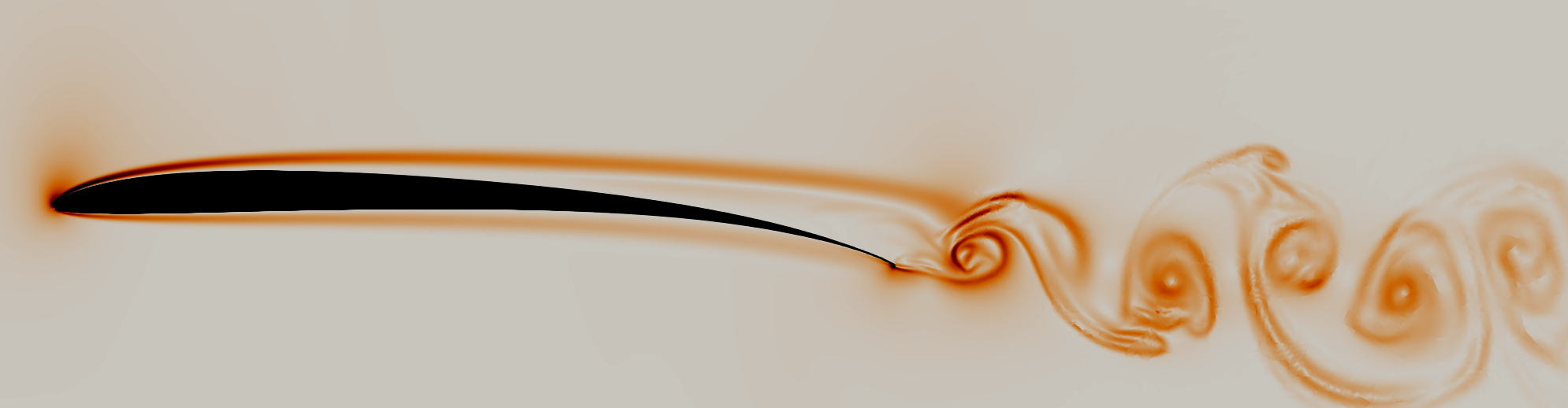}
    \end{subfigure}
    \caption*{$\alpha = 4^\circ$}
  \end{minipage}\\[1ex]

  % --------- AoA = 5
  \begin{minipage}{\linewidth}
    \centering
    \begin{subfigure}[t]{0.99\linewidth}
      \includegraphics[height=1.6cm]{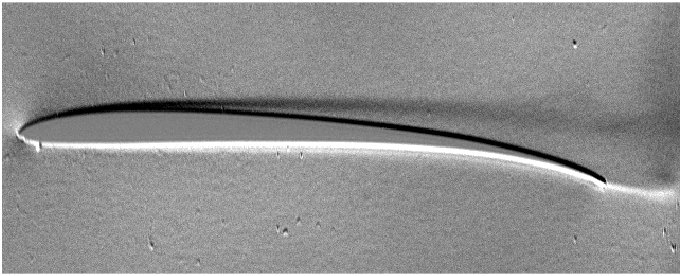}
      \includegraphics[height=1.6cm]{Data/OVERFLOW/Density_gradient/OF-clf5605-q3D-5deg.png}
      \includegraphics[height=1.6cm]{Data/PyFR/Density_gradient/rhograd-5deg-q3D-clf5605.png}
    \end{subfigure}
    \caption*{$\alpha = 5^\circ$}
  \end{minipage}\\[1ex]
  
  % --------- AoA = 5.5
  \begin{minipage}{\linewidth}
    \centering
    \begin{subfigure}[t]{0.99\linewidth}
      \hspace{3.9cm}
      \includegraphics[height=1.6cm]{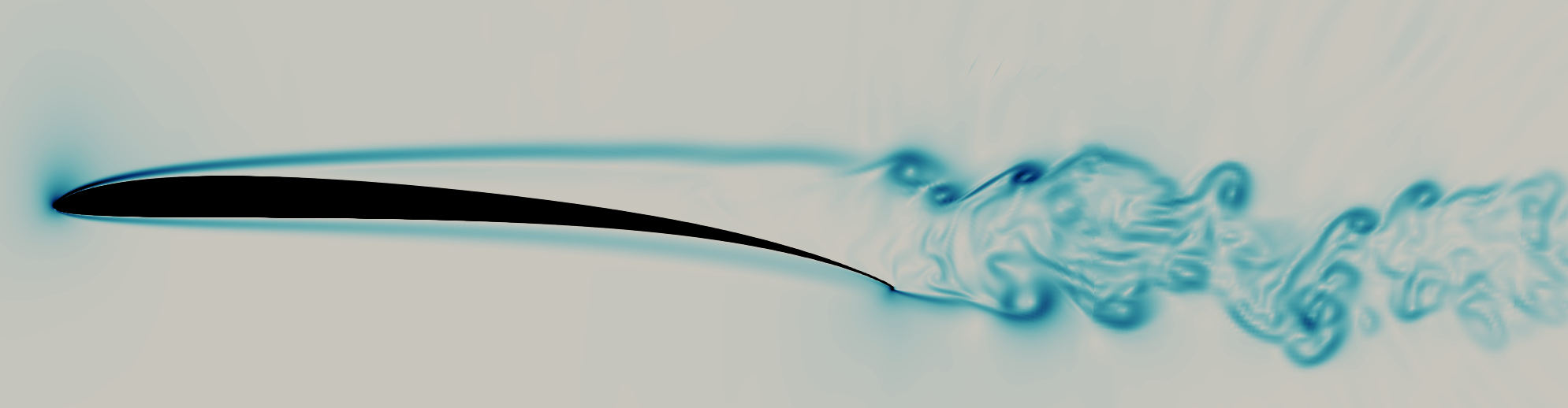}
      \includegraphics[height=1.6cm]{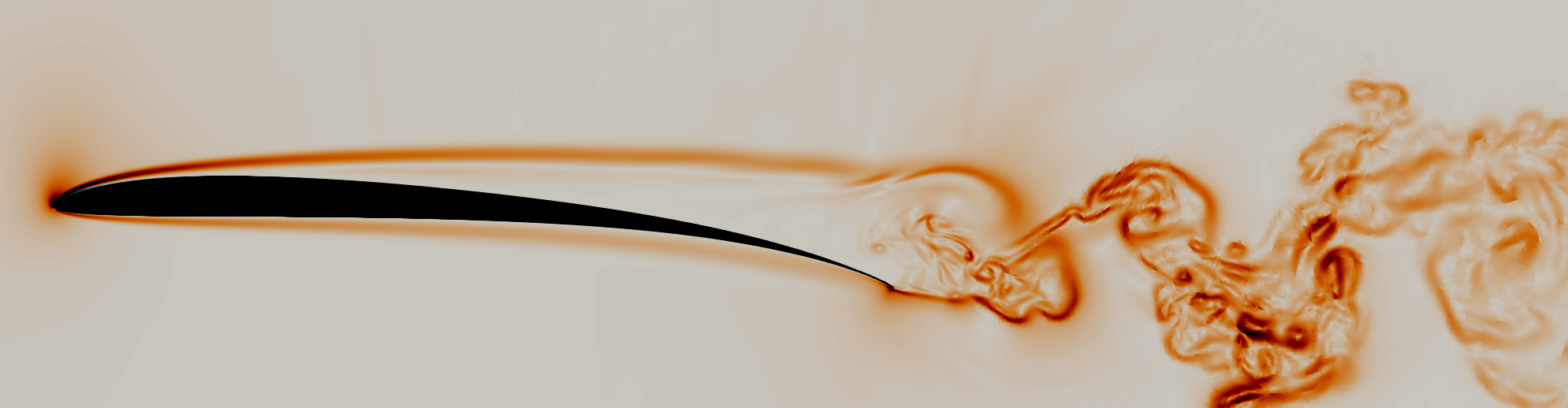}
    \end{subfigure}
    \caption*{$\alpha = 5.5^\circ$}
  \end{minipage}\\[1ex]

  % --------- AoA = 6
  \begin{minipage}{\linewidth}
    \centering
    \begin{subfigure}[t]{0.99\linewidth}
      \includegraphics[height=1.6cm]{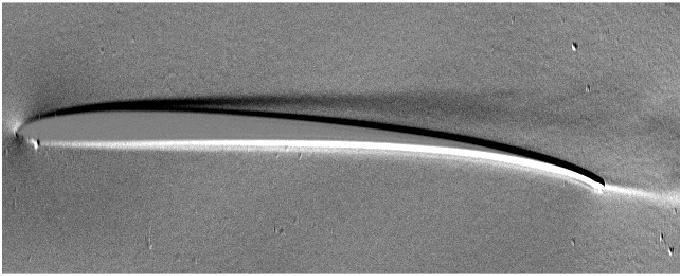}
      \includegraphics[height=1.6cm]{Data/OVERFLOW/Density_gradient/OF-clf5605-q3D-6deg.png}
      \includegraphics[height=1.6cm]{Data/PyFR/Density_gradient/rhograd-6deg-q3D-clf5605.png}
    \end{subfigure}
    \caption*{$\alpha = 6^\circ$}
  \end{minipage}

  \caption{Schlieren images for MWT experiments on model clf5605-us-s (left) and images of instantaneous density gradient magnitude $|\boldsymbol{\nabla}\rho|$ for 3D-SP OVERFLOW (middle) and PyFR (right) on CLF5605 for different $\alpha$.}
  \label{fig:clf5605-schlieren-allAoA}
\end{figure}

\begin{figure}[p]
  \centering

  % --------- AoA = -2
  \begin{minipage}{\linewidth}
    \centering
    \begin{subfigure}[t]{0.99\linewidth}
      \includegraphics[height=1.6cm]{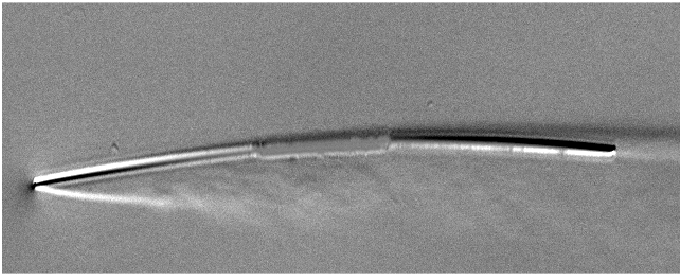}
      \includegraphics[height=1.6cm]{Data/OVERFLOW/Density_gradient/OF-roamx0201-q3D--2deg.png}
      \includegraphics[height=1.6cm]{Data/PyFR/Density_gradient/roamx0201_q3D_-2deg_densitygrad.png}
    \end{subfigure}
    \caption*{$\alpha = -2^\circ$}
  \end{minipage}\\[1ex]

  % --------- AoA = 0
  \begin{minipage}{\linewidth}
    \centering
    \begin{subfigure}[t]{0.99\linewidth}
      \includegraphics[height=1.6cm]{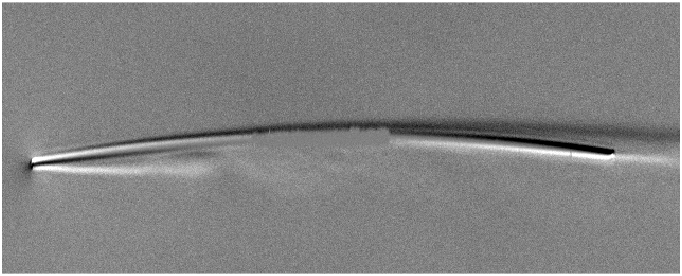}
      \includegraphics[height=1.6cm]{Data/OVERFLOW/Density_gradient/OF-roamx0201-q3D-0deg.png}
      \includegraphics[height=1.6cm]{Data/PyFR/Density_gradient/roamx0201_q3D_0deg_densitygrad.png}
    \end{subfigure}
    \caption*{$\alpha = 0^\circ$}
  \end{minipage}\\[1ex]

  % --------- AoA = 2
  \begin{minipage}{\linewidth}
    \centering
    \begin{subfigure}[t]{0.99\linewidth}
      \includegraphics[height=1.6cm]{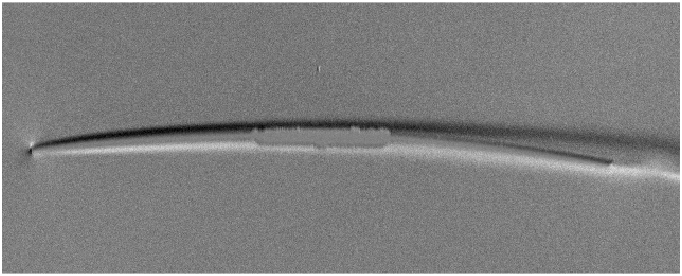}
      \includegraphics[height=1.6cm]{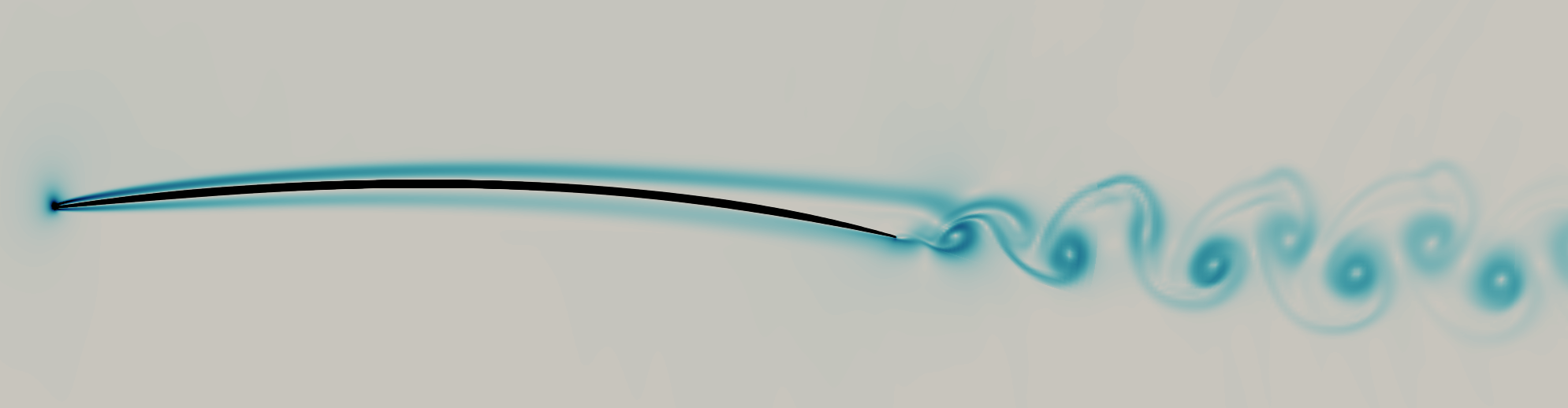}
      \includegraphics[height=1.6cm]{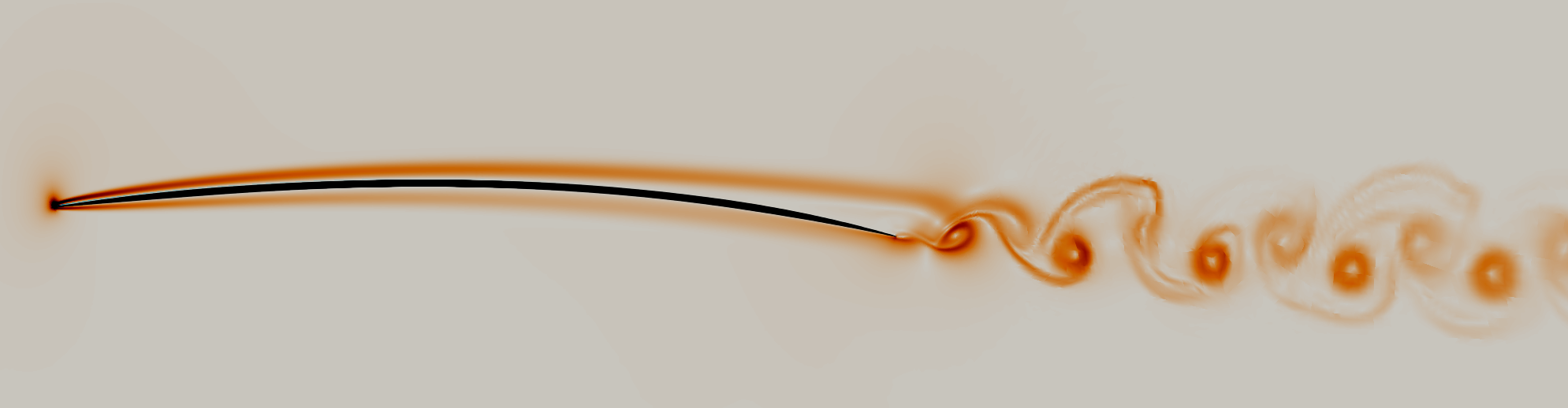}
    \end{subfigure}
    \caption*{$\alpha = 2^\circ$}
  \end{minipage}\\[1ex]

  % --------- AoA = 3
  \begin{minipage}{\linewidth}
    \centering
    \begin{subfigure}[t]{0.99\linewidth}
      \includegraphics[height=1.6cm]{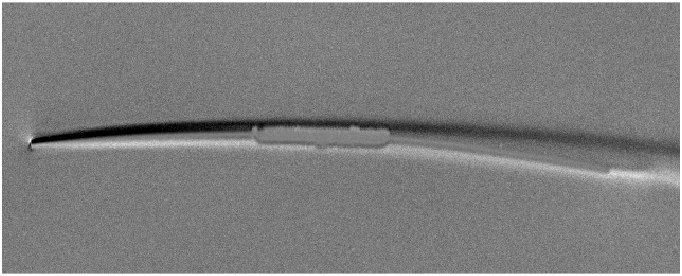}
      \includegraphics[height=1.6cm]{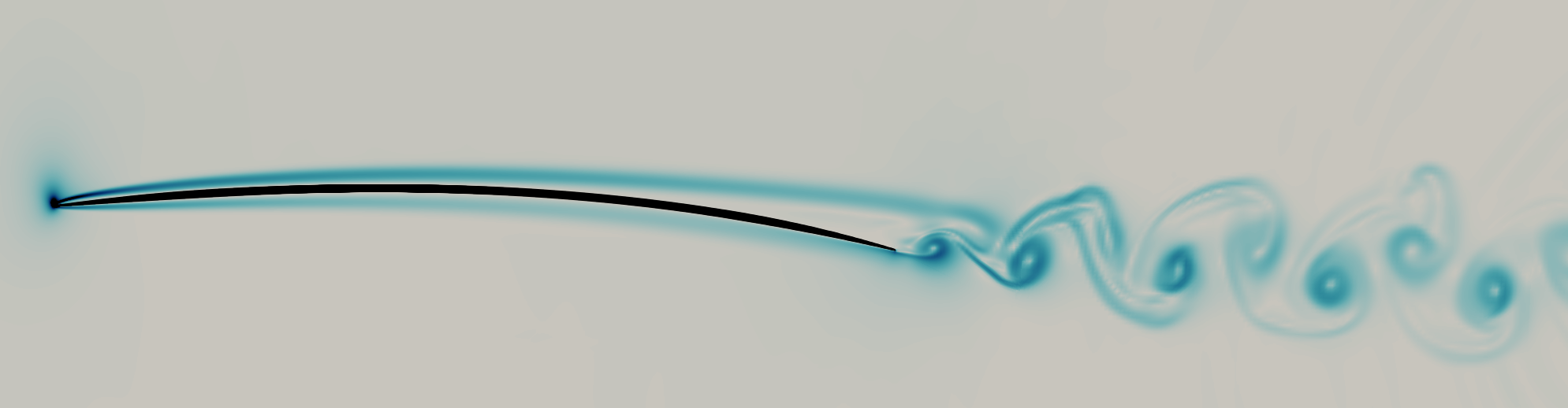}
      \includegraphics[height=1.6cm]{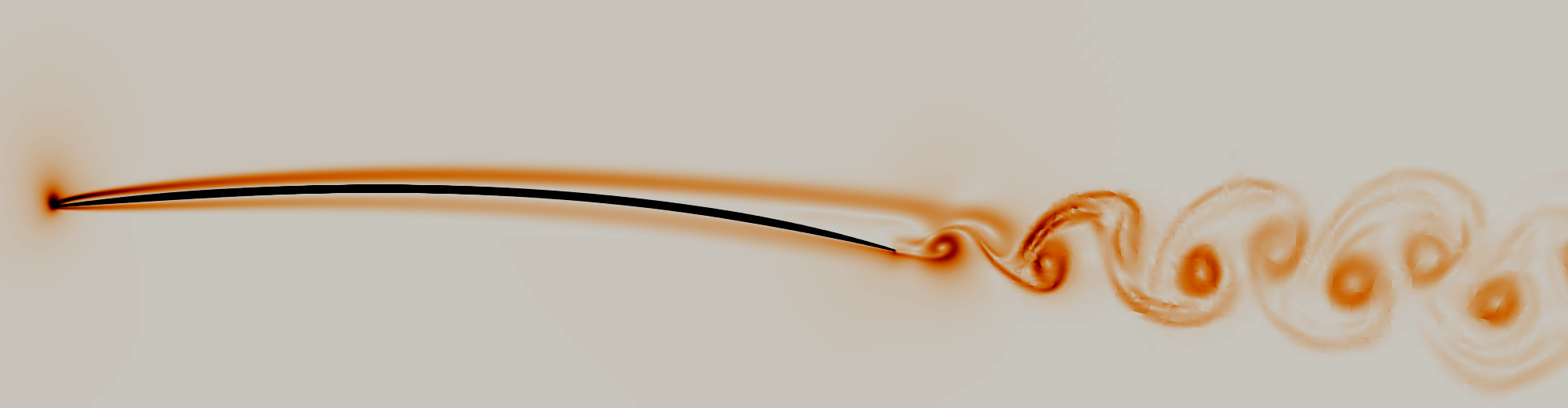}
    \end{subfigure}
    \caption*{$\alpha = 3^\circ$}
  \end{minipage}\\[1ex]

  % --------- AoA = 4
  \begin{minipage}{\linewidth}
    \centering
    \begin{subfigure}[t]{0.99\linewidth}
      \includegraphics[height=1.6cm]{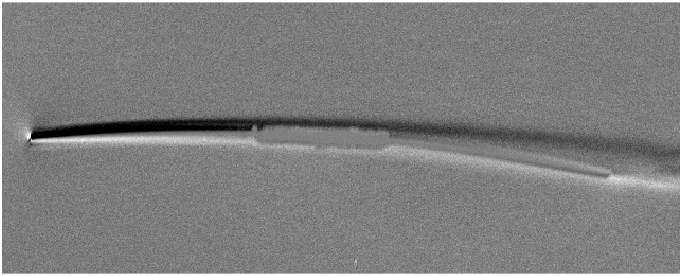}
      \includegraphics[height=1.6cm]{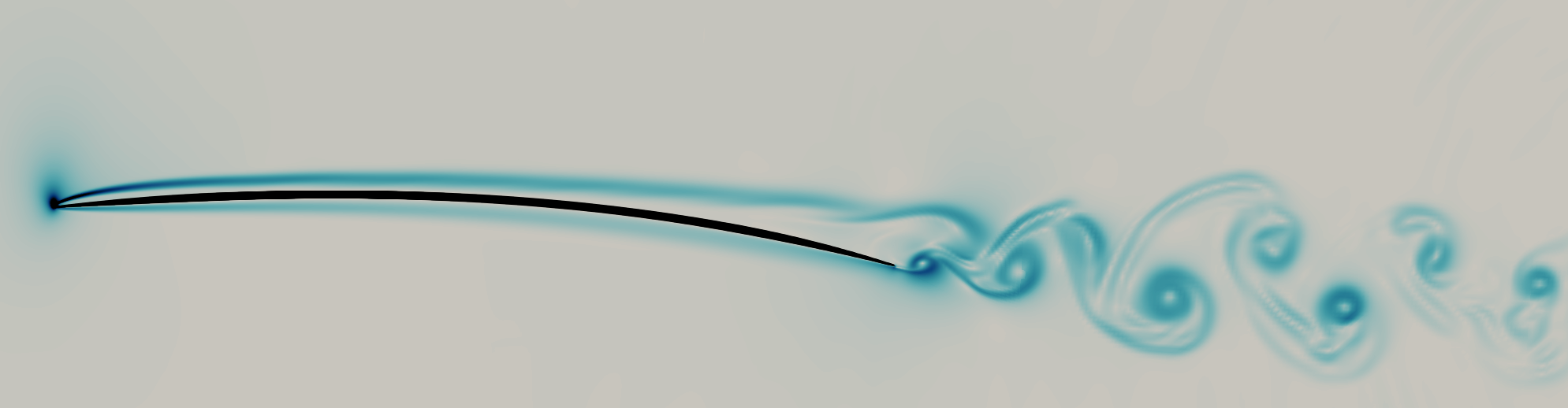}
      \includegraphics[height=1.6cm]{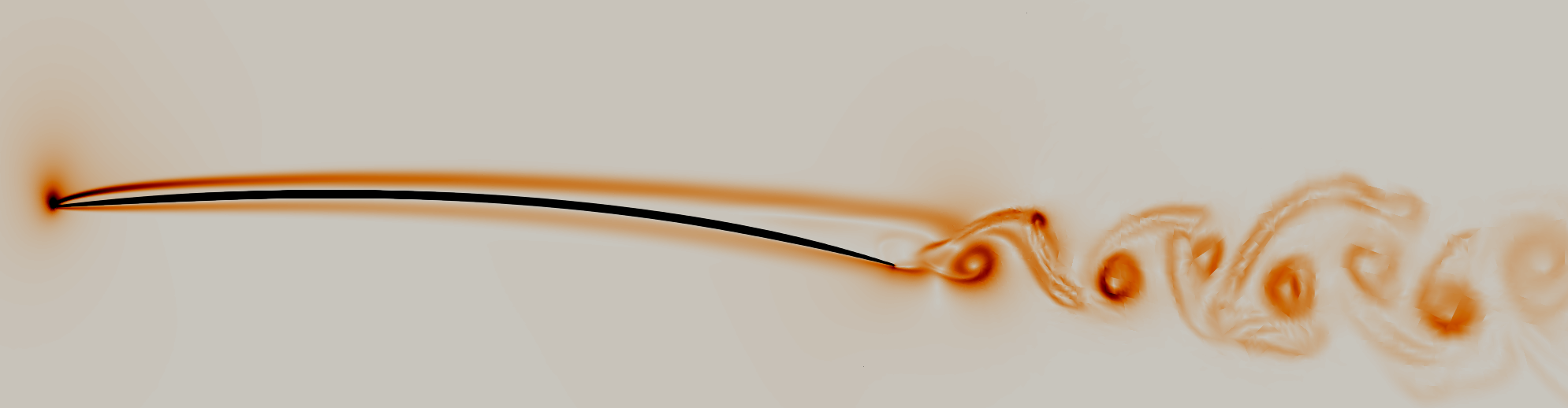}
    \end{subfigure}
    \caption*{$\alpha = 4^\circ$}
  \end{minipage}\\[1ex]
  
  % --------- AoA = 4.5
  \begin{minipage}{\linewidth}
    \centering
    \begin{subfigure}[t]{0.99\linewidth}
      \hspace{3.9cm}
      \includegraphics[height=1.6cm]{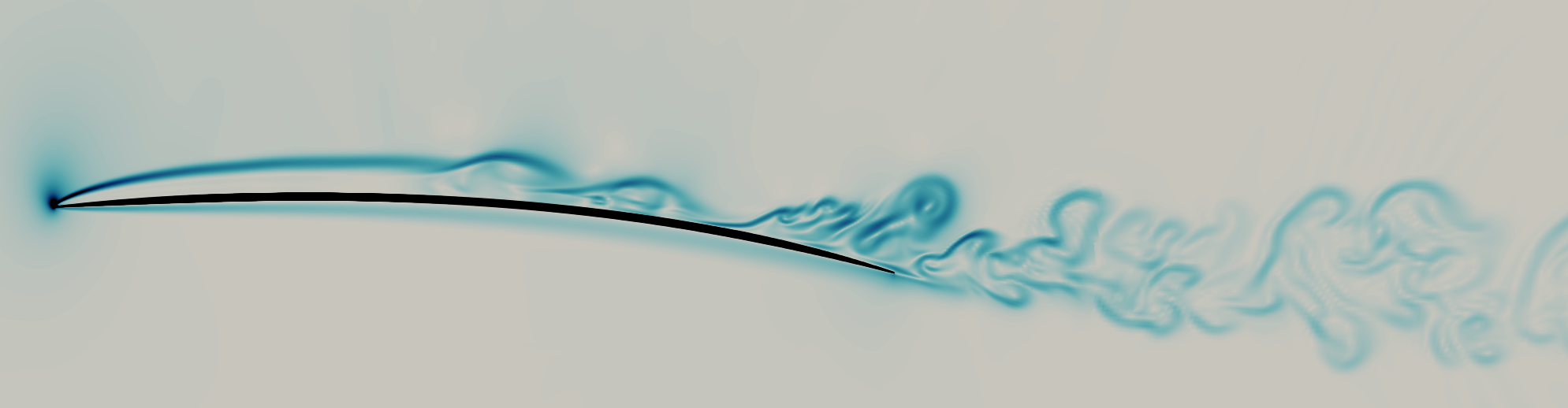}
      \includegraphics[height=1.6cm]{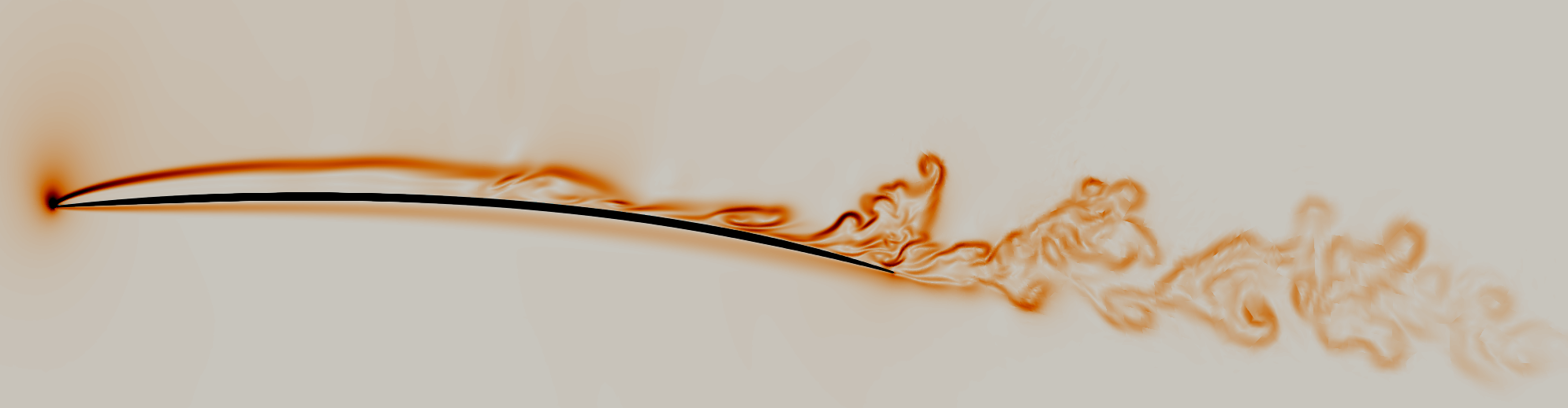}
    \end{subfigure}
    \caption*{$\alpha = 4.5^\circ$}
  \end{minipage}\\[1ex]

  % --------- AoA = 5
  \begin{minipage}{\linewidth}
    \centering
    \begin{subfigure}[t]{0.99\linewidth}
      \includegraphics[height=1.6cm]{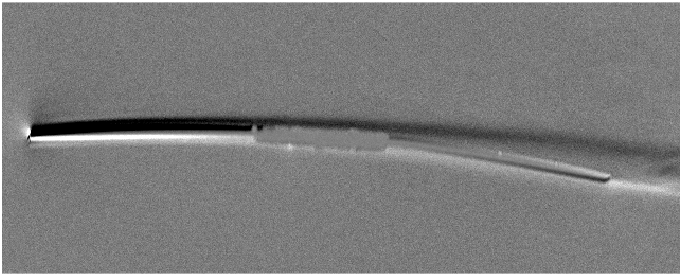}
      \includegraphics[height=1.6cm]{Data/OVERFLOW/Density_gradient/OF-roamx0201-q3D-5deg.png}
      \includegraphics[height=1.6cm]{Data/PyFR/Density_gradient/roamx0201_q3D_5deg_densitygrad.png}
    \end{subfigure}
    \caption*{$\alpha = 5^\circ$}
  \end{minipage}\\[1ex]

  % --------- AoA = 6
  \begin{minipage}{\linewidth}
    \centering
    \begin{subfigure}[t]{0.99\linewidth}
      \includegraphics[height=1.6cm]{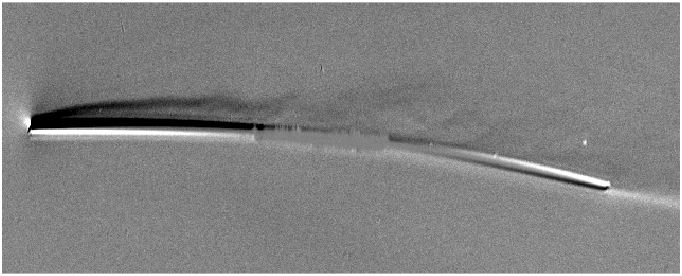}
      \includegraphics[height=1.6cm]{Data/OVERFLOW/Density_gradient/OF-roamx0201-q3D-6deg.png}
      \includegraphics[height=1.6cm]{Data/PyFR/Density_gradient/roamx0201_q3D_6deg_densitygrad.png}
    \end{subfigure}
    \caption*{$\alpha = 6^\circ$}
  \end{minipage}

  \caption{Schlieren images for MWT experiments on model roamx-0201-jp-s (left) and images of instantaneous density gradient magnitude $|\boldsymbol{\nabla}\rho|$ for 3D-SP OVERFLOW (middle) and PyFR (right) on roamx-0201 for different $\alpha$.}
  \label{fig:roamx0201-schlieren-allAoA}
\end{figure}